\documentclass [12pt]{iopart}
\usepackage{graphicx}
\usepackage{epsfig}
\usepackage[numbers,sort&compress,comma]{natbib}
\usepackage{psfrag}

\begin{document}

\article[Tuning microwave losses in superconducting resonators]{Topical Review}{Tuning microwave losses in superconducting resonators}

\author{Alex Gurevich}

\address{Department of Physics and Center for Accelerator Science, Old Dominion University, \\
4600 Elkhorn Avenue, Norfolk, Virginia 23529, United States of America}
\ead{gurevich@odu.edu}


\begin{abstract}
Performance of superconducting resonators, particularly cavities for particle accelerators and micro cavities and thin film resonators for quantum computations and photon detectors has been improved substantially by recent materials treatments and technological advances. As a result, the niobium cavities 
have reached the quality factors $Q\sim 10^{11}$ at 1-2 GHz and 1.5 K and the breakdown radio-frequency (rf) fields $H$
close to the dc superheating field of the Meissner state. These advances raise the question whether the state-of-the-art cavities are close to the  
fundamental limits, what these limits actually are, and to what extent the $Q$ and $H$ limits can be pushed   
by the materials nano structuring and impurity management. These issues are also relevant to many applications using high-Q thin film resonators, 
including single-photon detectors and quantum circuits. This topical review outlines basic physical mechanisms of the rf nonlinear surface 
impedance controlled by quasiparticles, dielectric losses and trapped vortices, as well as the dynamic field limit of the Meissner state. 
Sections cover ways of engineering an optimum quasiparticle density of states and superfluid density to reduce rf losses 
and kinetic inductance by pairbreaking mechanisms related to magnetic impurities, rf currents, and proximity-coupled metallic layers at the surface. 
A section focuses on mechanisms of residual surface resistance which dominates rf  losses at ultra low temperatures. Microwave losses 
of trapped vortices and their reduction by optimizing the concentration of impurities and pinning potential are also discussed.

\end{abstract}


\submitto{Superconductor Science and Technology}
\maketitle

\pagestyle{empty}
\tableofcontents


\section{Introduction}

Low-dissipative superconducting resonators are instrumental in many applications, particularly quantum computations \cite{qq}, single photon detectors \cite{spd1,spd2,spd3,caltech}, quantum memories \cite{qmem} and cavities for particle accelerators  \cite{cav1,cav2,cav3}. They employ fully gapped s-wave superconductors with extremely low electromagnetic losses at temperatures $T\ll T_c$ and microwave or radio frequencies (rf) $\hbar\omega\ll \Delta$, where $\Delta=1.76k_BT_c$ is the superconducting gap and $T_c$ is the critical temperature \cite{mb,nam,mbs,RsNb,hein}. The losses can be further reduced by encapsulating Josephson junctions in resonant cavities to eliminate radiation losses characteristic of thin film structures \cite{caltech}. RF losses are quantified by the quality factor $Q$ proportional to the ratio of electromagnetic energy in the cavity to a power dissipated in the cavity wall \cite{jackson}:
\begin{equation}
Q=\frac{\omega\mu_0\int_V |{\bf H}({\bf r})|^2dV}{\oint_AR_s|{\bf H({\bf r})|^2}dA},
\label{Q}
\end{equation}
where ${\bf H}({\bf r},\omega)e^{-i\omega t}$ is the magnetic field in the cavity mode with the circular eigenfrequency $\omega=2\pi\nu$ and  $R_s$ is the surface resistance.
Generally, $R_s[H({\bf r}),{\bf r}]$ varies along the surface due to trapped vortices, materials and topographic defects and the dependence of $R_s$ on the rf field amplitude $H_a({\bf r})$ for a particular resonant mode.  Since
$\omega\sim c/L$ is inversely proportional to the cavity size $L$, it is convenient to present Eq. (\ref{Q}) in the form
$Q(B_a)=Z_0/\langle R_s\rangle$, where $B_a=\mu_0H_a$,  $Z_0=c\mu_0\alpha$, $c$ is the speed of light, $\langle ... \rangle$ means averaging of $R_s$ over the cavity surface, and $\alpha\sim 1$ is a geometrical factor \cite{cav1}. The scale of $Z_0$ is set by the vacuum impedance $\mu_0c = 377\, \Omega$ .

The best Nb cavities can achieve extremely high quality factors $Q\sim 10^{10}-10^{11}$ corresponding to $R_s\sim 5$ n$\Omega$ and sustain accelerating fields up to $50$ MV/m at $T=1.5-2$ K and $\nu=1.3-2$ GHz \cite{rec1,rec2}. The peak fields $B_a\simeq 200$ mT at the equatorial surface of these cavities approach the thermodynamic critical field $B_c\approx 200$ mT of Nb at 2K \cite{cav1, cav2, cav3}. At $B_a\simeq B_c$ the screening rf current density flowing at the inner cavity surface is close to the depairing current density $J_c\simeq B_c/\mu_0\lambda$ - the maximum dc current density a superconductor can carry in the Meissner state \cite{tinkh,parment,bardeen,maki}, where $\lambda$ is the London penetration depth. Thus, the breakdown fields of the best Nb cavities have nearly reached the dc superheating field $B_s\simeq B_c$ \cite{bean,galaiko,matricon,chapman,catelani,lin}.  The Q factors can be increased by materials treatments such as high temperature ($600-800^o$ C) annealing followed by low temperature ($100-120^o$ C) baking which not only increase $Q(B_a)$ and the breakdown field but also reduce deterioration of $Q$ at high fields \cite{bak1,bak3,bak4,bak5,bak6}. High temperature treatments combined with the infusion of nitrogen, titanium or oxygen can produce an anomalous increase of $Q(B_a)$ with $B_a$ \cite{raise1,raise2,raise3,raise4,raise5,raise6,raise7,raise8,raise9} and  $Q\simeq (3-4)\cdot 10^{11}$ at 1.5 K and $\simeq 5\times 10^{10}$ at 2K and 1.3 GHz \cite{bak6}. These advances raise the question about the fundamental limits of rf losses and the breakdown fields in high-$Q$ resonators and the extent to which these limits can be pushed by surface nano-structuring and impurity management.  In high-power rf applications superconductors with high $T_c$ and $B_c$ (for example Nb$_3$Sn) can only be used in the form of thin film \cite{nb3snn} or multilayer \cite{ml,agsust} coatings of Nb cavities. Thin film superconducting resonators have been widely used in single-photon detectors, quantum memory and quantum computations \cite{qq,spd1,spd2,spd3,caltech,qmem}.    

The fact that microwave losses can be optimized by impurity management can be understood from the BCS theory, according to which a superconductor with no impurities and an ideal surface does not have the lowest surface resistance \cite{mb,nam}. For instance, $R_s$ of a type-II superconductor with a large Ginzburg-Landau (GL) parameter $\kappa=\lambda/\xi\gg 1$ at $T\ll T_c$ and $\hbar\omega\ll k_BT$ has the form  \cite{agsust}
\begin{equation}
R_s=\frac{\mu_0^2\omega^2\lambda^3\Delta}{\rho_sk_BT}\ln\left[\frac{C_1k_BT}{\hbar\omega}\right]\exp\left[-\frac{\Delta}{k_BT}\right]+R_i.
\label{rs0}
\end{equation}
The first term in the r.h.s. of Eq. (\ref{rs0}) is the BCS surface resistance $R_{BCS}(T)$ caused by absorption of low-energy $(\hbar\omega\ll k_BT)$ microwave photons  by a small density of quasiparticles resulting from thermal dissociation of Cooper pairs \cite{caltech,mb,nam,agsust}.  Here $\Delta$ is a superconducting gap, $\rho_s=1/\sigma_s$ is the resistivity in the normal state, $\xi$ is the coherence length, $C_1\approx 9/4$ and $k_B$ and $\hbar$ are the Boltzmann and Planck constants, respectively. The residual surface resistance $R_i$ in Eq. (\ref{rs0}) which remains finite or decreases much slower than $\exp(-\Delta/k_BT)$ at $T\to 0$ has been observed on many superconductors \cite{RsNb,hein} but it is not accounted for in the BCS model.

As follows from Eq. (\ref{rs0}), $R_{BCS}$ can be tuned by the materials disorder. For a spherical Fermi surface, $\Delta$ is independent of the  mean free path on nonmagnetic impurities $l_i$ \cite{tinkh}, so the dependence of $R_{BCS}$ on $l_i$ is determined by the factor $\lambda^3/\rho_s$. Here $\rho_s = p_F/n_0e^2l_i$ and $\lambda\simeq\lambda_0(1+\xi_0/l_i)^{1/2}$ \cite{tinkh}, where $\lambda_0=(m/n_0e^2\mu_0)^{1/2}$ and $\xi_0=\hbar v_F/\pi\Delta$ are the London penetration depth and the coherence length in a clean material at $T=0$, respectively, $p_F$ is the Fermi momentum, $n_0$ is the carrier density, and $e$ is the electron charge. Hence $\lambda^3/\rho_s\propto l_i(1+\xi_0/l_i)^{3/2}$ is minimum at $l_i = 0.5\xi_0$, which translates to the optimum $l_i \approx 20$ nm for Nb. Microscopic calculations \cite{mbs} have shown that $R_{BCS}(l)$ does have a minimum at $l_i\sim \xi_0$ and remains finite in the clean limit $l_i\gg\xi_0$ analogous to the anomalous skin effect in metals \cite{mb}. It turns out that $R_s$ can be further reduced by surface nanostructuring and magnetic impurities  \cite{gk,kg}.

The surface resistance is determined by multiple competing mechanisms so the materials treatments which reduce $R_s$ in a certain region of $T$, $\omega$ and $H$ can in turn increase $R_s$ outside that region.  For instance, Eq. (\ref{rs0}) suggests that s-wave superconductors with no nodes in the quasiparticle gap and the highest $T_c$ would have the lowest $R_{BCS}(T)$ at $T\ll T_c$. Yet the accelerating cavities are built of Nb with its modest $T_c=9.2$ K as compared to $T_c= 18.2$ K of Nb$_3$Sn or $T_c\approx 40$ K of MgB$_2$ or $T_c$ up to 55 K of iron pnictides  \cite{srfmat}. This is because materials with $T_c$ higher than $T_c^{Nb}=9.2$ K are type-II superconductors with the lower critical field $B_{c1}$ smaller than $B_{c1}\simeq 170-180$ mT of Nb which has the highest $B_{c1}$ among all superconductors \cite{srfmat,nb,smat}. This makes Nb best protected against penetration of vortices which can greatly increase $R_s$ at $B_a > B_{c1}$. Alloying a superconductor with impurities to reduce $R_s$ at the optimum $l_i\simeq \xi_0/2$ changes $\lambda\simeq \lambda_0(1+\xi_0/l_i)^{1/2}\to \sqrt{3}\lambda_0$ and $\xi\simeq \sqrt{l_i\xi_0}\to \xi_0/\sqrt{2}$, which decreases $B_{c1}=(\phi_0/4\pi\lambda^2)[\ln(\lambda/\xi)+1/2]$ by more than $50\%$ and reduces the superheating field $B_s$ at which the Meissner state becomes unstable \cite{bean,galaiko,matricon,chapman,catelani,lin}. This illustrates how a lower $R_s$ at weak fields is achieved at the expense of larger $R_s$ at strong fields. 

Reducing microwave losses in the vortex-free Meissner state requires optimization of the quasiparticle $R_s$, while widening the field region of the Miessner state. This could  be achieved by thin film or multilayer coating of Nb resonators with high-$T_c$ but low $B_{c1}$ superconductors \cite{ml}. Here the properties of such materials in the normal state become important. For instance, Nb$_3$Sn has high $T_c$ and low $R_s$ in weak rf fields \cite{nb3sn1,nb3sn2,nb3sn3}, but its thermal conductivity is some 3 orders of magnitude lower than that of clean Nb at 2K \cite{cody}. Thus, despite its better performance at low fields, Nb$_3$Sn is more prone to penetration of vortices and rf overheating which degrades $Q(B_a)$ at higher fields (even a few micron thick Nb$_3$Sn film on the inner surface of the Nb cavity can double the thermal impedance of the cavity wall \cite{cav3}). Another source of rf losses comes from second phase precipitates and weakly coupled grain boundaries in polycrystalline superconductors with short coherence lengths, such as  Nb$_3$Sn \cite{gb1,gb2,gb3} and iron pnictides \cite{gb4}.

At very low temperatures  the residual resistance $R_i$ becomes the dominant source of dissipation. It can result from subgap states at the quasiparticle energies $|\epsilon|<\Delta$ revealed by tunneling measurements \cite{dynes,JZ}. The subgap states have been attributed to multiple mechanisms but none of them has been unambiguously established as a prime source of $R_i$.  Besides the subgap states, $R_i$ has also been attributed to two-level states (TLS) in surface oxides \cite{caltech,tlsr1,tlsr2},  grain boundaries \cite{gb1,gb2,gb3,gb4} and non-superconducting second phase precipitates \cite{cav1,cav2}. A significant contribution to $R_i$ can come from trapped vortices \cite{tf1,tf2,tf3,tf4,tf5,tf6,tf7,tf8,tf9} which appear during the cooldown of a superconductor through $T_c$ in stray magnetic fields. This is also essential for thin films  in quantum circuits \cite{qq,ph1,ph2,ph3} in which vortices can be generated by very weak perpendicular stray fields as $B_{c1}$ is greatly reduced by demagnetizing effects  \cite{ehb,demag}. Spontaneous vortex-antivortex pairs and vortex loops can appear upon cooling with a finite temperature ramp rate \cite{kz1,kz2} or be produced by thermal fluctuations \cite{kz3}. Because of the extremely small $R_{BCS}(T)$ the losses in high-$Q$ resonators can be dominated by a small number of trapped vortices oscillating under rf field. Trapped vortices can bundle together, forming hotspots which have been revealed both in Nb cavities \cite{bak1,tmap,gc2} and thin film structures \cite{vortrf,anlage}. Reducing $R_i$ at low rf fields involves effective pinning of trapped vortices without degradation of quasiparticle and TLS surface resistance \cite{ph1,ph2,ph3}.   

This paper gives an overview of basic physical mechanisms which can control both the quasiparticle and residual surface resistance, including subgap states, TLS, nonlinear current pairbreaking, trapped vortices and a dynamic superheating field which determines the field limit of a nonequilibrium Meissner state.  Reducing $R_s$ by engineering an optimum quasiparticle density of states using pairbreaking mechanisms, such as magnetic impurities, rf currents and proximity-coupled metallic overlayers are discussed. Furthermore, mitigation of microwave losses by surface nanostructuring, impurity management and optimization of pinning of trapped vortices are considered. 
This review primarily focuses on superconducting parameters which can be tuned to enhance the performance of high-Q resonators while not addressing specific atomic mechanisms by which these parameters are affected by materials treatments. 
 
\section{Complex conductivity of superconductors}

The electromagnetic response of a superconductor in a weak field is described by the following relation for the Fourier components of the current density ${\bf J}({\bf k}, \omega)$ induced by the magnetic vector potential ${\bf A}({\bf k}, \omega)$ (in the gauge $\mbox{div}{\bf A}=0$) \cite{mb,tinkh}
\begin{equation}
{\bf J}({\bf k}, \omega) = -K({\bf k},\omega){\bf A}({\bf k}, \omega),
\label{em}
\end{equation}
where ${\bf B}=\nabla\times{\bf A}$, and the complex electromagnetic kernel $K(k,\omega)$ depends on the circular frequency $\omega = 2\pi\nu$ and the wave vector ${\bf k}$ of the rf field. The real part of $K$ describes the Meissner effect caused by the superconducting condensate, so that $K(0,0) = 1/\mu_0\lambda^2$ in the static local limit. The imaginary part of $K$ describes dissipative processes caused by quasiparticles driven by rf field. The BCS theory gives general formulas for $K(k,\omega,T,l_i)$ which also depends on the mean free path $l_i$ due to scattering of electrons on impurities \cite{mb,kopnin}. Although the BCS model captures the fundamentals of electrodynamics of superconductors, it can hardly be used for quantitative calculations of $K(k,\omega,T,l_i)$ for Nb or Pb or Nb$_3$Sn in which the electron-phonon coupling is not weak \cite{carbotte}.  Superconductors with strong electron-phonon interaction are described by the Eliashberg theory \cite{carbotte} in which $K(k,\omega,T,l_i)$ was obtained in Ref. \cite{nam}. Microwave conductivity of different superconductors described by the Eliashberg theory was calculated in Refs. \cite{eli1,eli2,eli3,eli4}.

The power $P$ dissipated per unit surface area of a superconductor is determined by the real part of the surface impedance $Z(\omega)=R_s+iX$: 
\begin{equation}
P=\frac{R_s}{2} H_a^2,
\label{p}
\end{equation}
where $H_a$ is the amplitude of the rf field $H(t)=H_a\cos\omega t$ at the surface. The surface resistance $R_s$ can be expressed in terms of $K(k,\omega,T,l_i)$ by  integral relations \cite{mb,nam,mbs,eli4} which depend on the way the electrons are scattering by the surface (specular or diffusive). Using these results, $Z(\omega,T)$ can be calculated numerically for arbitrary $T$, $\omega$ and $l_i$ for a particular material like Nb \cite{eli2,eli3}.  The situation simplifies at low temperatures $k_BT\ll \Delta$ and frequencies much lower than the gap frequency $\nu \ll \nu_g=\Delta/\pi\hbar$ at which high-$Q$ resonators operate. For instance, $\nu_g$(GHz) = $74T_c$ (K)$= 680$ GHz for Nb is much larger than the rf frequency domain $0.1-5$ GHz in which the absorption of single photons cannot break the Cooper pairs. In this case the superconducting condensate follows nearly instantaneously to the driving rf field, while the dissipative current  of thermally-activated quasiparticles can have much longer relaxation times determined by inelastic scattering on phonons  \cite{kopnin}.

In the most transparent case of $\lambda\gg \xi$, the rf magnetic and electric fields are confined in the layer of thickness $\simeq \lambda$ at the surface, where $\lambda$ is the static London penetration depth. Dissipation comes from a small "normal" component of the current density $J_n$ oscillating in-phase
with the driving electric field. Both in-phase and out-of phase components of ${\bf J}({\bf r},t)$ can be calculated from the Maxwell equations
combined with Eq. (\ref{em}), which generally gives a nonlocal integral relation between ${\bf J}({\bf r},\omega)$ and ${\bf A}({\bf r},\omega)$ in the coordinate space \cite{mb}. In the limit of  $\lambda\gg\xi$, the relation between ${\bf J}({\bf r},\omega)$ and ${\bf A}({\bf r},\omega)$ simplifies to the local "ohmic" form with a frequency-dependent complex conductivity $\sigma(\omega)=\sigma_1(\omega)-i\sigma_2(\omega)$:
\begin{equation}
{\bf J({\bf r}\omega)}=(\sigma_1-i\sigma_2){\bf E}({\bf r},\omega).
\label{sig}
\end{equation}
The reactive part $\sigma_2$ responsible for the Meissner effect is given by
\begin{equation}
\sigma_2 = 1/\omega\mu_0\lambda^2.
\label{s2}
\end{equation}
The surface impedance is calculated using the standard formula of the electromagnetic theory \cite{jackson} in the limit of weak Ohmic dissipation,
$\sigma_1\ll\sigma_2$:
\begin{equation}
Z=\left[\frac{i\mu_0\omega}{\sigma_1-i\sigma_2}\right]^{1/2}\simeq \frac{\mu_0\omega\sigma_1}{2\sigma_2^2}\sqrt{\frac{\sigma_2}{\mu_0\omega}}+i\sqrt{\frac{\mu_0\omega}{\sigma_2}}.
\label{zz}
\end{equation}
Substituting here Eq. (\ref{s2}) yields
\begin{equation}
2R_s=\mu_0^2\omega^2\lambda^3\sigma_1(\omega),\qquad X_s = \mu_0\omega\lambda.
\label{xr}
\end{equation}
The dissipative conductivity $\sigma_1(\omega)$ evaluated from the Mattis-Bardeen theory \cite{mb,nam} at $\hbar\omega\ll \Delta$, $T\ll T_c$ and $l_i<\xi_0$ 
takes the form \cite{caltech,agsust}:
\begin{equation}
\sigma_1=\frac{4\sigma_n\Delta}{\hbar\omega}\sinh\!\left[\frac{\hbar\omega}{2k_BT}\right]\!K_0\!\left[\frac{\hbar\omega}{2k_BT}\right]\!e^{-\Delta/k_BT},
\label{sigma1}
\end{equation}
where $K_0(x)$ is a modified Bessel function. At $\nu=1-2$ GHz and $T=1-2$ K  
we have $\hbar\omega/2k_BT\sim 10^{-2}$ so Eq. (\ref{sigma1}) can be expanded in $\hbar\omega/2k_BT\ll 1$ using $K_0(x)\simeq \ln(2/x)-\gamma_E$ at $x\ll 1$, where 
$\gamma_E=0.577$ is the Euler constant. Hence, $\sigma_1\simeq (2\sigma_n\Delta/k_BT)\ln(C_1k_BT/\hbar\omega)e^{-\Delta/k_BT}$, where $C_1=4e^{-\gamma_E}\approx 9/4$. This $\sigma_1$ combined with Eq. (\ref{xr}) yield the BCS surface resistance in Eq. (\ref{rs0}). 

In type-I superconductors such as Al, Sn, Ta or Pb the electromagnetic response becomes nonlocal and the screening current density does not decay exponentially 
over the London penetration depth $\lambda$. In the extreme Pippard limit $\xi_0\gg\lambda$, the effective field penetration depth 
$\tilde{\lambda}\simeq 0.65(\lambda^2\xi_0)^{1/3}$ can exceed $\lambda$ \cite{tinkh}. For instance, in clean Al with $\xi_0=1500$ nm and 
$\lambda=16$ nm we get $\tilde{\lambda}\approx 47$ nm, whereas Sn with $\xi_0=230$ nm and $\lambda=34$ nm has  
$\tilde{\lambda}\approx 42$ nm. Calculations of the surface impedance of Al and Sn films using the full 
Mattis-Bardeen electromagnetic kernel $K({\bf k},\omega)$ \cite{mb} has shown that the nonlocality makes $R_s$ dependent  
on the film thickness $d$ up to $d\sim \xi_0$ \cite{nimp}. 

\subsection{Subgap states}

\begin{figure}
\centering
\includegraphics[width=8cm]{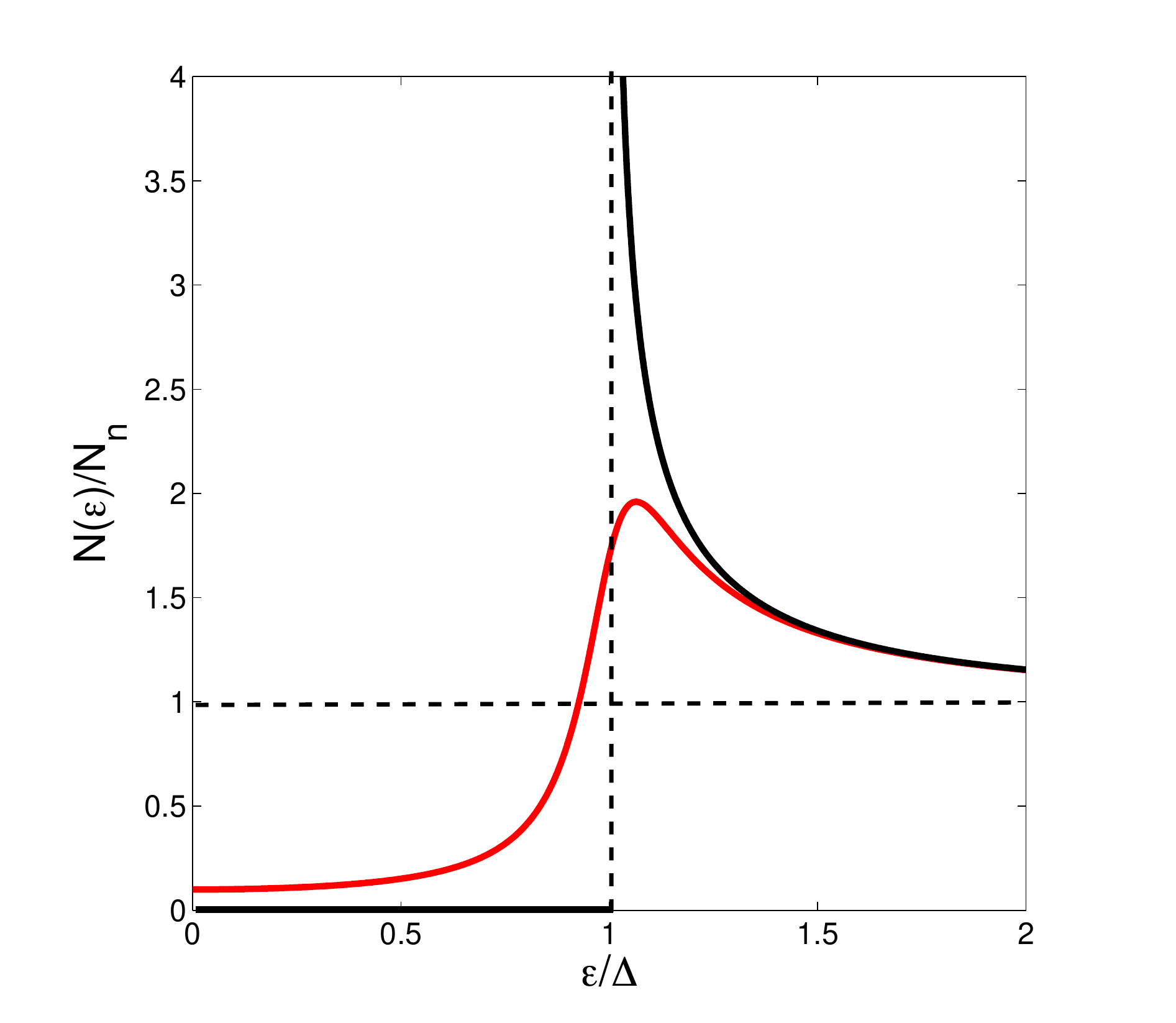}
\caption{$N(\epsilon)$ in the BCS model (black) and $N(\epsilon)$ described by Eq. (\ref{NE}) for $\Gamma/\Delta=0.2$ (red).}
\label{fig1}
\end{figure}

In the BCS model the quasiparticle density of states (DOS)   
$N(\epsilon)$ vanishes at energies $|\epsilon|<\Delta$, even if weak scattering on nonmagnetic impurities present  \cite{tinkh,kopnin,balatski}. It is the feature  of $N(\epsilon)$ which ensures the exponentially small $R_{BCS}(T)$ and $R_i = 0$ in Eq. (\ref{rs0}). Yet many tunneling measurements of $N(\epsilon)$ \cite{dynes,JZ} have shown that $N(\epsilon)$ differs from the idealized DOS which diverges at $\epsilon =\Delta$ and vanishes at $\epsilon<\Delta$, as shown in Fig. \ref{fig1}. In the observed $N(\epsilon)$ the gap singularities at $\epsilon=\Delta$ are smeared out and subgap states with a finite $N(\epsilon)$ appear at $\epsilon<\Delta$. Such DOS has been often described by the phenomenological Dynes model \cite{dynes} in which
\begin{equation}
N(\epsilon) = \mbox{Re}\frac{N_n(\epsilon-i\Gamma )}{\sqrt{(\epsilon-i\Gamma )^2 -  \Delta^2}}.
\label{NE}
\end{equation}
Here the damping parameter $\Gamma$ quantifies a finite lifetime of quasiparticles $\sim \hbar/\Gamma$, resulting in a finite DOS $N(0)\simeq \Gamma N_n/\Delta$ at the Fermi level.  Tunneling conductance measurements on Nb \cite{anlm1,anlm2} and Nb$_3$Sn \cite{anlm3} have indeed revealed a finite $N(\epsilon)$ at $\epsilon <\Delta$ (a review of tunneling measurements of $N(\epsilon)$ and applications of Eq. (\ref{NE}) are given in Ref. \cite{JZ}).   
The physics of subgap states is not fully understood (see e.g., reviews \cite{JZ,feigel} and the references therein). Many mechanisms of subgap states have been suggested in the literature, including inelastic scattering of quasiparticles on phonons \cite{inelast}, Coulomb correlations \cite{coulomb}, anisotropy of the Fermi surface \cite{anis}, inhomogeneities of the BCS pairing constant \cite{larkin}, magnetic impurities \cite{balatski}, spatial correlations in impurity scattering \cite{balatski,meyer}, diffusive surface scattering \cite{arnold} or quasiparticles trapped by inhomogeneities of $\Delta({\bf r})$ \cite{qptls}.  The phenomenological Eq. (\ref{NE}) captures the observed broadening of the DOS peaks at $\epsilon\approx \Delta$ but it can hardly describe  low-energy tails in $N(\epsilon)$ due to, for example, energy-dependent electron-phonon relaxation times \cite{kaplan}. Details of exponential or power-law energy tails in $N(\epsilon)$ at $|\epsilon|\ll \Delta$ can be essential for the behavior of $R_i(T)$ at ultra low temperatures \cite{qptls}. Yet the widely used Eq. (\ref{NE}) in which all microscopic mechanisms are included in a single parameter $\Gamma$ is rather useful to address qualitative effects of the DOS broadening on $R_s$. 

The broadening of the DOS gap peaks reduces $T_c$ and $\Delta$ in a way similar to the pairbreaking effect of magnetic impurities \cite{balatski}. For instance, $T_c$  decreases with $\Gamma$ and vanishes at $\Gamma_c=\Delta_0/2$, where $\Delta_{0}$ is the gap at $T=0$ and $\Gamma=0$ ~\cite{gk,hlub}. For weak DOS broadening $\Gamma\ll k_BT_{c0}$, we have \cite{gk,hlub}  
\begin{equation}
T_{c}=T_{c0}-\frac{\pi\Gamma}{4},
\label{tcc} 
\end{equation}
\begin{equation}
\Delta\simeq\Delta_{0}-\Gamma-\frac{\pi^{2}\Gamma T^2}{6\Delta_0^{2}}, \quad T\ll T_c.
\label{del0}
\end{equation}
A finite DOS at $\epsilon=0$ in the Dynes model yields a quadratic temperature dependence of $\Delta(T)$ instead of the BCS behavior of $\Delta(T)\simeq \Delta_0-\sqrt{2\pi k_BT\Delta_0}\exp(-\Delta_0/k_BT)$ at $T\ll T_{c}$ \cite{kopnin}. The DOS broadening increases the magnetic penetration depth $\lambda$ at $T\ll T_c$ in the dirty limit as following \cite{gk,hlub}:
\begin{equation}
\frac{1}{\lambda^{2}}=\frac{2\mu_{0}\Delta}{\hbar\rho_{n}}\tan^{-1}\frac{\Delta}{\Gamma},
\label{lam0}
\end{equation}
This reproduces the BCS result $\lambda^2=\hbar\rho_n/\pi\mu_0\Delta$ at $T=\Gamma=0$ \cite{kopnin}.
 The dependence of $\lambda$ on the mean free path $l_i$ at $T=\Gamma=0$ is given by  \cite{nam}:
\begin{eqnarray}
\frac{1}{\lambda^2}=\frac{1}{a\lambda_0^2}\left[\frac{\pi}{2}-\frac{\cos^{-1}(a)}{\sqrt{1-a^2}} \right],\quad a < 1
\label{l1} \\
\frac{1}{\lambda^2}=\frac{1}{a\lambda_0^2}\left[\frac{\pi}{2}-\frac{\cosh^{-1}(a)}{\sqrt{a^2-1}} \right], \quad a > 1
\label{l2}
\end{eqnarray}
where $\lambda_0=(m/\mu_0n_0e^2)^{1/2}$ and the parameter $a=\pi\xi_0/2l_i$ quantifies scattering on impurities, so that $a\ll 1$ and $a\gg 1$ correspond to the clean and the dirty limits, respectively. 
As $l_i$ decreases, $\lambda(l_i)$ increases  from $\lambda_0$ in the clean limit $l_i \gg \xi_0$ to $\lambda \simeq \lambda_0 (\xi_0/l_i)^{1/2}$ in the dirty limit, $l_i < \xi$. In turn, the coherence length $\xi(l_i)$ decreases from $\xi_0$ at $l_i \gg \xi_0$ to $\xi\simeq \sqrt{l_i\xi_0}$ if $l_i \ll\xi_0$ \cite{tinkh}. In the BCS model weak scattering of electrons on nonmagnetic impurities does not affect  $\Delta(T)$, $T_c$ and $B_c(T)$ of a superconductor with a spherical Fermi surface \cite{tinkh,balatski}. This is no longer the case for strong impurity scattering and anisotropic Fermi surface \cite{balatski}. Strong electron-phonon coupling gives rise to a power-law temperature dependence of $\lambda(T)$ at $T\ll T_c$ due to the contribution of low-energy phonons \cite{elil}.     

\subsection{Residual surface resistance}

The observed temperature dependence of $R_s(T)$ of s-wave superconductors follows Eq. (\ref{rs0}) with $\Delta = c_1 k_BT_c$,
where $c_1$ is slightly higher than the BCS prediction $c_1 = 1.76$ due to the effects of strong electron-phonon coupling  \cite{carbotte}.  
Many extrinsic mechanisms of  the 
residual surface resistance have been pointed out in the literature, including lossy oxides or metallic hydrides in Nb \cite{cav1,cav2}, trapped vortices \cite{tf1,tf2,tf3,tf4,tf5,tf6,tf7,tf8,tf9}, grain boundaries \cite{gbi1,gbi2,gbi3}, or two-level states \cite{caltech,tls1,tls2}. The effect of metallic hydrides has been well documented for Nb cavities \cite{nbh1,nbh2} and films \cite{film1,film2,film3}. Formation of metallic hydride precipitates from over-saturated solid solution of H interstitials is characteristic of Nb \cite{hydrides}. In films $R_i$ can be increased by the surface roughness and absorption of noble gases  \cite{film1,film2,film3}.  These intrinsic factors can be mitigated by high temperature (600-800$^o$) annealing  \cite{cav2} of by pushing trapped vortices out by  temperature gradients \cite{tf3,gc2,broom,laser}. 

\begin{figure}
\centering
\includegraphics[width=11cm]{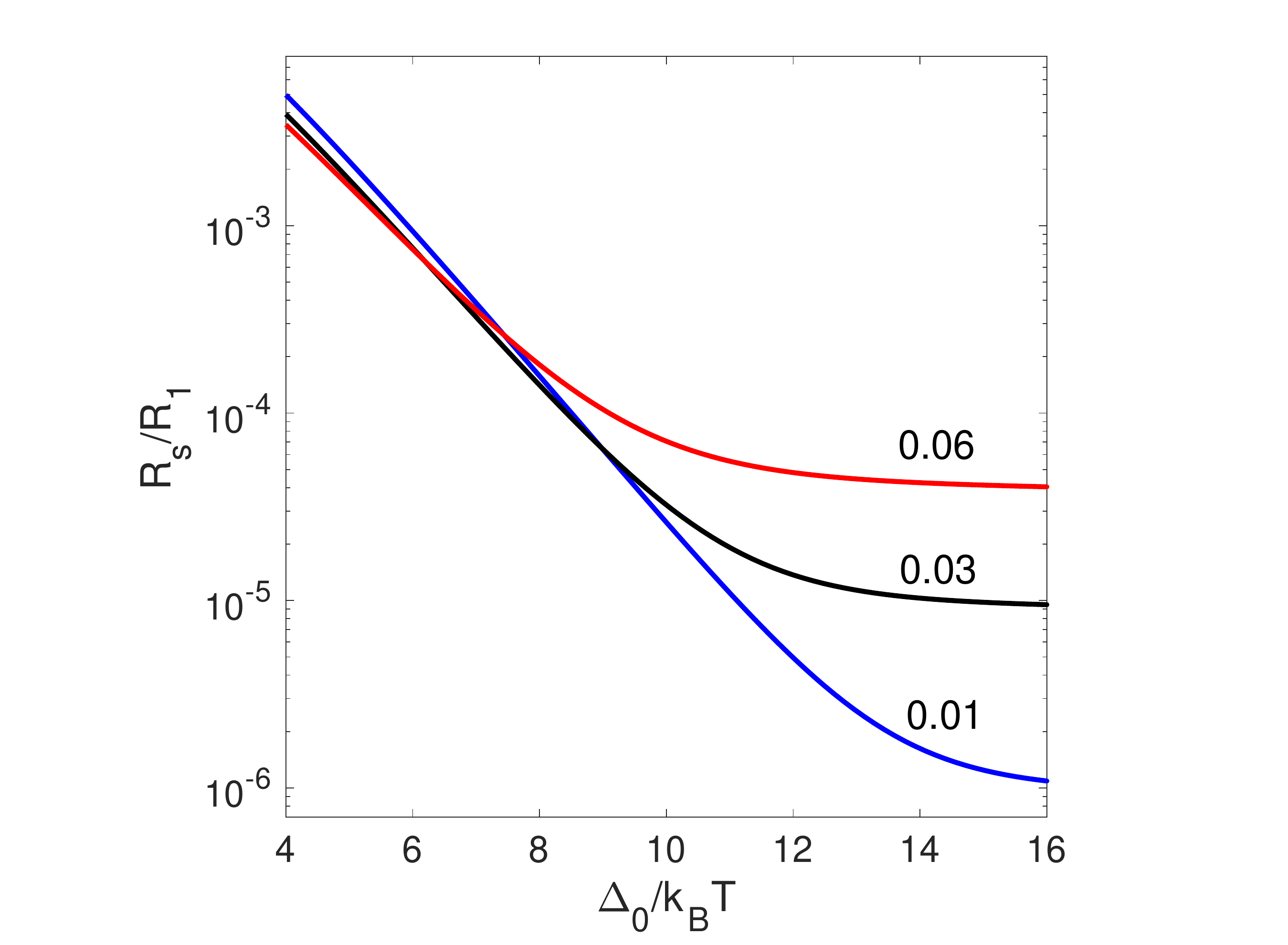}
\caption{Arrhenius plot for $R_s(T)$ calculated from Eq. (\ref{sg}) for Nb with $\Delta = 17.5$ K at 1.5 GHz and different ratios of $\Gamma/\Delta$.
Reproduced from Ref. \cite{gk}.}
\label{fig2}
\end{figure}

A residual resistance produced by the subgap states can be obtained by incorporating the Dynes DOS into the BCS expression for $R_s(T)$ \cite{agsust,gk}:
\begin{eqnarray}
R_s=R_1\sinh\left(\frac{\hbar\omega}{2k_BT}\right)\int_{-\infty}^\infty 
\frac{[n(\hbar\omega+\epsilon)n(\epsilon)+m(\hbar\omega+\epsilon)m(\epsilon)]d\epsilon}{\cosh(\epsilon/2k_BT)\cosh[(\epsilon+\hbar\omega)/2k_BT]},
\label{sg} \\
n(\epsilon)=\mbox{Re}\frac{(\epsilon-i\Gamma )}{\sqrt{(\epsilon-i\Gamma )^2 -  \Delta^2}},\qquad m(\epsilon)=\mbox{Re}\frac{\Delta}{\sqrt{(\epsilon-i\Gamma )^2 -  \Delta^2}},
\label{nm}
\end{eqnarray}
where $R_1=\mu_0^2\omega\lambda^3/4\hbar\rho_s$ and $n(\epsilon)$ and $m(\epsilon)$ are real parts of retarded normal and anomalous quasiclassical Green's functions \cite{kopnin}. Equation (\ref{sg}) with $\Gamma=0$ reproduces $R_{BCS}(T)$ in Eq. (\ref{rs0}). Complex conductivity in the Dynes model and arbitrary mean free path was calculated in Ref. \cite{hlub}. In additional to the BCS part $R_{BCS} \propto \exp(-\Delta/k_BT)$, Eq. (\ref{sg}) accounts for a residual $R_i(T)$ that is not exponentially small at $T\ll T_c$  \cite{gk}:
\begin{equation}
R_i(T)=\frac{\mu_0^2\omega^2\lambda^3 \Gamma^2}{2\rho_n(\Delta^2+\Gamma^2)}\biggl[1+\frac{4\pi^2k_B^2T^2\Delta^2}{3(\Delta^2+\Gamma^2)^2}\biggr],\quad k_BT<\Gamma
\label{rit} 
\end{equation}
where $\Delta(\Gamma)$ is given by Eq. (\ref{del0}). According to Eq. (\ref{rit}), getting $R_i\simeq 4$ n$\Omega$ at 1.5 GHz for Nb with   
$\lambda = 40$ nm and $\rho_s=1$ n$\Omega\cdot$m requires $\Gamma \simeq 0.03\Delta$. The finite $R_i$ at $T=0$ results from the Dynes model 
assumption that $\Gamma$ is independent of $\epsilon$ and $T$. In the case of a power law  
$N_s(\epsilon)$ at $\epsilon\ll \Delta$ (see, e.g., Ref. \cite{feigel,qptls}), one could expect a power law $R_s\propto T^n$ at ultralow temperatures.  
Shown in Fig. \ref{fig2} is the Arrhenius plot of $R_s(T)$ calculated from Eqs. (\ref{sg})-(\ref{nm}) for different ratios of $\Gamma/\Delta_0$. At higher temperatures $\ln R_s(T)$ follows the BCS linear dependence on $\Delta/k_BT$ and levels off as $T$ decreases. The latter is indicative of a residual resistance resulting from the broadening of the DOS gap peaks. One can see that at low temperatures $R_s(T,\Gamma)$ increases as $\Gamma$ increases but at higher $T$ this trend reverses and $R_s(T,\Gamma)$ decreases as $\Gamma$ increases.

The nonmonotonic dependence of $R_s$ on $\Gamma$ shown in Fig. \ref{fig2} can be understood by noticing that $R_i$ in Eq. (\ref{rit}) increases with $\Gamma$.  In turn, the reduction $R_s$ with $\Gamma$ at higher $T$ comes from the reduction of the logarithmic factor in Eq. (\ref{rs0}). In the BCS model with $\Gamma=0$ the factor $\ln(C_1k_BT/\hbar\omega)$ in $R_{BCS}$ occurs because the square root singularities in the DOS at $\epsilon=\Delta$ and $\epsilon=\Delta\pm\hbar\omega$ merge at $\omega\to 0$ and produce a pole in the integrand of Eq. (\ref{sg}). However,  if $\Gamma>0$, the gap singularities in $n(\epsilon)$ and $m(\epsilon)$ broaden into peaks of width $\sim\Gamma$ cutting off the divergence in Eq. (\ref{sg}) at $\omega=0$. At $\Gamma >\hbar\omega$ but $\Gamma\ll k_BT$ integration of these peaks in Eq. (\ref{sg}) at $\epsilon\simeq\Delta$ yields a logarithmic term similar to that in Eq. (\ref{rs0}) but with the energy cutoff $\Gamma$ instead of $\hbar\omega$.  The smearing of the DOS gap peaks can be qualitatively taken into account by the following replacement in Eq. (\ref{rs0}):
  \begin{equation}
\ln\frac{k_BT}{\hbar\omega} \to \ln\frac{k_BT}{\Gamma}.
\label{rule}
\end{equation}   
Hence, broadening the peaks in the DOS reduces $R_s(T)$ at temperatures at which $R_i$ is negligible. At 2 K and 1 GHz we have $k_BT/\hbar\omega \sim 10^2$, so even weak broadening with $\Gamma = 0.02\Delta$ causes a two-fold reduction of $R_s$. Broadening the peaks in the DOS can be used to reduce the rf losses by pairbreaking mechanisms, as discussed below.  

\section{Reduction of $R_s$ by pairbreaking mechanisms}

In this section we discuss the ways by which $R_s$ can be reduced by tuning the DOS by pairbreaking mechanisms related to magnetic impurities, proximity-coupled metallic overlayers and rf current. The latter results in a microwave suppression of $R_s(B_a)$ and its nonmonotonic dependence on the rf field amplitude.  

\subsection{Magnetic impurities}

\begin{figure}[ht]
\centering
\includegraphics[width=11cm]{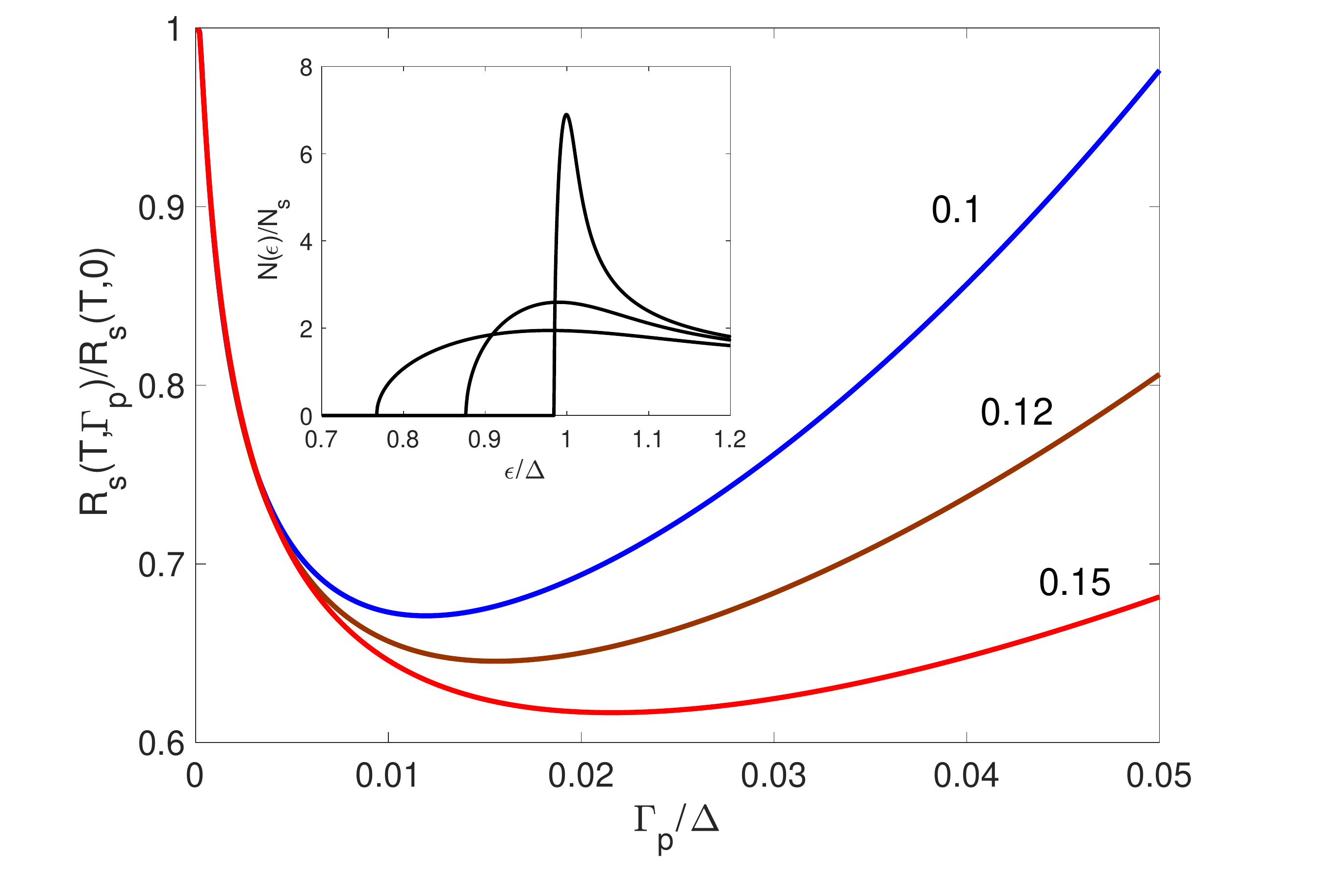}
\caption{Minimum in the surface resistance $R_s(\Gamma_p)$ as a function of the spin-flip pairbreaking parameter $\Gamma_p$ calculated from Eqs. (\ref{sg}) and (\ref{usas}) at $\hbar\omega=0.005\Delta$ and $k_BT/\Delta=0.1,~ 0.12,~ 0.15$. Inset shows $N(\epsilon)$ calculated at $\Gamma_p/\Delta= 0.001,~0.02,~0.05$. Reproduced from Ref. \cite{gk}.}
\label{fig3}
\end{figure}

It is unclear how the Dynes parameter $\Gamma$ could be tuned,  but $R_s$ can be reduced by magnetic impurities the concentration of which can be varied by materials treatments \cite{gk,glaz,khariton}. The spin-flip scattering on magnetic impurities reduces $T_c$, smears the gap singularities in the DOS and decreases the quasiparticle gap \cite{maki}:
\begin{equation}
\epsilon_g=\left(\tilde{\Delta}^{2/3}-\Gamma_p^{2/3}\right)^{3/2}.
\label{gaps}
\end{equation}  
Here $\Gamma_p= \hbar/2\tau_s$, where the spin-flip scattering time $\tau_s$ is inversely proportional to the 
volume density of magnetic impurities \cite{balatski,maki}. If the Dynes broadening of the DOS is disregarded   
and only the effect of magnetic impurities is taken into account, the quasiparticle gap $\epsilon_g$ in Eq. (\ref{gaps}) is smaller than the order parameter 
$\tilde{\Delta}$  \cite{maki}:
\begin{equation}
\tilde{\Delta} = \Delta_0-\frac{\pi}{4}\Gamma_p, \qquad \Gamma_p\ll\Delta.
\label{delts}
\end{equation}    
Here $\Delta_0$ is the order parameter in the absence of magnetic impurities. 
The broadening of the DOS peaks increases with $\Gamma_p$ as shown in the inset of Fig. \ref{fig3}.  
 
The microwave conductivity and the factors $n(\epsilon)=\mbox{Re}\cosh\theta$ and $m(\epsilon)=\mbox{Re}\sinh\theta$ in Eq. (\ref{sg}) were calculated in Refs. \cite{gk,glaz,khariton} by solving the Usadel equation which takes into account the magnetic pairbreaking  \cite{kopnin,balatski}:
\begin{equation}
\epsilon\sinh\theta+i\Gamma_p\cosh\theta\sinh\theta=\tilde{\Delta}\cosh\theta.
\label{usas}
\end{equation}

The effect of magnetic impurities on $R_s$ was calculated in Ref. \cite{gk}. The results are
shown in Fig. \ref{fig3}, where $R_s(\Gamma_p)$ is plotted as a function of $\Gamma_p$ at different temperatures. 
There is a minimum in $R_s(\Gamma_p)$ resulting from interplay of the DOS broadening which reduces $R_s$ as $\Gamma_p$ increases, and the reduction of the quasiparticle gap $\epsilon_g$ which increases $R_s$ with $\Gamma_p$. The position of the minimum in $R_s(\Gamma_p)$ shifts to larger $\Gamma_p$ as $T$ increases. Thus, incorporation of a small density of magnetic impurities in a superconductor can noticeably (by $\sim 30-40\%$) decrease the surface resistance at low temperatures. The conditions of the minimum in $R_s(\Gamma_p)$ can be evaluated using the Abrikosov-Gor'kov theory of weak magnetic scattering in which $T_c$ vanishes at $\Gamma_p =\hbar/2\tau_s=\Delta_0/2$ \cite{balatski}. The latter implies that magnetic impurities suppress superconductivity if the spin flip mean free path $l_s =v_F\tau_s$ becomes of the order of the size of Cooper pair, $\xi_0=\hbar v_F/\pi\Delta_0$. The values of $\Gamma_p \simeq (0.01-0.02)\Delta$ in Fig. \ref{fig3} correspond to $l_s\sim 10^2\xi_0$ if no bound states on magnetic impurities occur \cite{balatski}. Magnetic impurities associated with oxygen vacancies in the native surface oxide of Nb have been revealed by tunneling measurements \cite{anlm1}. 

\subsection{Proximity-coupled normal layer at the surface}

\begin{figure}[ht]
\centering
\includegraphics[width=12cm, trim={50mm 60mm 20mm 65mm},clip]{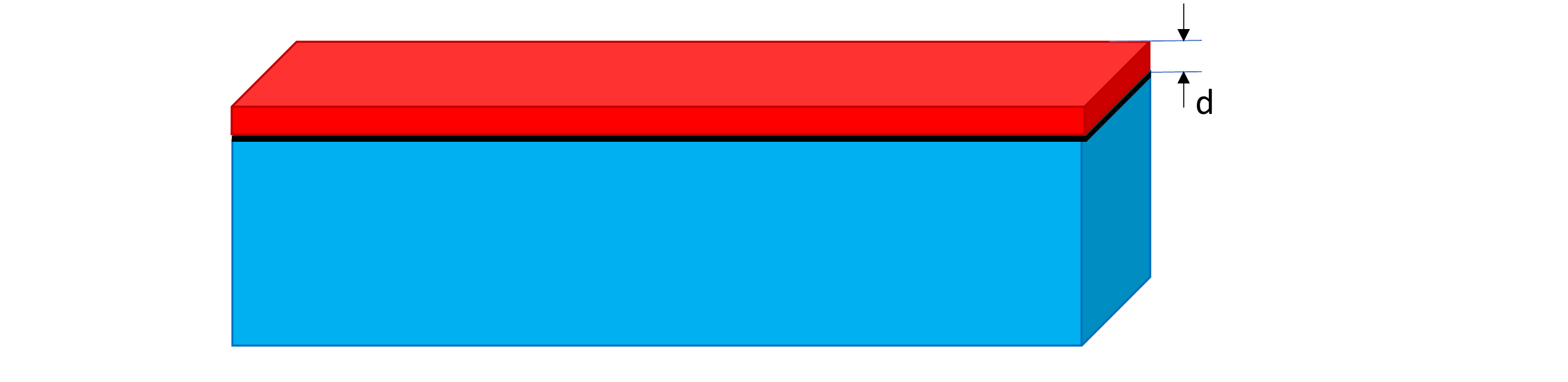}
\caption{A metallic (N) layer of thickness $d$ (red) coupled by the proximity effect to the bulk superconducting (S) substrate (blue). The black layer between N and S 
depicts a resistive interface barrier.}
\label{fig4}
\end{figure}

Another tunable pairbreaking mechanism of reducing $R_s$ involves a thin metallic (N) layer coupled to the bulk superconductor (S) by the proximity effect as shown in Fig. \ref{fig4}. Such N layer models a generic surface oxide structure of superconducting materials, particularly a 1-2 nm thick metallic NbO suboxide sandwiched between the dielectric NbO$_2$ - Nb$_2$O$_5$ oxides at the surface and the bulk Nb. This model was investigated in Refs. \cite{gk,kg} in which the Usadel equations were solved to calculated a position-dependent quasiparticle density of states $N_n(\epsilon, x)$ across a thin N layer coupled to the bulk superconductor, and their effect on the surface resistance. The DOS profile in the N layer can be inferred from STM measurements \cite{nick,eric}.  

The DOS profile and $R_s$ are controlled by the parameters $\alpha$ and $\beta$ which quantify the thickness of N layer and a resistive N-S interface barrier, respectively:
\begin{equation}
\alpha=\frac{dN_n}{\xi_sN_s},\qquad \beta =\frac{4e^2}{\hbar}R_BN_n\Delta d.
\label{ab}
\end{equation}
Here $d$ is the thickness of N layer, $\xi_s=(D_s/2\Delta)^{1/2}$ is the coherence length in the bulk superconductor with nonmagnetic impurities, $D_s$ is the electron diffusivity proportional to the conductivity $\sigma_{n,s}=2e^2N_{n,s}D_{n,s}$ in the normal state, $N_s$ and $N_n$ are the respective DOS at the Fermi surface in the normal state, the subscripts $n$ and $s$ label the parameters of the N layer and the S substrate, respectively, and $R_B$ is a contact resistance between N and S. The properties of the structure shown in Fig. \ref{fig4} can be tuned by materials treatments which change the thickness and conductivity of N layer and the interface resistance $R_B$.  For instance, $R_B(T)$ can either increase or decrease with $T$ depending on the heat treatment which can change $R_B$ by several orders of magnitude, as was shown for the YBCO-Ag interface \cite{int1,int2}. Complexities of the Schottky barrier between different materials are not fully understood \cite{barrier}, but the dependence of $R_s(T)$ on the interface resistance could be used to optimize $R_s$.   
\begin{figure}[ht]
\centering
   \includegraphics[scale=0.32]{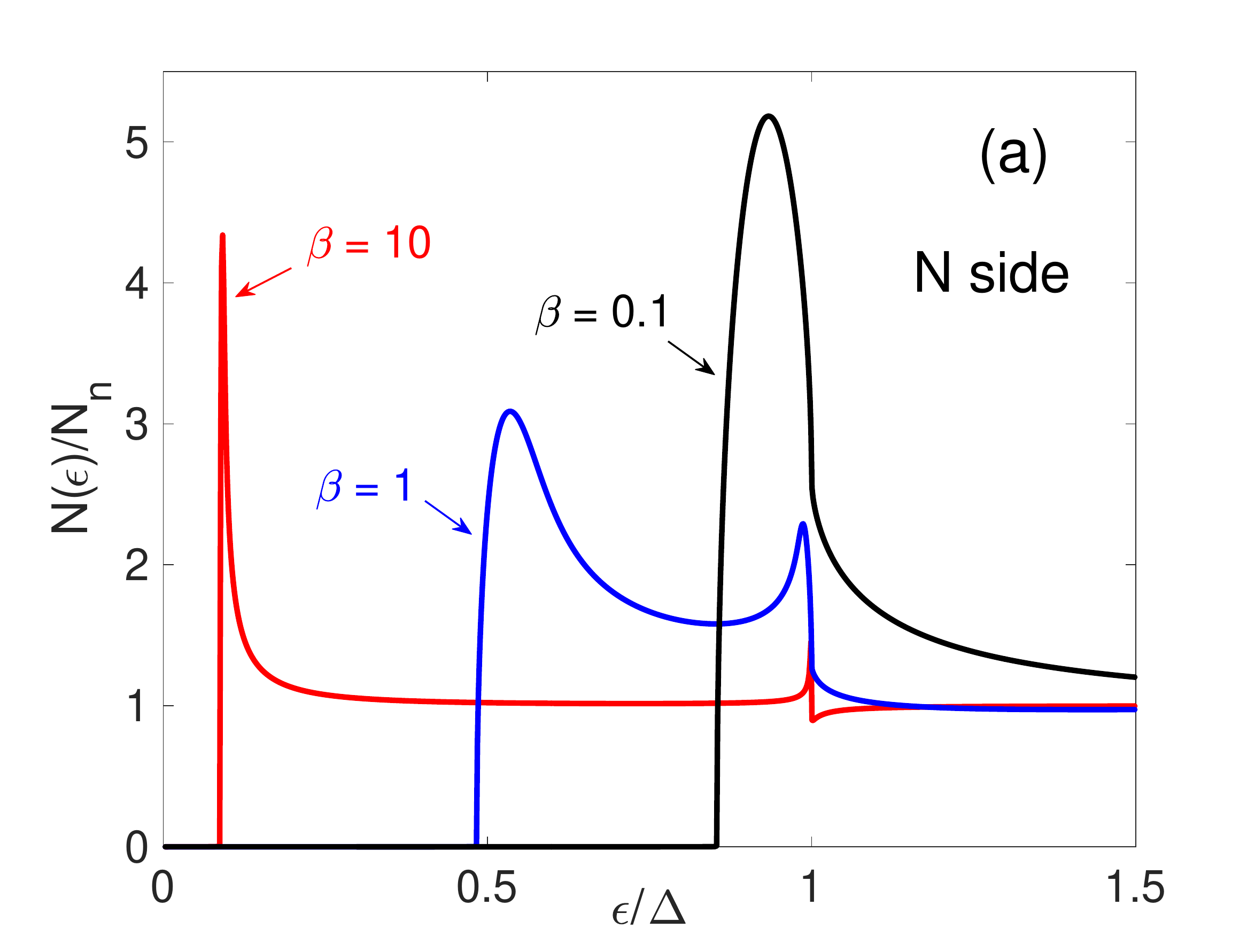}
    \includegraphics[scale=0.32]{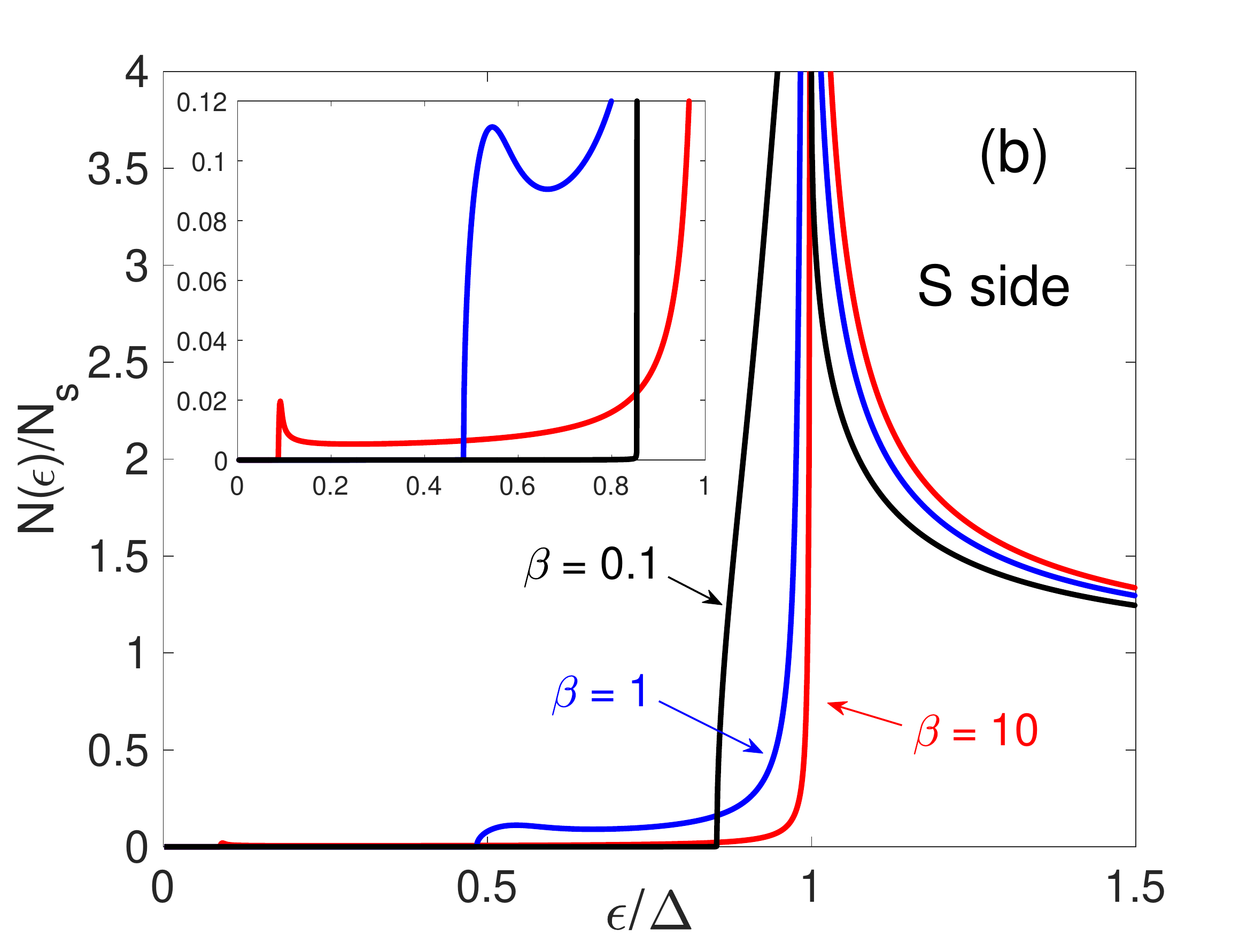}
   \caption{
Densities of states at (a) N side and (b) S side of the interface calculated for  
$\alpha=0.05$, $\Gamma = 0.05\Delta$, $k_B\Theta=11\Delta$ and $\beta=0.1,~1,~10$, where $\Theta$ is the Debye temperature.
Inset in (b) shows $N(\epsilon)$ at $\epsilon\ll \Delta$. Reproduced from Ref. \cite{gk}.
   }
   \label{fig5}
\end{figure}

The thickness of N layer and the interface resistance which control the strength of proximity coupling of the N layer with the S substrate, strongly affect the DOS profile at the surface.   Shown in Fig. \ref{fig5} are the DOS in the N layer much thinner than the proximity length $\xi_n=(\hbar D_n/2\Delta)^{1/2}$ and the DOS at the S side of the N-S  interface calculated in Ref. \cite{gk} for different values of $\beta$ at $\Gamma=0$.  For strong coupling $\beta\ll 1$, the DOS in the N layer has a sharp peak at $\epsilon\simeq \Delta$ and drops to zero below the minigap energy $\epsilon_0<\Delta$ characteristic of N-S proximity-coupled structures  \cite{mg1,mg2}. If $\beta\ll 1$ the proximity effect makes the N layer superconducting with $\epsilon_0\approx \Delta$. As $R_B$ increases, the minigap in the N layer decreases and the DOS approaches $N_n$ for a decoupled N layer at $\beta\to\infty$.  Here $\epsilon_0(\beta)$ at $\Gamma=0$ is determined by the equation \cite{gk}: 
\begin{equation}
\beta=\frac{\Delta}{\epsilon_0}\left(\frac{\Delta-\epsilon_0}{\Delta+\epsilon_0}\right)^{1/2}.
\label{be}
\end{equation} 
Hence, $\epsilon_0$ decreases as $\beta$ increases:  
$\epsilon_0\simeq (1-2\beta^2)\Delta$ at $\beta \ll 1$ and 
$\epsilon_0 = \Delta/\beta$, at  $\beta\gg 1$. In a weakly-coupled N layer $(\beta\gg 1)$ the minigap $\epsilon_0=\Delta/\beta$ expressed in terms of $R_B$  using Eq. (\ref{ab}) is independent of superconducting parameters:
\begin{equation}
\epsilon_0=\frac{\hbar}{4e^2N_ndR_B},\qquad \beta\gg 1.
\label{mgap}
\end{equation}
The account of the Dynes parameter $\Gamma$ smoothens the peaks in $N_n(\epsilon)$ and $N_s(\epsilon)$ shown in Fig. \ref{fig5} and produces   
low-energy tails in $N_n(\epsilon)$ extending to the region $0<\epsilon<\epsilon_0$ of "mini subgap" states in the N layer \cite{gk}.      
Here a thin metallic layer can significantly affect $N_n(\epsilon)$ and $N_s(\epsilon)$, resulting in a variety of temperature dependencies of $R_s(T)$.  

\begin{figure}[tb]
\centering
   \includegraphics[scale=0.4]{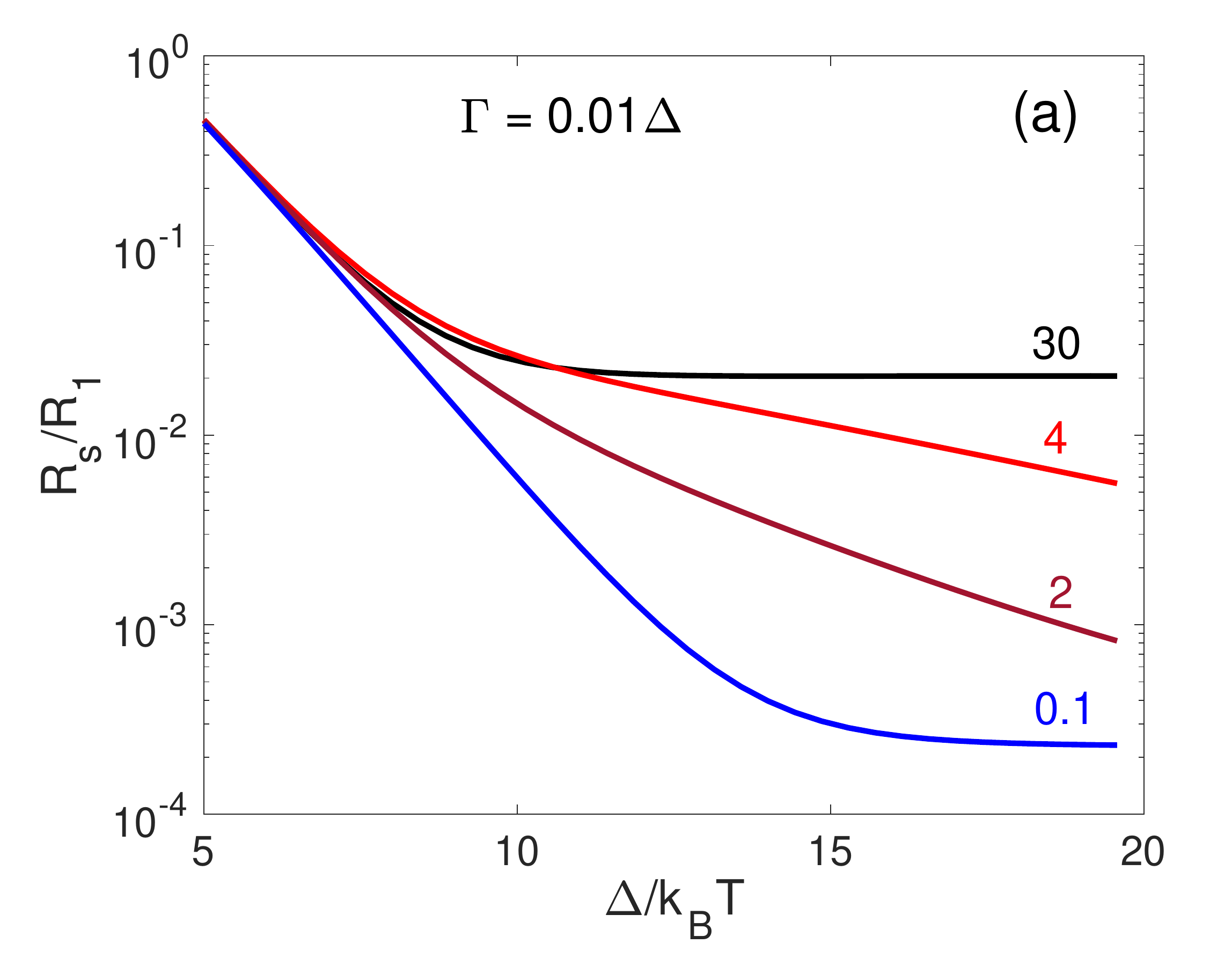}
   \caption{Arrhenius plots calculated for $\alpha=0.05$ (which corresponds to $d\simeq 1$ nm for Nb), $\lambda=4\xi_s$,  $k_B\Theta= 11\Delta$, $D_n=D_s/3$,  $\beta=0.1,~2,~4,~30$, and $\Gamma=0.01\Delta$.  Here $R_1= \mu_0^2\omega^2\xi_s\lambda^2/2\rho_s$ and $\Theta$ is the Debye temperature. Reproduced from Ref. \cite{gk}.}
   \label{fig6}
\end{figure}

Shown in Fig. \ref{fig6} are the Arrhenius plots of $\ln R_s(T)$ as functions of $\Delta/k_BT$ calculated in Ref. \cite{gk} for a thin N layer 
with $\alpha=0.05$, $\hbar\omega<\Gamma$ and different values of $\beta$ varying from $\beta=0.1$ (weak resistive barrier) 
to $\beta=30$ (strong resistive barrier). In the limits of $\beta\ll 1$ and $\beta\gg 1$ the qualitative behaviors of $\ln R_s(T)$ are similar to those shown 
in Fig. \ref{fig2}: as $T$ decreases, the slope of $\ln R_s(T)$ changes from the bulk $\Delta$ at high $T$ to zero at low $T$, the residual resistance 
at $\beta\gg 1$ being much larger than at $\beta\ll 1$. The latter reflects higher rf losses in the normal layer as the 
proximity-induced superconductivity in the N layer weakens with the increase of $R_B$. As a result, $R_i$ at $\beta\gg 1$ is dominated by the N layer fully decoupled from the S substrate, whereas at $\beta\ll 1$, the N layer is strongly coupled with the S substrate and the structure shown in Fig. \ref{fig4} behaves as a single superconductor with an effective $\Gamma$.   At intermediate values of $\beta=2-4$, a change in the slope of $\ln R_s(T)$ occurs around $\Delta/k_B\simeq 8-10$ due to switching from a thermally-activated resistance controlled by the bulk gap $\Delta$ at high $T$ to $R_s(T)$ controlled by the smaller minigap $\epsilon_0$ in the thin N layer. As the temperature decreases further, $R_s(T)$ approaches a temperature-independent residual resistance.  

\begin{figure}[ht]
\centering	
   \includegraphics[scale=0.33]{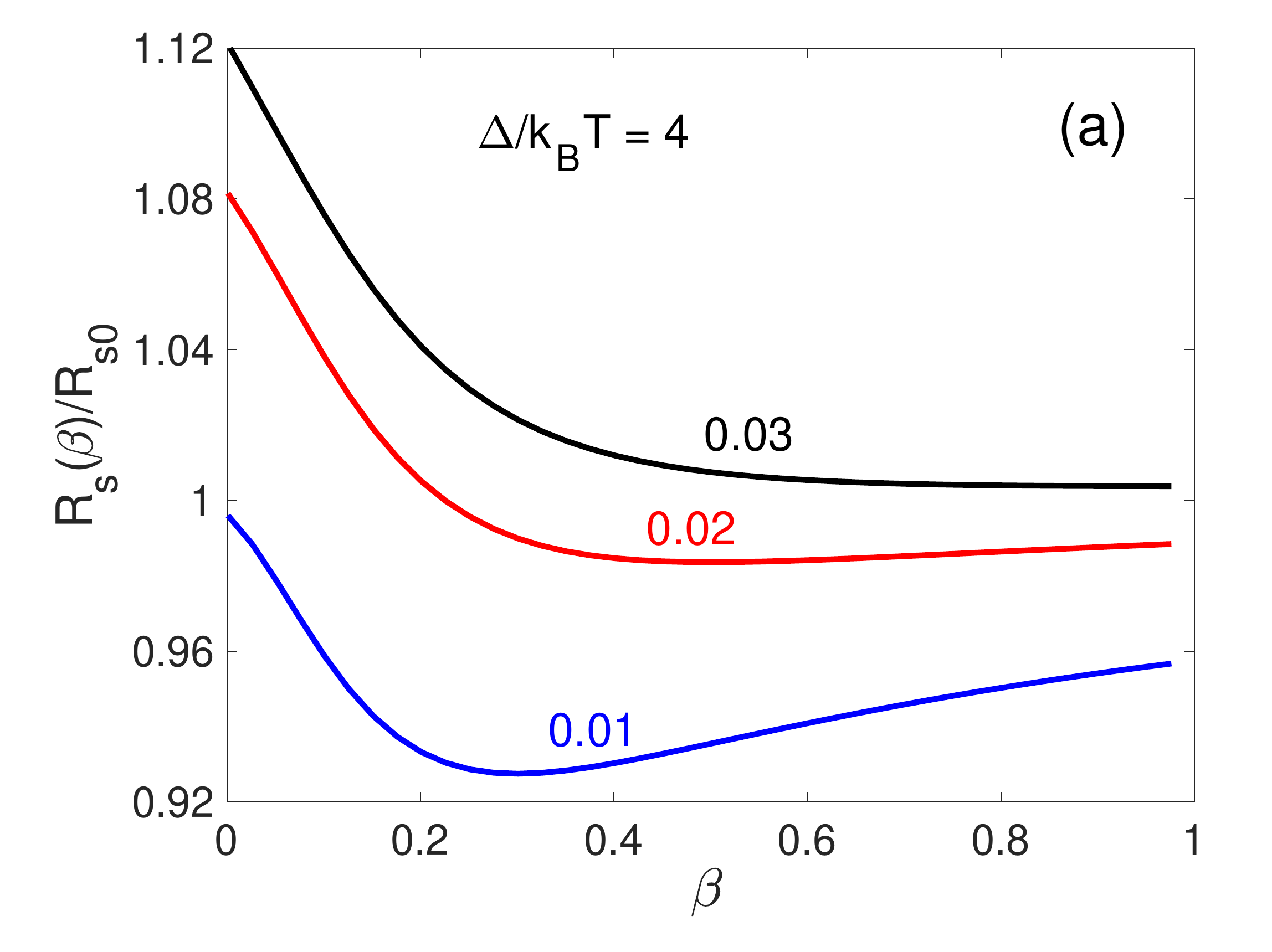}
   \includegraphics[scale=0.33]{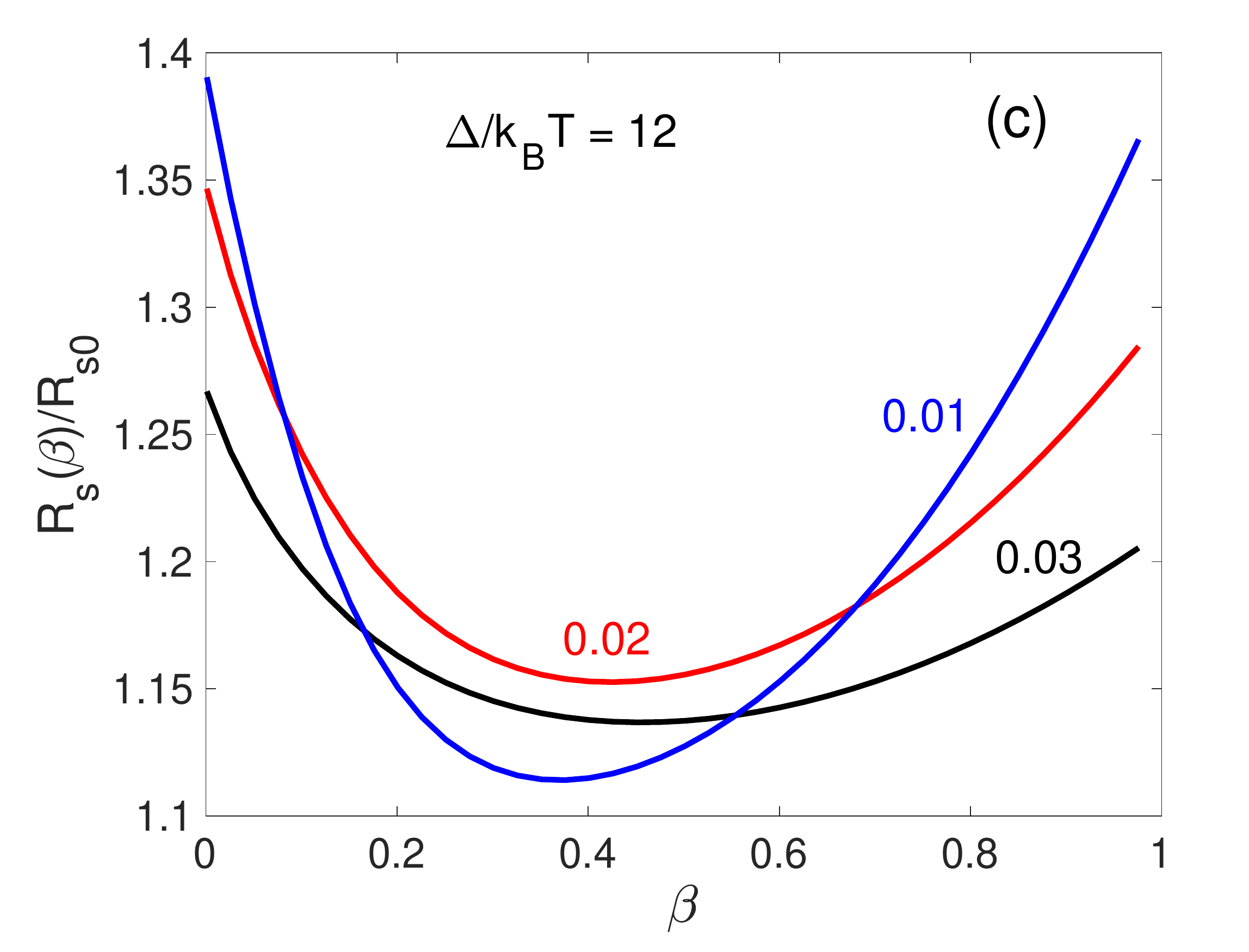}
   \caption{Minima in $R_s(\beta)$ as a function of the 
   interface barrier parameters $\beta$ calculated at $\Delta/k_BT=4,~8,~12,~20$, $\Gamma=0.01,~0.02, ~0.03$,  $\lambda=4\xi_S$,
   $D_n=0.1D_s$, $\alpha=0.05$, and $k_B\Theta = 11\Delta$. Here $R_s(T,\beta,\Gamma)$ are normalized to their respective 
   values $R_{s0}(T,\Gamma)$ for each $\Gamma$ in the absence of the N layer and $\Theta$ is the Debye temperature. Reproduced from Ref. \cite{gk}.}
   \label{fig7}
\end{figure}

The surface resistance can be reduced by tuning the parameters of N layer. For example, Fig. \ref{fig7} shows $R_s(T,\Gamma,\beta)$ for a thin dirty N layer at two temperatures and different values of $\Gamma$.   The minimum in $R_s(\beta)$ results from interplay of two effects. The first one causing the increase of $R_s$ with $\beta$ results from decreasing the proximity-induced minigap $\epsilon_0$ in the N layer as the interface resistance $R_B$ increases. The second effect causing the initial decrease of $R_s$ with $\beta$ results from the change in the DOS around the N layer as shown in Fig. \ref{fig5}.  Here a moderate broadening of the DOS peaks due to the combined effect of finite quasiparticle lifetime $\hbar/\Gamma$ and the metallic layer reduces  $R_s$ in a way similar to Eq. (\ref{rule}). Moreover, in the case of $\Delta/k_BT = 4$ shown in Fig. \ref{fig7}(a), the minimum in $R_s$ at $\Gamma=0.01\Delta$ with the N layer is below $R_{s0}$, suggesting that one can engineer an optimal DOS to reduce $R_s$ below its value for an ideal surface.

\subsection{Nonlinear electromagnetic response}

\begin{figure}[ht]
\centering
\includegraphics[width=8cm]{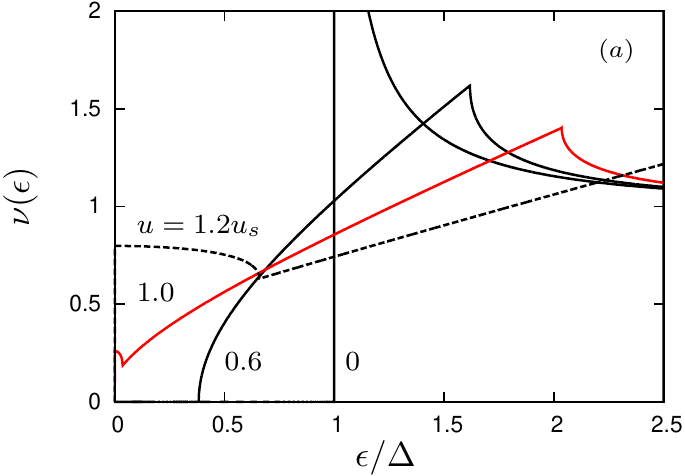}
\\
\includegraphics[width=8cm]{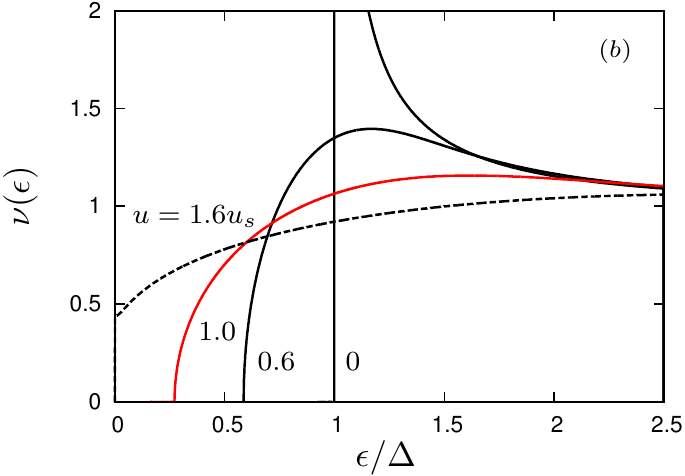}
\\
\includegraphics[width=8cm]{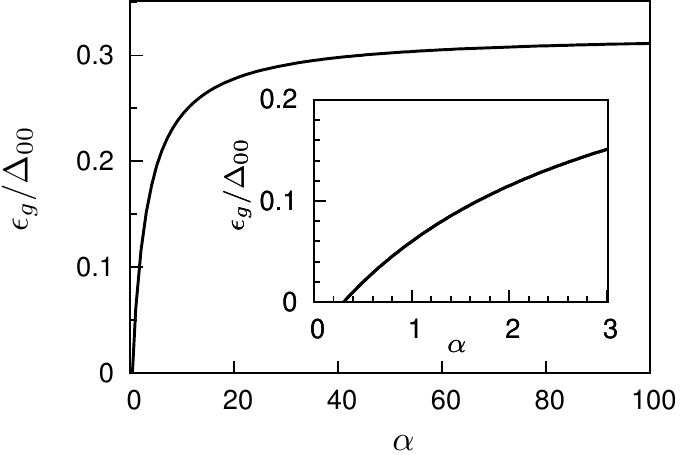}
\caption{Effect of Meissner current on the normalized density of states $\nu(\epsilon)=N(\epsilon)/N_n$ at different ratios $H/H_s$ for:
(a) clean limit at $a=0$, (b) dirty limit at $a=3.6$. The red lines show $\nu(\epsilon)$ at the superheating field. Dependence of the 
quasiparticle gap on the mean free path (c). Reproduced from Ref. \cite{lin}}
\label{fig8}
\end{figure}

The pairbreaking effect of current on the DOS was addressed theoretically in the sixties \cite{parment,bardeen,maki,fulde} and then observed by tunneling measurements  \cite{denscurr}.  Shown in Fig. \ref{fig8}a is the DOS in the clean limit $l_i\gg\xi_0$ and $\Gamma=0$ calculated in Ref. \cite{lin}. Here the current turns the DOS singularity at $\epsilon = \Delta$ into a finite peak and reduces the quasiparticle gap $\epsilon_g$ at which $N(\epsilon_g)=0$.  The gap $\epsilon_g(J)$ is smaller than $\Delta$ and decreases with the current density $J$, whereas $\Delta^2$ proportional to the superfluid density is
{\it independent} of $J$ at $J<J_g$ and $T=0$, where $J_g=en_0\Delta_0/p_F$ is the critical current density at which $\epsilon_g$ vanishes  
\cite{parment,bardeen,maki}. In a clean superconductor the gap $\epsilon_g$ is anisotropic and depends on the angle between $\textbf{J}$ and the momentum of a quasiparticle. 

In the current-carrying state impurities round the cusps in $N(\epsilon)$ as shown in Fig. \ref{fig8}b. Here the quasiparticle gap decreases as $J$ increases but {\it remains finite} as $J$ reaches the depairing current density $J_c$. Thus, impurities preserve the gapped state all the way to the pairbreaking limit, unlike the case shown in Fig. \ref{fig8}a. There is a critical concentration of impurities above which the gap $\epsilon_g$ at $J=J_c$ opens, so that a superconductor at $J=J_c$ is in the gapless state if $a=\pi\xi_0/l<a_c$ and in a gapped state at $a=\pi\xi_0/l>a_c$. The critical value of $a_c=0.36$ corresponds to the mean free path smaller than $l_c=\pi \xi_0/a_c \approx 8.72\xi_0$ \cite{lin}. The dependence of $\epsilon_g$ on the mean free path at $a>0.36$ is shown in Fig. \ref{fig8}c. The critical gap $\epsilon_g(a)$ at $J=J_c$ increases monotonically with $a$ and approaches $\epsilon_g(\infty)=0.323\Delta_0$ at $l_i\ll \xi_0$. Scattering of quasiparticles on impurities makes $\epsilon_g$ isotropic and independent of the direction of ${\bf J}$.

The dependencies of $\epsilon_g$ and $\Delta$ on $J$ in the dirty limit are given by \cite{maki} 
\begin{eqnarray}
\epsilon_g^{2/3}=\Delta^{2/3}-\left(\frac{J}{J_0}\right)^{4/3}\Delta_0^{2/3}
\label{dd} \\
\Delta(J)=\left(1-\frac{\pi J^2}{4J_0^2}\right)\Delta_0,\qquad J\ll J_0,
\label{ed}
\end{eqnarray}
where $J_0=\phi_0/\sqrt{2}\pi\mu_0\lambda^2\xi\sim J_c$. The current-induced DOS broadening makes the surface resistance dependent on $J$ in a way similar to the dependence of $R_s$ on other pairbreaking parameters considered above.
At $T\ll T_c$ and $\hbar\omega\ll\Delta$, the exponentially small density of quasiparticles affects neither $T_c$ nor the dynamics of superconducting condensate, but the rf superflow causes temporal oscillations of the DOS and $\epsilon_g(t)$, resulting in a field dependence of $R_s(B_a)$. Usually $R_s$ is expected to increase with the rf field due to current pairbreaking, electron overheating, penetration of vortices, etc. \cite{hein,oates}.   Yet a significant reduction of $R_s$ by the rf field has been observed in alloyed Nb cavities  \cite{raise1,raise2,raise3,raise4,raise5,raise6,raise7,raise8,raise9} in which the decrease of $R_s$ by $\simeq 20-60\%$ at 2K extends to the fields $B_a\simeq 90-100$ mT at which the density of screening currents at the surface reaches  $\simeq 50\%$ of the pairbreaking limit, $J_c\simeq B_c/\mu_0\lambda$.  The reduction of $R_s$ by a  microwave field is not unique to Nb: similar effects have been observed on thin films \cite{nrf1,nrf2,nrf3}, and a reduction of $R_s(B_0)$ with the dc field $B_0$ superimposed onto a low-amplitude rf field in Pb, Sn, Tl, and Al has been known since the fifties \cite{nr1,nr2,nr3,nr4,nr5,nr6,nr7}. 

The microwave suppression of $R_s$ can be understood as follows \cite{agsust}.  The rf field $B(t)=B_a\cos\omega t$ causes temporal oscillations of the DOS between the singular $N(\epsilon)$ at $B(t)=0$ to a broadened $N(\epsilon, B_a)$ at the peak field $B(t)=B_a$, as shown in Fig. \ref{fig8}. Thus, the peak in the DOS $\langle N(\epsilon)\rangle $ averaged over the rf period is smeared out within the energy range $\epsilon_g < \epsilon < \Delta_0$.  In the dirty limit Eq. (\ref{dd}) yields the width of the averaged gap peak $\delta\epsilon = \Delta_0-\epsilon_g \sim (B_a/B_c)^{4/3}\Delta_0$ at $B_a\ll B_c$. As a result, the current-induced broadening of the DOS can be roughly accounted for by replacing the materials-related broadening parameters $\Gamma$ or $\Gamma_p$ with the current-induced broadening $\delta\epsilon$ in Eq. (\ref{rule}).  This yields a logarithmic decrease of $R_s$ with $B_a$:
\begin{equation}
R_s(B_a) \sim \frac{\mu_0^2\omega^2\lambda^3\Delta}{\rho_nk_BT}\ln\left[\frac{CT B_a^{4/3}}{T_cB_c^{4/3}}\right]e^{-\Delta/k_BT},
\label{qual}
\end{equation}  
where $C\sim 1$. This qualitative picture gives an insight into a mechanism of microwave reduction of $R_s(B_a)$ in the region of the parameters $(\Gamma/\Delta)^{3/4}B_c\ll B_a\ll B_c$ and $\mbox{max}(\Gamma,\hbar\omega)\ll k_BT$  relevant to the experiments \cite{raise1,raise2,raise3,raise4,raise5,raise6,raise7,raise8}. The behavior of $R_s(B_a)$ is affected by materials treatments, yet the qualitative Eq. (\ref{qual}) describes well $Q(B_a)$ observed on Ti-doped Nb cavities \cite{raise3}. The effect of current-induced DOS broadening on $R_s$ was pointed out by Garfunkel \cite{garf} who calculated $R_s(H)$ biassed by a strong parallel dc field $H$ in 
the BCS clean limit.  

A theory of nonlinear surface resistance $R_s(B_a)$ in strong microwave field must take into account both the temporal DOS oscillations and nonequilibrium effects \cite{kopnin,lon,belzig,tdgl,jim}.  Many previous works have focused on nonequilibrium states caused by absorption of high-frequency photons at weak fields $B_a\ll (\omega/\Delta)^{3/4}B_c$ and $\hbar\omega > k_BT$ for which the effect of rf current on $N(\epsilon)$ is weak \cite{mooij,dmitriev,semen}.  In this case $\sigma_1(B_a)$ can decrease with $B_a$ as the quasiparticle population spreads to higher energies $\epsilon > k_BT$ \cite{nrf3}. This mechanism similar to that of stimulated superconductivity \cite{eliashb,tikh} can explain the reduction of $\sigma_1$ with $B_a$ observed on Al films at 5.3GHz at $350$ mK \cite{nrf3}.  However, this approach is not applicable to low-frequency and  high-amplitude rf fields with $\hbar\omega\ll k_BT$ and $B_a > (\omega/\Delta)^{3/4}B_c$ for which $R_s(B_a)$ is determined by the time-dependent DOS and a nonequilibrium distribution function $f(\epsilon,t)$ driven by oscillating superflow.  A nonlinear surface resistance $R_s(B_a)$ in the dirty limit and $\lambda\gg\xi$ and $\hbar\omega \ll k_BT$ was calculated in Ref. \cite{ags} by solving the time-dependent Usadel equations, taking into account temporal oscillations of $N(\epsilon,t)$, $\epsilon_g(t)$ and $f(\epsilon,t)$ under strong rf field. Here $R_s$ is obtained from the relation $H_a^2R_s/2=\int_0^\infty\langle J(A(x,t)E(x,t)\rangle dx$, where $J[A(x,t)]$ is the current density calculated for exact solutions of the Usadel equations, $\langle ... \rangle$ denotes averaging over rf period and $E(x,t)=-\dot{A}= B_a\omega\lambda e^{-x/\lambda}\sin\omega t$. This theory, in which $R_s(B_a)$ can decrease with $B_a$ even for the equilibrium Fermi distribution of quasiparticles, captures the field dependence of $R_s(B_a)$ observed on Nb cavities  \cite{raise1,raise2,raise3,raise4,raise5,raise6,raise7,raise8,raise9}.

Strong rf fields can drive quasiparticles out of thermodynamic equilibrium making $f(\epsilon,t)$ different from the Fermi-Dirac distribution  
$f_0(\epsilon)=(e^{\epsilon/k_BT}+1)^{-1}$. Generally, $f(\epsilon,t)$ is determined by kinetic equations taking into account current pairbreaking and scattering of quasiparticles on phonons and impurities \cite{kopnin}. The deviation from equilibrium depends upon the rate $1/\tau_\epsilon$ with which the rf power absorbed by quasiparticles is transferred to the crystal lattice. The time $\tau_\epsilon$ is determined by inelastic scattering of quasiparticles on phonons and recombination of quasiparticles into Cooper pairs \cite{kaplan}. If $\omega\tau_\epsilon\ll 1$, quasiparticles adiabatically follow the temporal DOS oscillations so $f(\epsilon,t)\to f_0(\epsilon)$, but the density of quasiparticles $n(t)=2\int_0^\infty f_0(\epsilon) N[\epsilon,J(t)]d\epsilon$ varies in response to the instant changes of $N[\epsilon,J(t)]$ shown in Figs. \ref{fig8}.  If $\omega\tau_\epsilon\gg 1$, quasiparticles do not have enough time to equilibrate so their density does not change much during rapid oscillations of $N(\epsilon,t)$. Due to slow power transfer between electrons and phonons at $\omega\tau_\epsilon\gg 1$, quasiparticles become hotter than the lattice, at it has been established in thin film electronic applications, for example, superconducting bolometers \cite{hotsc}. 

The rf periods $\sim 0.1-1$ ns at are typically much longer than the electron-electron scattering time $\tau_{ee}$ and the condensate relaxation time $\tau_\Delta\sim\hbar/\Delta$. In this case the quasiparticle energy relaxation time $\tau_\epsilon$ is limited by inelastic electron-phonon collisions for 
which $\tau_\epsilon(T)$ at $T\approx T_c$ is given by \cite{kopnin}:   
\begin{equation}
\tau_\epsilon=\frac{8\hbar}{7\pi\zeta(3)\gamma k_B T_F}\left(\frac{c_s}{v_F}\right)^2\left(\frac{T_F}{T}\right)^3.
\label{tau}
\end{equation}
Here $\gamma$ is the electron-phonon coupling constant, $c_s$ is the speed of longitudinal sound, $T_F=\epsilon_F/k_B$ is the Fermi temperature and $\zeta(3)\approx 1.2$. The time $\tau_\epsilon(T)$ increases rapidly as $T$ decreases. For Nb$_3$Sn with $c_s/v_F\simeq 10^{-3}$, $T_F\sim 10^5$ K, $T_c=17$ K and $\gamma\simeq 1.5$  \cite{orlando}, Eq. (\ref{tau}) yields $\tau_\epsilon\sim 10$ ps at $T_c$ and $\tau_\epsilon\sim 6$ ns at 2 K. Hence Nb$_3$Sn at 1 GHz is in a quasi-equilibrium state near $T_c$ but can be in a non-equilibrium state at 2K. For Al with $c_s\simeq 5.1$ km/s, $v_F\simeq 2030$ km/s, $T_F=1.36\times 10^5$ K, $T_c=1.2$ K and $\gamma=0.43$, \cite{carbotte,ashkroft}, one obtains $\tau_\epsilon\sim 0.4\, \mu$s at $T_c$. Generally,  $\tau_\epsilon$ depends on energy $\epsilon$, which becomes essential at low temperatures \cite{kaplan}. The electron-phonon relaxation time $\tau_\epsilon$ can be reduced by nonmagnetic and magnetic impurities \cite{ep1,ep2,ep3} or by a thin proximity coupled metallic suboxide layer which reduces the quasiparticle minigap at the surface and allows more effective energy transfer from the quasiparticles to phonons. These effects can expand the temperature range of quasi-equilibrium state.    

The nonlinear conductivity controlled by the nonequilibrium kinetics of quasiparticles in strong electromagnetic fields is beyond the scope of this review. Here we focus on the field dependence of $R_s$ due to the temporal current broadening of the DOS affected by the Dynes parameter $\Gamma$, magnetic impurities or a proximity coupled N layer at quasi-equilibrium, $\omega\tau_\epsilon< 1$. Interplay of the current and materials broadening of the DOS can produce a multitude of field dependencies of $R_s(B_a)$ \cite{gk,kg,kub1,kub2,kub3,kub4}. Unlike the Dynes parameter in the bulk, tuning the layer thickness $d$ and conductivity $\sigma_n$ of the metallic suboxide, the contact resistance $R_B$ and the bulk conductivity $\sigma_s$ by materials treatments can be used to optimize $R_s(B_a)$.

\begin{figure}[ht]
   \centering
   \includegraphics[width=0.45\linewidth]{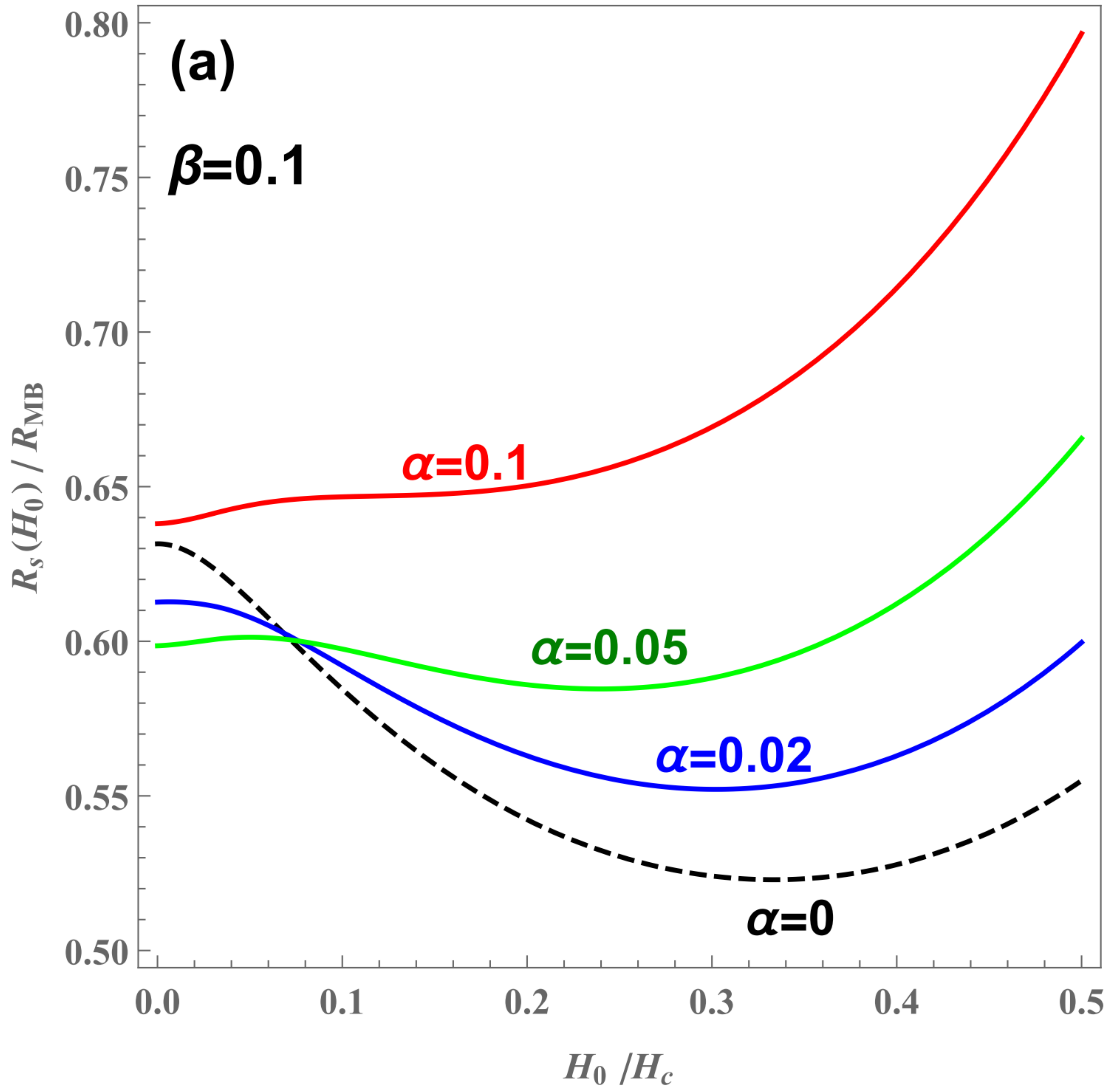}
   \includegraphics[width=0.45\linewidth]{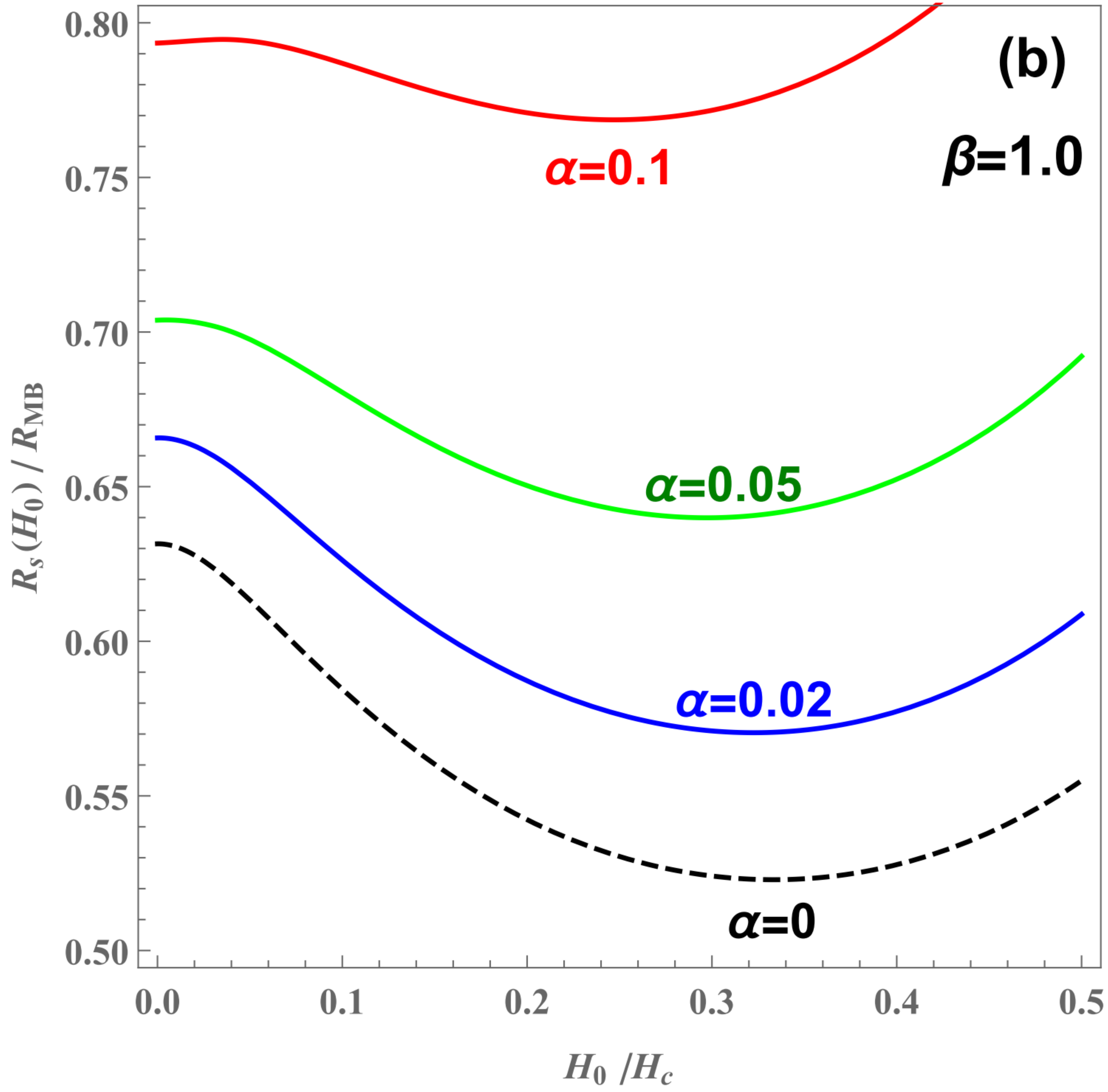}
   \caption{
$R_s(H_0)$ as a function of the field amplitude $H_0$ calculated at different N layer thicknesses:  $\alpha=0.02, 0.05, 0.1, 0.2$, 
$\beta=0.1, 1.0$, $\Gamma_n/\Delta= \Gamma_s/\Delta = 0.005$, $\Gamma_p=0$, 
$k_B\Theta = 11\Delta$, $D_n=0.5 D_s$, $\lambda=10\xi_s$, 
$\omega/\Delta = 0.001$, and $T/\Delta = 0.11$, where $\Theta$ is the Debye temperature. 
The dashed line shows $R_s(H_0)$ calculated for $d=0$, $\Gamma/\Delta=0.005$. All $R_s(H_0)$ curves are normalized to 
the ideal BCS surface resistance $R_{MB}$ at $\Gamma=0$. Reproduced from Ref. \cite{kg}.}
   \label{fig9}
\end{figure}

Shown in Fig. \ref{fig9} are examples of $R_s(B_a)$ curves calculated in Ref. \cite{kg} for different thicknesses of the N layer and two interface barrier parameters $\beta=0.1$ and $\beta=1$ being around the minimum of $R_s$ in Fig. \ref{fig7}. The dashed line shows the microwave suppression of $R_s(B_a)$ caused by the current broadening of the DOS for an ideal surface with $d=0$ \cite{agsust}. For $\beta=0.1$, the dip in $R_s(B_a)$ gets less pronounced as the N layer thickness increases, $R_s(B_a)$ increasing with field at $\alpha>0.1$. This is consistent with weakening the induced superconductivity and reduction of the minigap in the N layer as it becomes thicker. Yet the crossover of $R_s(B_a,\alpha)$ curves at low fields in Fig. \ref{fig9}a imply that removing the N layer increases $R_s(B_a)$, in agreement with Fig. \ref{fig7}. This crossover disappears at a larger contact resistance shown in Fig. \ref{fig9}b.

\begin{figure}[ht]
   \centering
   \includegraphics[width=0.55\linewidth]{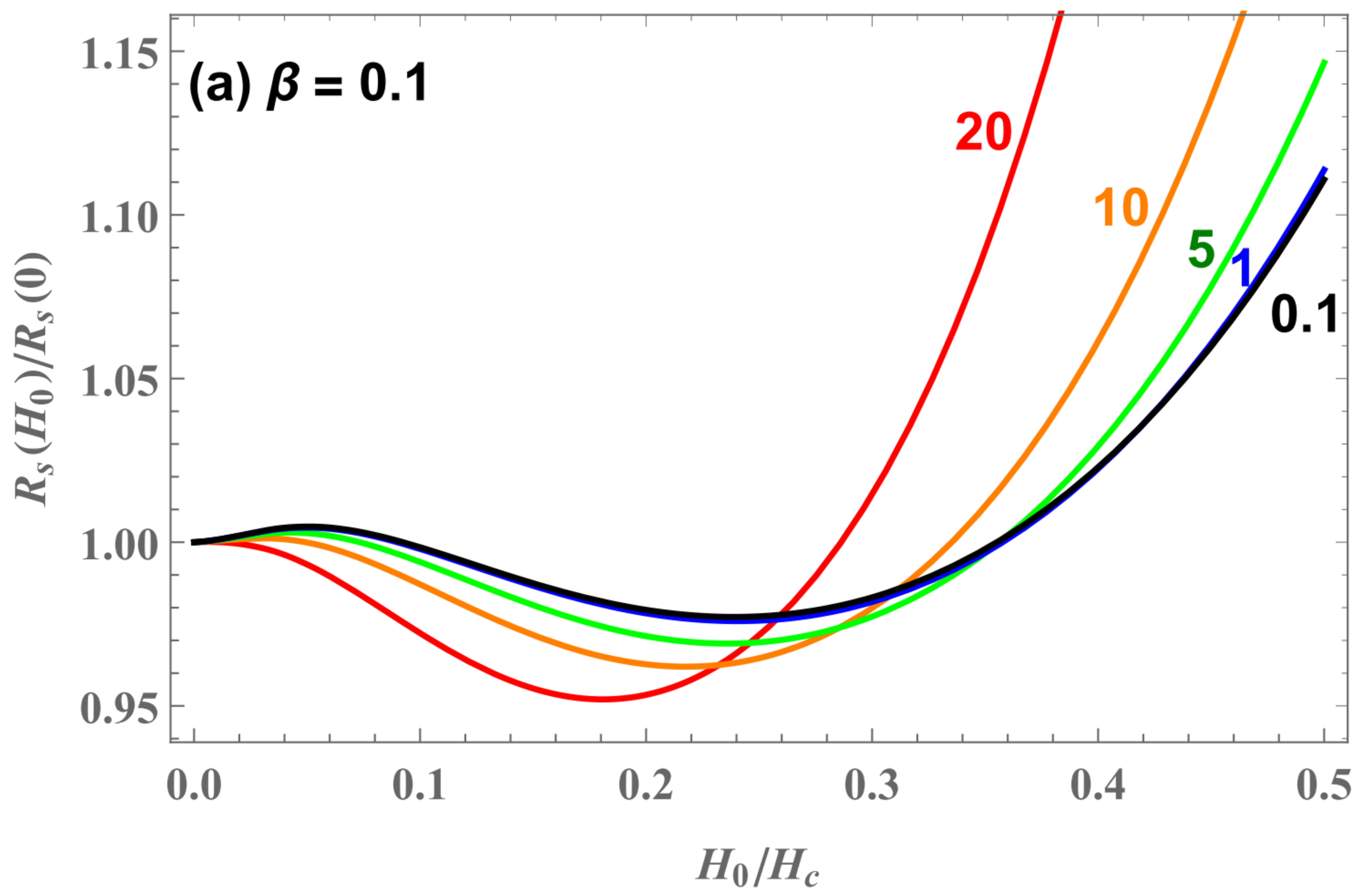}
   \\
   \includegraphics[width=0.55\linewidth]{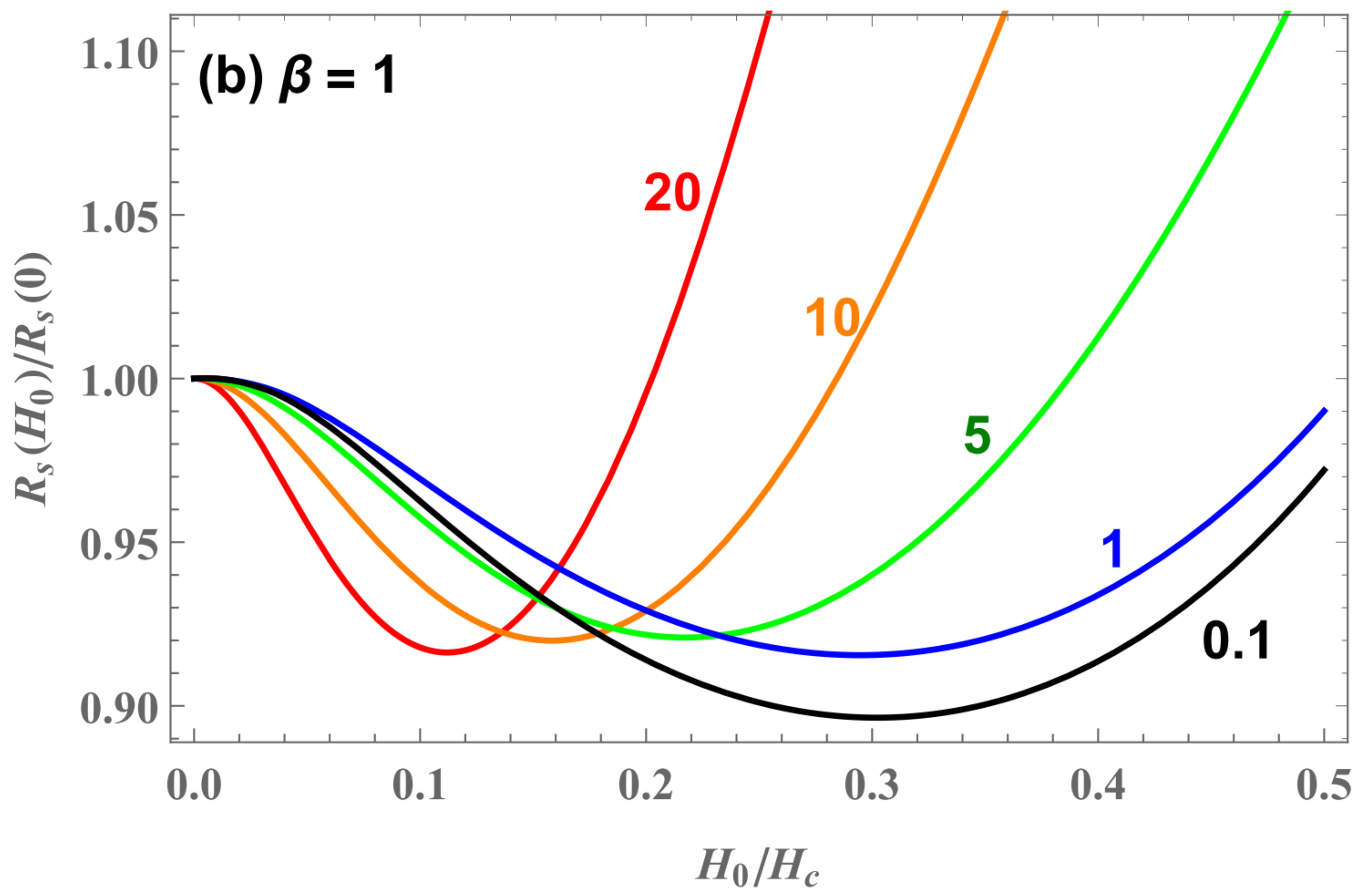}
   \\
   \includegraphics[width=0.55\linewidth]{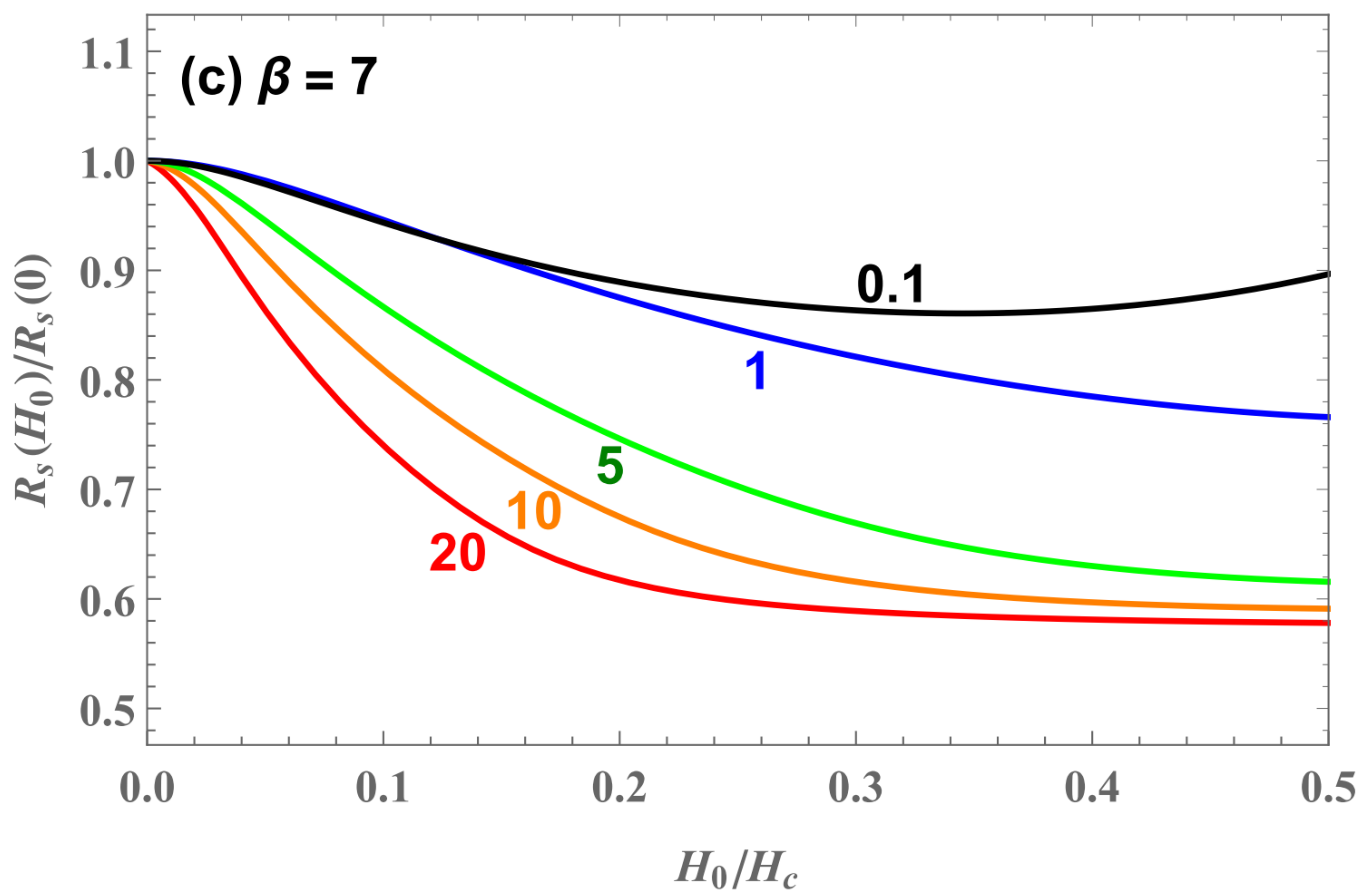}
   \caption{
$R_s(H_0)$ as a function of the rf amplitude $H_0$ calculated for $D_n/D_s = 0.1, 1, 5, 10, 20$, 
$\alpha=0.05$, $\beta=0.1, 1, 7$, $\Gamma_n= \Gamma_s = 0.005\Delta$, $\Gamma_p=0$,  
$k_B\Theta/\Delta = 11$, $\lambda=10\xi_s$, 
$\hbar \omega /\Delta= 0.001$, and $k_B T /\Delta = 0.11$, where $\Theta$ is the Debye temperature. 
Reproduced from Ref. \cite{kg} }
\label{fig10}
\end{figure}
Figure \ref{fig10} shows how the field dependence of $R_s(B_a)$ changes as the conductivity ratio $\sigma_n/\sigma_s$ is varied at a fixed thickness of the N layer for three values of the interface barrier parameter $\beta$ \cite{gk}.  Given that $R_B$ and $\sigma_n/\sigma_s$ can be changed significantly by heat treatments \cite{int1,int2,barrier}  and alloying with nonmagnetic impurities, the results shown in Figs. \ref{fig9} and \ref{fig10} may account for the variability of the Q-factors of Nb cavities.    

\subsection{Tuning $R_s$ by alloying }

Reduction of microwave losses by optimizing the DOS using pairbreaking effects may shed light on the mechanisms behind the improvement of the rf performance of Nb cavities after a low-temperature baking (by 100-200 C for 2 hrs) \cite{bak1,bak3,bak4,bak5,bak6}, medium temperature baking at $290 - 390$ C for 3h \cite{bak6}, high-temperature (800 C) annealing \cite{cav1} and infusion of Nb with impurities. The latter has caused much interest since the discovery of microwave reduction of $R_s(B_a)$ after  alloying the Nb cavities with nitrogen, titanium, oxygen and other impurities \cite{raise1,raise2,raise3,raise4,raise5,raise6,raise7,raise8,raise9}.  Alloying with nonmagnetic impurities could reduce the high-field rf losses since the quasiparticle gap $\epsilon_g$ which ensures an exponentially small $R_s$ does not close up to $B_a=B_s$. This may also pertain to the baking effect which reduces the high-field $Q$-slope by diffusive redistribution of impurities, particularly interstitial oxygen or hydrogen a thin $\sim 20$ nm layer at the surface \cite{bak1,bak3,bak4,bak5,bak6}. The length $L=(Dt)^{1/2}$ over which impurities diffuse from the oxide surface layer to the bulk during the time $t$ gives $L\simeq 17$ nm for the interstitial oxygen at 120$^o$ C and $t=48$ hrs taking the diffusivity $D$ from Ref. \cite{diff}. Uncovering the mechanisms by which materials treatments affect superconducting properties requires compositional analysis of the Nb surface using multiple tools such as TEM, APS, XPS, EELS and  atom probe microscopy \cite{cav2,aps,xps1,xps2,xps3,aprobe1,aprobe2,eels,roma} combined with STM and $Q(B_a)$ measurements to reveal the effects of different treatments on the DOS and $R_s$. 

There are several scenarios by which infusion of impurities over a few $\mu$m from the surface could contribute to the field-induced reduction of $R_s(B_a)$. 1. Impurities mostly reduce the DOS broadening in the entire layer of rf field penetration $\simeq 2\lambda\sim 100$ nm, which reveals the microwave reduction of $R_s(B_a)$ characteristic of the BCS model \cite{agsust,ags}. 2. The impurity infusion primarily modifies the surface oxides, for instance, by shrinking the metallic suboxide layer \cite{agsust}. 3. The appearance of magnetic impurities and two-level states in the oxide layer and N-S interface \cite{caltech,tls1,tls2,tls3}. To determine which of these scenarios is more relevant, the $R_s$ measurements are to be combined with scanning tunneling spectroscopy (STM) to measure the DOS at the surface and link it with the behavior of $R_s(B_a)$. This has been implemented by several groups, starting with the pioneering work \cite{raise1} which showed that Ti infusion significantly reduces the lateral distribution of local values of $\Delta$. The effect of N-doping on the DOS at the Nb surface was addressed in Refs. \cite{nick,eric}. Particularly, the analysis of STM spectra in Ref. \cite{eric} using the model of Ref. \cite{gk} gave an insight into the effect of N infusion on the properties of the metallic suboxide.  

\begin{figure}[ht]
\centering
\includegraphics[scale=0.7, trim={55mm 30mm 50mm 30mm},clip]{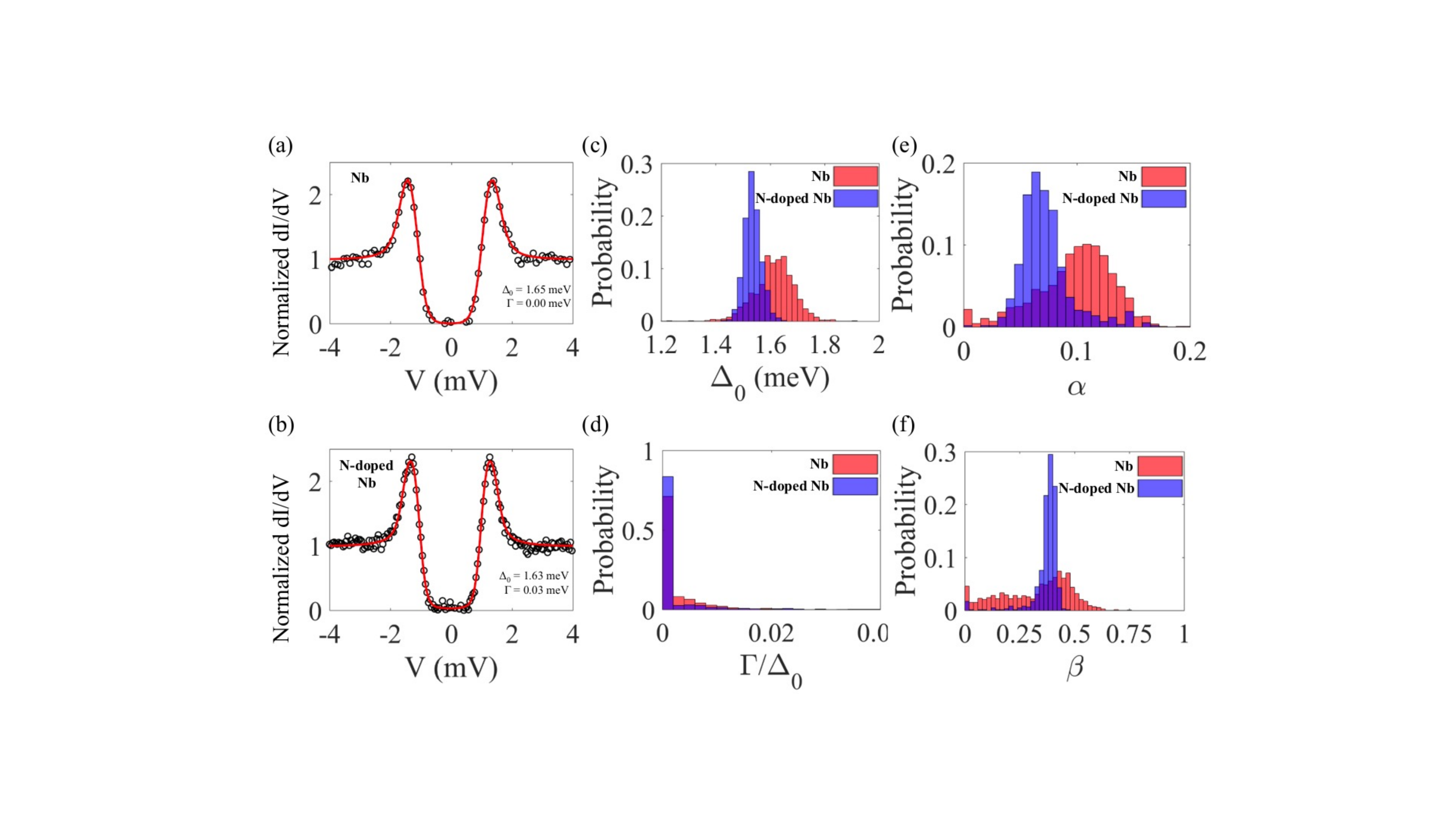}
\caption{Typical tunneling spectra (dots) acquired on Nb and N-doped Nb surfaces. The red lines are the fits with the model of Ref. \cite{gk}. (c),(d) Histogram comparison of the fit parameters $\Delta_0$ and $\Gamma/\Delta_0$ respectively. (e),(f) Histogram comparison of the fit parameters $\alpha$ and  $\beta$ respectively. For Nb samples the number of spectra analyzed is $N = 1440$ (red) and for N-doped samples $N = 576$ (blue). Spectra were taken $32.6$ nm away from each other at $T=1.5$ K.  Reproduced from Ref. \cite{eric} }
\label{fig11}
\end{figure}

The results of Ref. \cite{eric} summarized in Fig. \ref{fig11} indicate that the effect of the nitrogen infusion gives rise to the following: 1. Slightly reduces the average superconducting gap $\bar{\Delta}$ while significantly reducing spatial inhomogeneities of $\Delta$ and the Dynes parameter $\Gamma$. 2. Reduces the thickness of metallic suboxide from $\approx 2$ nm down to $\approx 1.2$ nm. 3. Reduces spatial inhomogeneities of the Nb suboxide thickness and the interface contact resistance parameter $\beta$ close to the optimal values $0.3-0.4$ corresponding to a minimum $R_s$ shown in Fig. \ref{fig7}.  It seems that these effects of the nitrogen doping brings the DOS toward its optimum which minimizes $R_s$, while reducing the effect of such extrinsic factors as the lateral inhomogeneity of superconducting parameters characteristic of the surface of Nb resonators.  

Materials mechanisms which cause the modifications of the oxide structure in Nb require further investigations. It was found that heat treatment in a temperature range sufficient to dissociate the natural surface oxide not only causes a significant reduction of the residual resistance down to $R_i\simeq 1$ n$\Omega$ but also reduces the thermally-activated BCS part of $R_s$ \cite{bak6}. This observation seems consistent with the theoretical results of Ref. \cite{gk} which show that both $R_i$ and $R_{BCS}$ can be reduced if $d_n$ and $R_B$ are reduced from their respective values on the right side of the minimum in Fig. \ref{fig9}. Furthermore, recent TEM investigation indicate that nitrogen doping passivates the Nb surface by introducing a compressive strain close to the Nb/air interface, which impedes the diffusion of oxygen and hydrogen atoms and reduces surface oxide thickness \cite{roma}. This conclusion seems consistent with the analysis of STM data shown in Fig. \ref{fig11}.  

\section{Dielectric losses and kinetic inductance}
This section focuses on two contribution to the electromagnetic response not related to quasiparticles. The first one gives rise to a microwave suppression of the residual surface resistance due to two level states (TLS), and the second one pertains to tuning the kinetic inductance of N-S bilayers and nanowires by the proximity effect.

\subsection{Two-level states}

Microwave losses in amorphous dielectrics at low temperatures can be dominated by the presence of
TLS in the material \cite{tlsr1,tlsr2}. These defects exist in a glassy state in which 
light atoms or trapped electrons or dangling atomic bonds can tunnel between two neighboring positions in a disordered atomic structure.  
Such TLS can occur in amorphous oxide layers on the surface of Al and Nb as well as interfaces between a superconducting film and a dielectric substrate. 
TLS have attracted much attention as a source of noise and decoherence in superconducting
quantum devices at low temperatures. Here TLS can not only contribute to the residual surface resistance but also 
result in decreasing $R_i(E)$ with the rf electric field $E$  \cite{tls1,tls2,tls3,tls4}: 
\begin{equation}
R_i^{TLS}(E)=R_{i0}^{TLS}\frac{\tanh(\hbar\omega/2k_BT)}{[1+(E/E_c)^2]^q}, 
\label{tls}
\end{equation}
where $q$ is close to $1/2$ in the standard tunneling model \cite{tlsr1,tlsr2},
although the $q$ values well below $1/2$ have also been observed \cite{tls3}. The factor $R_{i0}^{TLS}$ is proportional to  
the TLS density of states and depends on the resonator geometry, $E/E_c=\omega_R\sqrt{\tau_1\tau_2}$, where $\omega_R=2d_0E/\hbar$ is the Rabi frequency, $d_0\simeq ea_0$ is a dipole moment proportional to the tunneling distance $a_0$, and $\tau_1(T)$ and $\tau_2(T)$ are the TLS energy relaxation and dephasing times, respectively. One can see that  $R_i^{TLS}(E)$ decreases with $E$ at $E>E_c=\hbar\sqrt{\tau_1\tau_2}/2d_0$ as the TLS become saturated by the microwave field.    

TLS losses in superconducting films can depend strongly on the dielectric substrate, for instance, the losses in Nb on SiO$_2$ are significantly higher than in Nb on Al$_2$O$_3$ \cite{tlsr2}. Of particular interest are intrinsic TLS losses coming from the native oxides in Nb or Al separated from the losses in external  
dielectric components \cite{tir0,tlr1,tlr2}. Fitting $R_i(B_a)$ measured on Nb cavities at $1.5-2$ K and $1.3$ GHz with Eq. (\ref{tls}) gave 
$E_c\approx 0.2$ MV/m \cite{tls2}, much higher than the parallel electric field at the equatorial cavity surface $E\simeq \omega B_a\lambda\simeq 3\times(10^{-2}-1)$ V/m at $B_a=0.1- 10$ mT. The decrease of $R_i(B_a)$ with $B_a$ observed in Ref. \cite{tls2} was attributed to TLS on the inner cavity surfaces near the orifices, where the perpendicular rf electric field can reach a few MV/m.  Yet a much lower $E_c\simeq 56$ V/m was observed on a 150 nm thick Nb stripline resonator grown on a Si substrate and coated with aluminum oxide  \cite{tls1}.

TLS in AlO$_x$ or Nb$_2$O$_5$ oxides have been commonly associated with dangling atomic bonds and oxygen vacancies  \cite{tlr1,tlr2}, 
although the true atomic origin of TLS has not been fully understood \cite{tlsr2}.     
TLS may also result from common environmental impurities such as nitrogen, carbon or hydrogen  
which get dissolved in the material during the film growth and deposition \cite{cav2}. For instance, formation of metallic hydride precipitates in Nb and their contribution to rf losses is well-known \cite{cav2,nbh1,nbh2,film1,film2,film3}.  If the hydrogen bound to the oxygen vacancy in Nb$_2$O$_5$ does behave as TLS \cite{vcox}, one may expect $R_i$ to increase after neutron or proton irradiation \cite{irrad} which produces hydrogen irradiation defects and lattice disorder.  In any case, the manifestations of TLS and quasiparticle contributions to the surface resistance are quite different. At GHz frequencies and $T=1-2$ K the TLS residual resistance $R_i^{TLS}\propto \tanh(\hbar\omega/2k_BT)$ increases with $\omega$ and becomes independent of $T$ at mK temperatures.  This distinguishes the TLS microwave reduction of $R_i^{TLS}(B_a)$ from that of the quasiparticle surface resistance $R_s\propto \exp(-\Delta/k_BT)$ which decreases exponentially as $T$ decreases.    

\subsection{Tuning the kinetic inductance by the proximity effect}

In addition to the reduction of dissipative conductivity $\sigma_1$, the proximity effect can be used to tune the inductive conductivity $\sigma_2$ and ether increase of decrease the kinetic inductance $L_k$ which defines the energy $L_kI^2/2$ of flowing supercurrent $I$.  Here large $L_k$ are desirable in transmons \cite{qq} and kinetic inductance photon detectors \cite{spd1,spd2,spd3,caltech}, whereas small $L_k$ can be useful to reduce the readout or reset times $\tau_r=L_k/R_0$ in quantum memories, qubits and photon detectors, where $R_0$ is a resistance of the readout circuit. The influence of the proximity effect on $L_k$ is most transparent for thin-film S-N bilayers or nanowires  which have been used in single photon detectors \cite{bl1,bl2,bl3}. 

Consider a N-S bilayer shown in Fig. \ref{fig4} with the thicknesses $d_n$ and $d_s$ smaller than the respective coherence lengths $\xi_n$ and $\xi_s$. This corresponds to the Cooper limit \cite{cooper,degennes} in which the superconducting order parameters $\psi_{N,S}$ are uniform through the N and S layers, although $\psi_N$ can be different from $\psi_S$ due to the decoupling effect of $R_B$. Such N-S bilayers have been studied in the literature (see e.g, a review \cite{golub} and the references therein). In the case of $R_B=0$ and $\Gamma\ll k_BT_c$ the critical temperature of the bilayer decreases exponentially with the N layer thickness \cite{cooper}:
\begin{equation}
T_{c}=T_{c0}\exp\left(-d_nN_n/\gamma_s d_sN_s\right),
\label{Tc}
\end{equation}
where $T_{c0}=1.13\Theta\exp(-1/\gamma_s)$ is the critical temperature of S layer, $\gamma_s$ is the BCS pairing constant and $\Theta$ is the Debye temperature. At $R_B=0$ a dirty N-S bilayer has a uniform superconducting order parameter determined by the Usadel equation (\ref{usas}) with $\Gamma_p=0$, $\epsilon\to \epsilon+i\tilde{\Gamma}$ and the composite parameters \cite{agbl}:
\begin{equation}
\tilde{\Delta}=\frac{d_sN_s\Delta_s+d_nN_n\Delta_n}{d_sN_s+d_nN_n},\quad \tilde{\Gamma}=\frac{d_sN_s\Gamma_s+d_nN_n\Gamma_n}{d_sN_s+d_nN_n}.
\label{DG}
\end{equation}
The pairing potential $\Delta_s$ is nonzero in the S layer and vanishes $(\Delta_n=0)$ in the N layer \cite{degennes,golub} The composite $\tilde{\Delta}$ is determined by the BCS gap equation with the effective coupling constant
\begin{equation}
\tilde{\gamma}=\frac{d_sN_s\gamma_s+d_nN_n\gamma_n}{d_sN_s+d_nN_n}.
\label{gam}
\end{equation} 
Eqs. (\ref{DG})-(\ref{gam}) describe both the N-S bilayer with $\gamma_n=\Delta_n=0$ and a bilayer of two different superconductors S and S$^\prime$ for which the index $n$ refers to the S$^\prime$ layer with nonzero $\gamma_n$ and $\Delta_n$. For the N-S bilayer with $\tilde{\Gamma}\ll k_BT_c$, Eq. (\ref{gam}) yields Eq. (\ref{Tc}) because  $T_c=1.13\Theta\exp(-1/\tilde{\gamma})$ and $\gamma_n=0$. Unlike the dissipative conductivity $\sigma_1$, the effect of weak Dynes broadening of the DOS on $\sigma_2$ and $L_k$ is negligible \cite{agbl}. 

The current $I$ flowing along a strongly coupled bilayer of width $w\ll \lambda_s^2/d_s$ in response to the vector potential $A$ is a sum of phase-locked currents in both films: 
$I=-\mu_0w\left(d_n/\lambda_n^2+d_s/\lambda_s^2\right)A$, where $\lambda_{s,n}^{-2}=(\pi\mu_0\sigma_{s,n}\tilde{\Delta}/\hbar)\tanh(\tilde{\Delta}/2k_BT)$ in the dirty limit \cite{agbl}. The kinetic inductance per unit length $L_k=-A/I$ is then:
\begin{equation}
L_k=\frac{\hbar\coth(\tilde{\Delta}/2k_BT) }{\pi w\tilde{\Delta}(d_s\sigma_s+d_n\sigma_n)}
\label{lk}
\end{equation}
For a single S film, Eq. (\ref{lk}) yields $L_{k0}=\mu_0\lambda_s^2/wd_s$ \cite{caltech}.  

\begin{figure}[ht]
\centering
\includegraphics[scale=0.6, trim={20mm 70mm 40mm 75mm},clip]{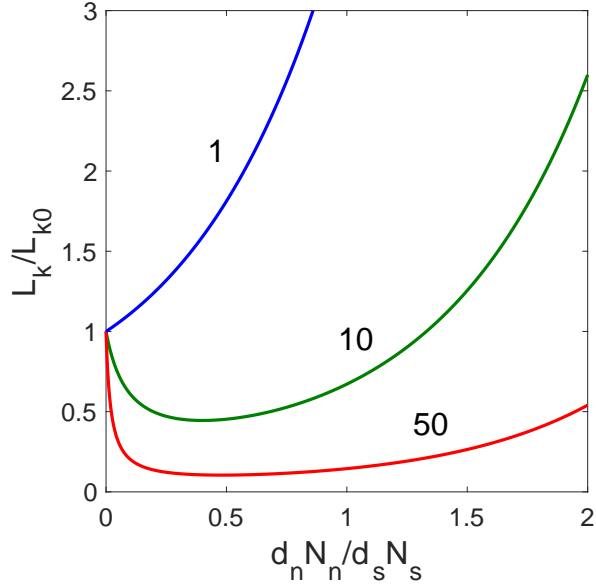}
\caption{The kinetic inductance of a N-S bilayer as a function of the N layer thickness calculated from Eq. (\ref{lk}) at $\gamma_s=0.5$, $T\ll T_c(d_n)$ 
and different diffusivity ratios $D_n/D_s=1, 10, 50$.  }
\label{fig12}
\end{figure}

Deposition of the N layer can either decrease or increase $L_k$. Shown in Fig. \ref{fig12} is $L_k(d_n)$ calculated from Eq. (\ref{lk}) for different diffusivity ratios $D_n/D_s=\sigma_nN_s/\sigma_sN_n$ at $\gamma_s=0.5$, $T\ll T_c$ and $\tilde{\Delta}(d_n)=1.74k_BT_c(d_n)$, where $T_c(d_n)$ is given by Eq. (\ref{Tc}).  If $D_n<D_s/\gamma_s$ the kinetic inductance monotonically increases with $d_n$ due to suppression of $T_c$ by the proximity effect. For a more conductive N layer with $D_n>D_s/\gamma_s$ the dependence of $L_k$ on $d_n$ becomes nonmonotonic due to interplay of two effects: 1. The decrease of $L_k$ with $d_n$ due to additional inertia of Cooper pairs induced in the N layer (the term $d_n\sigma_n$ in  the denominator of Eq. (\ref{lk})). 2. Increase of $L_k$ with $d_n$ due to the proximity effect reduction of $T_c(d_n)$. If $D_n\gg D_s$ the optimum thickness $\tilde{d}_n$ and the minimum inductance $\tilde{L}_k$ at $d_n=\tilde{d}_n$ are given by:
\begin{equation}
\tilde{d}_n\simeq d_s\frac{\gamma_sN_s}{N_n},\qquad \tilde{L}_k\simeq \frac{3L_{k0}D_s}{\gamma_sD_n}\ll L_{k0}.
\label{minL}
\end{equation}  
Deposition of highly conductive Ag or Cu films onto a superconducting film can significantly decrease $L_k$. For instance, a thin Cu film of thickness $d_n=\tilde{d}_n$ on top of a NbN film could decrease $L_k$ at $J\ll J_d$ by some two orders of magnitude because $\sigma_{NbN}/\sigma_{Ag}\sim 10^{-2}-10^{-3}$. The effect of the mean free path and subgap states on $L_k$ was investigated in Ref. \cite{kub4}. Increasing the contact resistance weakens the S-N proximity coupling and reduces the contribution of the N layer to the kinetic inductance, so that $L_k\to L_{k0}$ at $\beta\gg 1$.  Both $\sigma_2(J)$ and $L_k(J)$ increase with $J$ due to current pairbreaking \cite{caltech,kub1,kub4,kind,dynH}, which manifests itself in a nonlinear Meissner effect and intermodulation \cite{oates,nme1,nme2,nme3}. Nonlinear screening of a dc parallel field and the breakdown of superconductivity in N-S bilayers has been investigated in Refs. \cite{dc1,dc2,dc3,dc4}. 

The kinetic inductance can increase greatly in polycrystalline films with weak-linked grain boundaries, particularly granular Al films which have been used in transmons and photon detectors \cite{gral1,gral2,gral3}.  Weakly-coupled grain boundaries characteristic of polycrystalline Nb$_3$Sn \cite{gb1,gb2,gb3} can increase $L_k$ and amplify the nonlinear Meissner effect in Nb$_3$Sn coplanar thin film resonators \cite{junki}. Here a polycrystalline film can be regarded as a network of weakly-coupled planar Josephson junctions, each of which adding a nonlinear kinetic inductance $L_J=\phi_0/2\pi I_c\cos\chi $ inversely proportional to the tunneling Josephson critical current $I_c$ and depending on the superconducting phase difference $\chi$ on the junction \cite{BP}. 

\section{Dynamic superheating field}

The Meissner state becomes unstable as the applied magnetic field $H$ exceeds a dc superheating field $H_s(T)$ at which a transition to a dissipative state occurs. In type-II superconductors with $\kappa>1/\sqrt{2}$, the superheating field lies between the lower critical field $H_{c1}(T)=(\phi_0/4\pi\mu_0\lambda^2)[\ln\kappa+c(\kappa)]$ and the upper critical field $H_{c2}(T)=\phi_0/2\pi\mu_0\xi^2$. Here $c(\kappa)\approx 1/2+(1+\ln 2)/(2\kappa-\sqrt{2}+2)$ which approximates the GL calculations better than $1\%$ \cite{hc1} decreases from $c\approx 1.35$ at $\kappa=2^{-1/2}$ to $c\to 1/2$ at $\kappa\gg 1$.  At $H_{c1}<H < H_{c2}$ a superconductor is in a metastable state because the Bean-Livingston barrier prevents penetration of vortices through an ideal surface \cite{bean}.  Clean Nb is a marginal type-II superconductor with $\kappa\approx 0.85$,  $B_{c1}(0)\approx (\ln\kappa+c)B_c/\kappa\sqrt{2}\approx 180$ mT and the thermodynamic critical field $B_c(0)\approx 204$ mT \cite{nb,smat,carbotte}, where  $B_c(T)=\phi_0/2^{3/2}\pi\lambda\xi$ at $T\approx T_c$ and $B_c(0)=(\mu_0N_s)^{1/2}\Delta_0$. In the BCS model electron scattering on nonmagnetic impurities increases $H_{c2}$, does not affect $H_c$ and decreases $H_{c1}$ \cite{tinkh}. In the case of anisotropic Fermi surface impurity scattering decreases $T_c$ and $H_c$  \cite{balatski}. 

The GL calculations of $H_s$ at $T\approx T_c$ \cite{matricon,chapman} have shown that $H_s(\kappa)$ decreases monotonically with $\kappa$ from $H_s=2^{-1/4}\kappa^{-1/2}H_c$ at $\kappa \ll 1$ to $H_s\approx 1.2H_c$ at $\kappa\approx 1$ and to  $H_s= (\sqrt{5}/3)H_c$ at $\kappa\gg 1$.   The superheating field at $T\ll T_c$ has been usually evaluated by extrapolating the GL results to low temperatures, where the GL theory is invalid. Microscopic calculations of $H_s(T)$ in the entire range $0<T<T_c$ have only been done for $\kappa\to \infty$ in which case $H_s=0.84H_c$ at $T=0$ exceeds the GL extrapolation $H_s= 0.745H_c$ \cite{galaiko}. Calculations of $H_s(T)$ using the Eilenberger equations in the clean limit revealed a maximum in $H_s(T)$ at low $T$ \cite{catelani}, indicating that $H_s(T)$ at low $T$ can hardly be extrapolated from the GL results near $T_c$. Furthermore, the effect of impurities on $H_s(T)$ at $T\ll T_c$ is different from the GL results at $T\approx T_c$, particularly in the clean limit in which the Meissner currents do not affect $\lambda$ until the superfluid velocity $v_s = J/n_se$ reaches the critical value, $v_s = v_c =\Delta_0/p_F$ \cite{parment,bardeen,maki}. If $\kappa\gg 1$ and $T=0$ the gap $\epsilon_g$ closes at the field $H_g(2/3)^{1/2}H_c \approx 0.816 H_c<H_s$ \cite{parment,lin}.

The effect of nonmagnetic and magnetic impurities on $H_s(T)$ calculated from the Eilenberger equations at $\kappa\gg 1$ and $0<T<T_c$ was addressed in Ref. \cite{lin}. The results show that nonmagnetic impurities do not affect $H_s$ near $T_c$ where $H_s=0.745H_c$. At low temperatures $H_s(a)$ has a small maximum as a function of the impurity parameter $a = \pi\xi_0/l_i$, the maximum in $H_s(a)$ washes out as $T$ decreases. The effect of nonmagnetic impurities on $H_s$ at $\kappa\gg 1$ is weak: $H_s$ varies from $0.84H_c$ in the clean limit $(a = 0)$ to $H_s\simeq 0.812H_c$ in the dirty limit $(a = 20)$. By contrast, pairbreaking magnetic impurities diminish $T_c$ and $H_s$ \cite{lin}. The effect of subgap states on $H_s$ was calculated in Ref. \cite{kub1,kub3,kub4}. The full temperature dependence of the dc depairing current density $J_c(T,l_i)$ was calculated both from the Eilenberger equations for arbitrary mean free path \cite{jd} and from the Eliashberg equations \cite{jde}.

\begin{figure}[ht]
\centering
\includegraphics[scale=0.6, trim={40mm 70mm 40mm 48mm},clip]{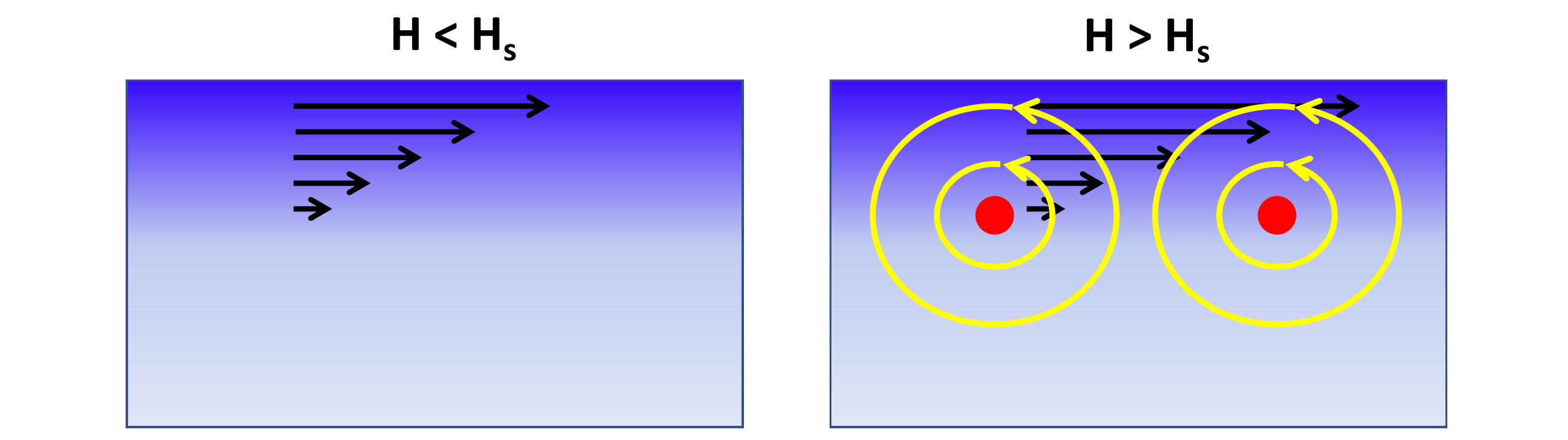}
\caption{An illustration of how penetration of vortices at $H>H_s$ prevents the breakdown of superconductivity by Meissner current. Left: screening  Meissner currents at $H$ slightly below $H_s$. Right: counterflow produced by penetrating vortices against the screening currents at $H$ slightly above $H_s$. Vortex cores are shown by red dots.}
\label{fig13}
\end{figure}

An important question is whether the dc superheating field is the true field limit of the Meissner state under strong microwaves \cite{q1}. The answer depends on the relation between the time formation of the vortex core $\tau_v$ and the rf period: if $\nu\tau_v\gg 1$ vortices do not have enough time to form and the dynamic superheating field $H_d$ may exceed the static $H_s$.  The core formation time $\tau_v$ can be evaluated as $\tau_v\sim \xi/v_d\simeq (v_F/\pi v_d)\hbar/\Delta$, where $v_d$ is the terminal velocity of a vortex penetrating through the surface at $H=H_s$. Measurements of $v_d$ in Pb \cite{vel1} and Nb-C \cite{vel2} gave 
$v_d\sim 10-20$ km/s. Taking $\xi\sim 40$ nm for Nb yields $\tau_v\sim (2-4)\times 10^{-12}$ s and $\nu\tau_v\sim 10^{-2}-10^{-3}$ at GHz frequencies. In this case  penetration of vortices occurs nearly instantaneously once $H(t)$ exceeds $H_s$. It is the penetration of vortices which preserves the superconducting state at $H(t)>H_s$, as illustrated in Fig. \ref{fig13}. At $H<H_s$ the Meissner state is metastable but at $H(t)>H_s$ the screening current density $J(0)$ at the surface exceeds $J_c$ so to prevent the breakdown of superconductivity, vortices penetrate and produce current counterflow which keeps $J(0)$ below $J_c$. Here a delay with penetration of vortices at $\nu\tau_v\ll 1$ and $H>H_s$ would destroy superconductivity but not to extend the field region of the Meissner state. Measurements on Pb, In, InBi and SnIn at 90-300 MHz  gave the breakdown field $H_b(T)\approx H_s(T)$ \cite{q1}. Pulse rf measurements on Nb gave $H_b(T)\approx H_s(T)$ at all $T$ \cite{q2,q3}. However, for Nb$_3$Sn, it was found that $H_b(T)\approx H_s(T)$ only near $T_c$ but becomes smaller than $H_s(T)$ at $T\ll T_c$ \cite{q2,q3}.     

The breakdown of the Meissner state at the static superheating field $H_s$ at $\nu\tau_v\ll 1$ implies that nonequilibrium quasiparticles, for which the electron-phonon relaxation time $\tau_\epsilon(T)$ can exceed the rf period at $T\ll T_c$, do not play a significant role.  This may happen in a dirty superconductor  
in which the quasiparticle gap $\epsilon_g$ remains finite at $H_a=H_s$ (see Fig. \ref{fig8}c). Here the density of thermally-activated quasiparticles $n_{qp}\propto\exp(-\epsilon_g/k_BT)$ at $H_a=H_s$ and $k_BT\ll \epsilon_g$ is much smaller than the superfluid density $n_s$, so their slow relaxation has a negligible effect on the breakdown of superconducting condensate and $H_d\to H_s$.  However, in a clean superconductor with $l_i>8.7\xi_0$ the gap $\epsilon_g$ closes at $H_a<H_s$ so the breakdown of superconductivity is affected by slow relaxation of quasiparticles with $n_{qp}(B_s)\sim n_s$, and the dynamic superheating field $H_d$ can be different from $H_s$.    
               
Calculation of a dynamic superheating field $H_d(T,\omega)$ or a dynamic depairing current density $J_d(T,\omega)$ requires solving equations of nonequilibrium superconductivity which account for the effects of rf current pairbreaking on the DOS and $\Delta(t)$ and the energy relaxation due to inelastic scattering of quasiparticles on phonons \cite{kopnin}.  In the case of slow temporal and spatial variations of the order parameter $\psi(\mathbf{r},t)=\Delta({\bf r},t)e^{i\chi({\bf r},t)}$ and $\mathbf{J}(\mathbf{r},t)$ at $T\approx T_c$, these equation can be reduced to the time-dependent Ginzburg-Landau (TDGL) equations for a dirty gapped superconductor \cite{kopnin,tdgl}:
\begin{eqnarray}
\hspace{-23mm} \frac{\tau_{GL}}{\sqrt{1+(2\tau_\epsilon\Delta/\hbar)^2}}\left(\hbar\frac{\partial}{\partial t}+2ie\varphi+\frac{2\tau_\epsilon^2}{\hbar}\frac{\partial \Delta^2}{\partial t}\right)\!\psi =\left(1-\frac{\Delta^2}{\Delta_0^2}\right)\!\psi+\xi_s^2\left(\mathbf{\nabla}-2ie\mathbf{A}\right)^2\!\psi,
\\ \label{gtdgl}
\mathbf{J}=-\frac{\pi\sigma_s}{4e\hbar k_BT_c}\Delta^2(\nabla\chi+2e{\bf A})-\sigma_s\left(\mathbf{\nabla}\varphi+\frac{\partial \mathbf{A}}{\partial t}\right),
\label{j}
\end{eqnarray}
Here $\xi_s=[\pi\hbar D_s/8k_B(T_c-T)]^{1/2}$, $\tau_{GL}=\pi\hbar/8k_B(T_c-T)$, $\varphi({\bf r},t)$ is the electric potential,
$\Delta_0^2=8\pi^2k_B^2T_c(T_c-T)/7\zeta(3)$, and $\sigma_s=2e^2D_sN_s$.   Equations (\ref{gtdgl}) and (\ref{j}) were derived from the kinetic BCS  theory, assuming that $\mathbf{A}(\mathbf{r},t)$ and $\Delta(\mathbf{r},t)$ vary slowly over $\xi_s(0)$, the diffusion length $L_\epsilon=(D_s\tau_\epsilon)^{1/2}$ and $\tau_\epsilon$ ~\cite{kopnin,tdgl}.

\begin{figure} [ht]
\includegraphics[scale=0.56]{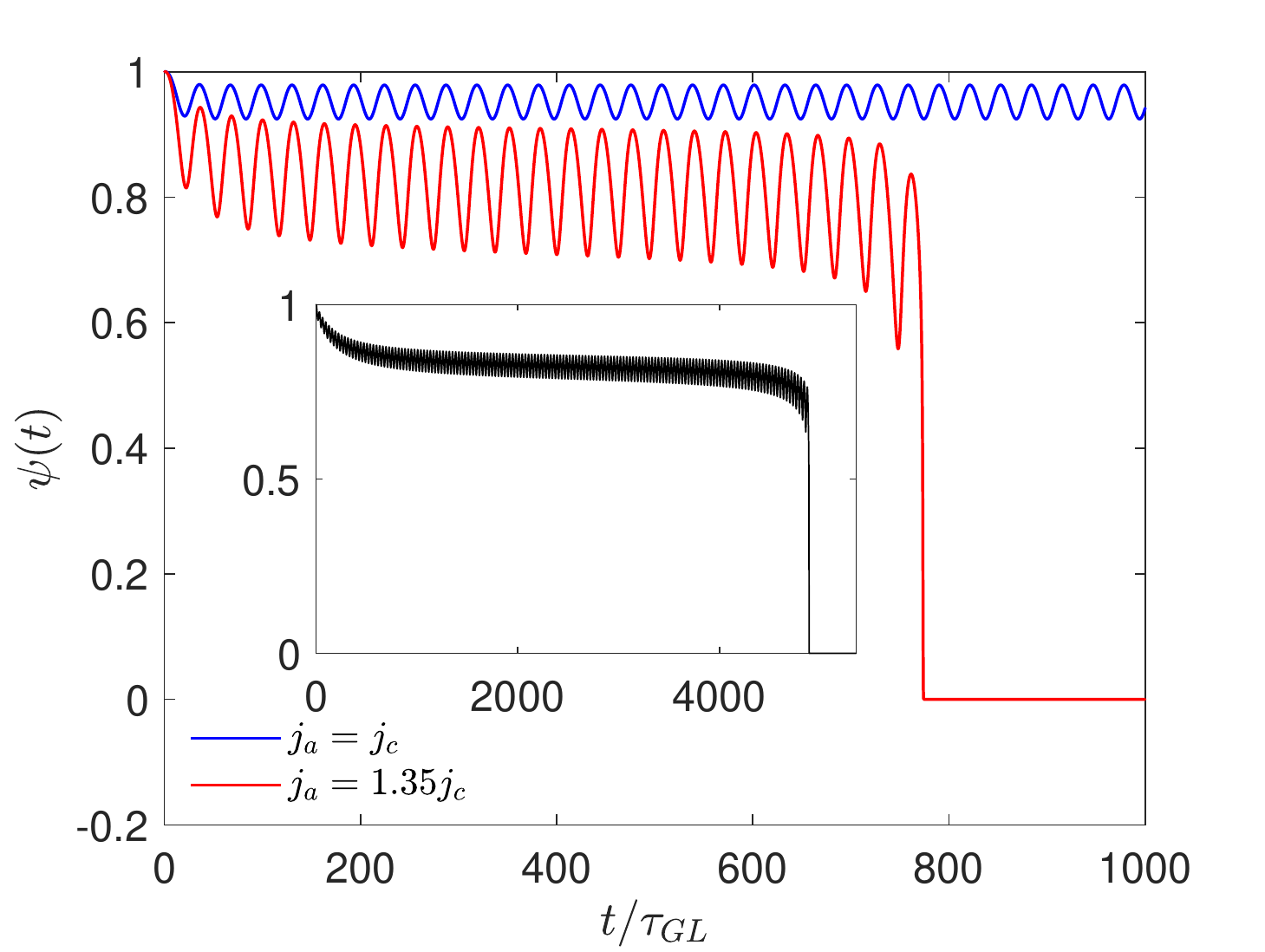} 
\includegraphics[scale=0.56]{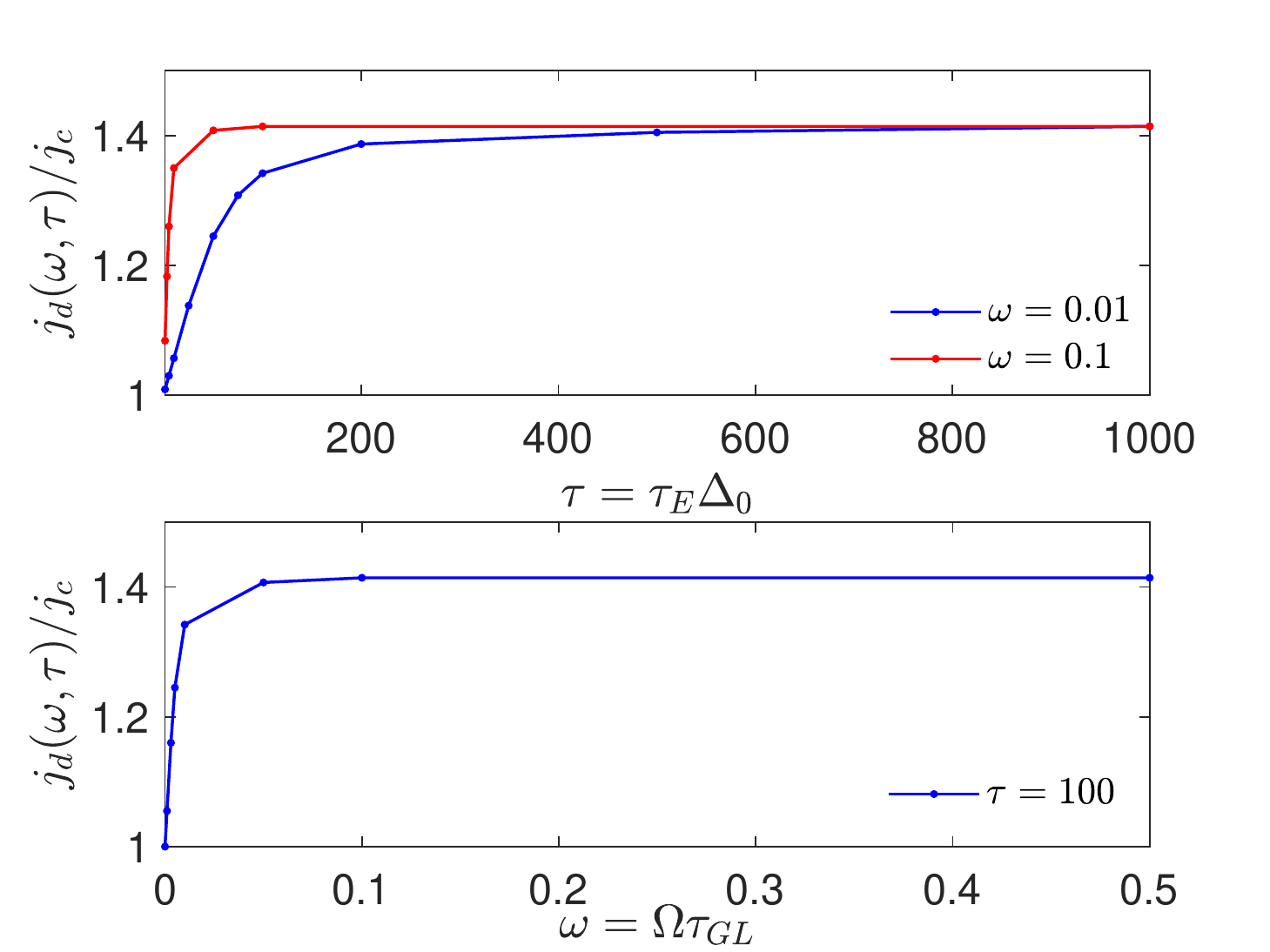}
\caption{Left: Time dependence of $\psi(t)=\Delta(t)/\Delta_0$ calculated at $J=J_a\sin \omega t$, $\hbar\omega=0.1\Delta_0$, $\tau=\tau_E\Delta_0/\hbar=10$, $J_a=J_c$ 
and the dynamic depairing current $J_d=1.35J_c$ at which $\psi(t)$ vanishes abruptly. The inset shows $\psi(t)$ calculated at $\tau=100$ and $J_a=\sqrt{2}J_c$. 
Right: dependencies of $J_d(\omega,\tau)$ on $\tau$ and $\omega$. Here $J_d$ levels off at $\sqrt{2}J_c$ at $\Omega\tau_E\gg 1$, where $\Omega$ is the rf frequency  Reproduced from Ref. \cite{dynH}. }
\label{fig14}
\end{figure}

The dynamic depairing current density $J_d(T,\omega)$ was calculated in Ref. \cite{dynH} by solving both the TDGL equations and a full set of nonequilibrium equations for a dirty superconductor at $T\approx T_c$ \cite{tdgl}. Both approaches gave qualitative similar results illustrated in Fig. \ref{fig14}. The left panel  of Fig. \ref{fig14} shows how $J_d$ is defined: the simulations start from an initial superconducting state until the steady state oscillations of $\Delta(t)$ set in at $J_a<J_d$.  Once $J_a$ is increased to $J_d$, $\Delta(t)$ drops to zero after a transient period. The dependencies of $J_d(\omega,\tau_\epsilon)$ on $\omega$ and the electron-phonon relaxation time $\tau_\epsilon$ are shown in the right panels. One can see that $J_d(\omega)$ increases with $\omega$ and levels off at $J_d(\omega)\to \sqrt{2}J_c$ at $\omega\tau_\epsilon\gg 1$, which can be understood as follows.  

Near $T_c$ the relaxation time constant $\tilde{\tau}=\tau_{GL}\sqrt{1+(2\Delta_0\tau_\epsilon/\hbar)^2}$ of $\Delta(t)$ in Eq. (\ref{gtdgl}) depends on both the GL time $\tau_{GL}$ and the electron-phonon time $\tau_\epsilon$. At high frequencies $\omega\tilde{\tau}\gg 1$ the order parameter cannot follow rapid oscillations of the magnetic drive $A(t)$ so $\Delta(t)$ undergoes small-amplitude temporal oscillations around a mean value $\bar{\Delta}$, as one can see in Fig. \ref{fig14}.  Here $\bar{\Delta}$ is determined by Eq. (\ref{gtdgl}) in which the pairbreaking term $4e^2A^2(t)\propto J_a^2\sin^2\omega t$ is replaced with its value $\langle J_a^2\sin^2\omega t\rangle=J_a^2/2$ averaged over the rf period. As a result, the pairbreaking term at $\omega\tilde{\tau}\gg 1$ is reduced in half as compared to low frequencies $\omega\tilde{\tau}\ll 1$ at which $\Delta(t)$ follows $J(t)$ adiabatically and the superconductivity breakdown occurs once the peak value of $J(t)$ exceeds $J_c$.  Thus, the dynamic depairing current density $J_d$ at $\omega\tilde{\tau}\gg 1$ is by the factor $\sqrt{2}$ larger than the static $J_c$. The resulting enhancement of $H_d\to \sqrt{2}H_s$ at $\omega\tilde{\tau} \gg 1$ was obtained in Ref. \cite{dynH} by solving both the TDGL equations and a full set of dynamic equations for $\Delta(t)$ and kinetic equation for the nonequilibrium distribution function derived in Ref. \cite{tdgl}. In the dirty limit at $\kappa\gg 1$ the dynamic superheating field is related to $J_d$ by $H_d=J_d\lambda$, where the field dependence of $\lambda$ due to the nonlinear Meissner effect can be neglected. Calculations of $H_d(T,\omega,l_i)$ at a finite $\kappa$,  particular in the clean limit at low temperatures, have not yet been done.

\section{Surface nanostructuring}

\begin{figure}
\centering
\includegraphics[scale=0.45, trim={30mm 30mm 50mm 35mm},clip]{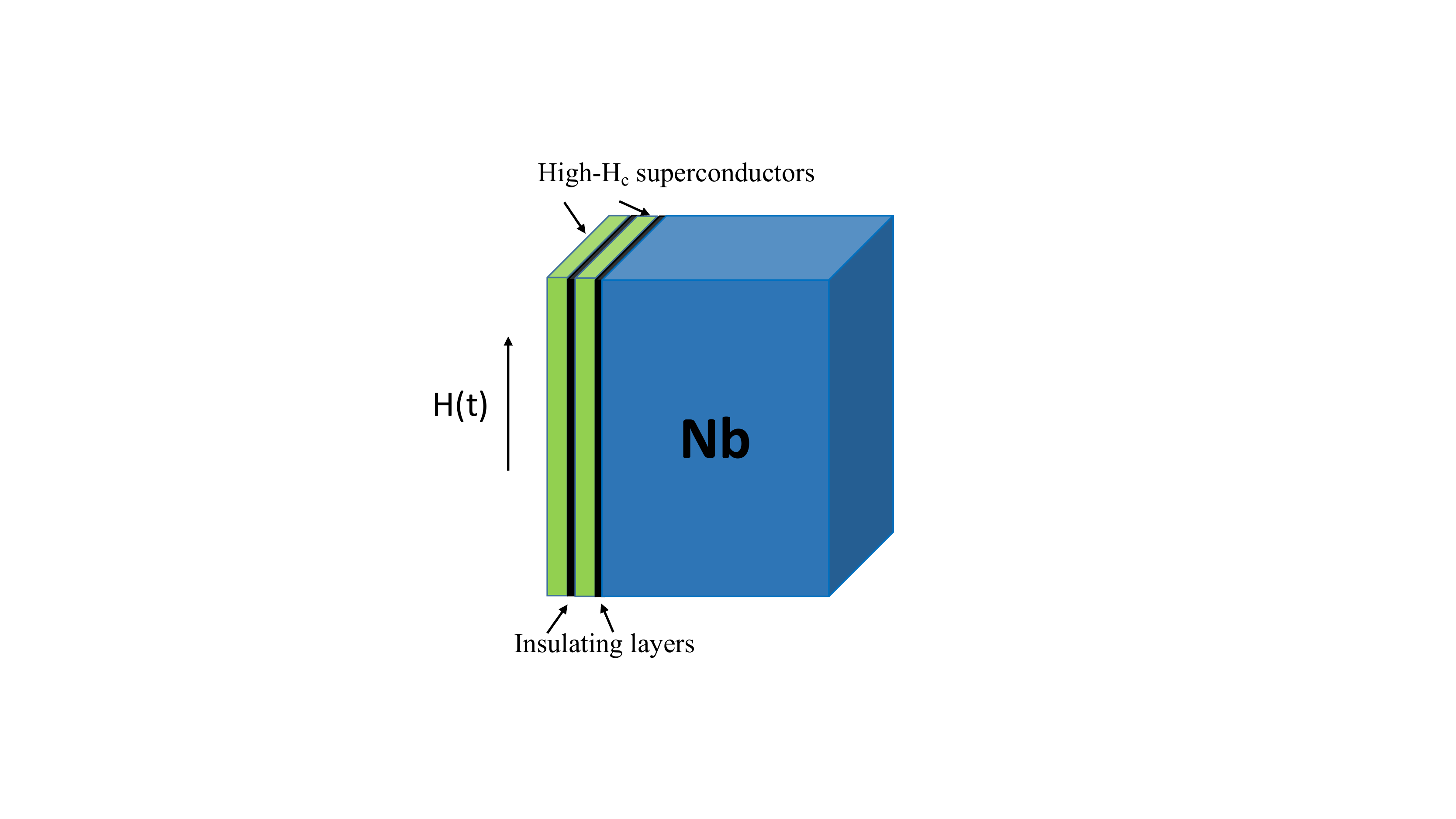}
\includegraphics[scale=0.4]{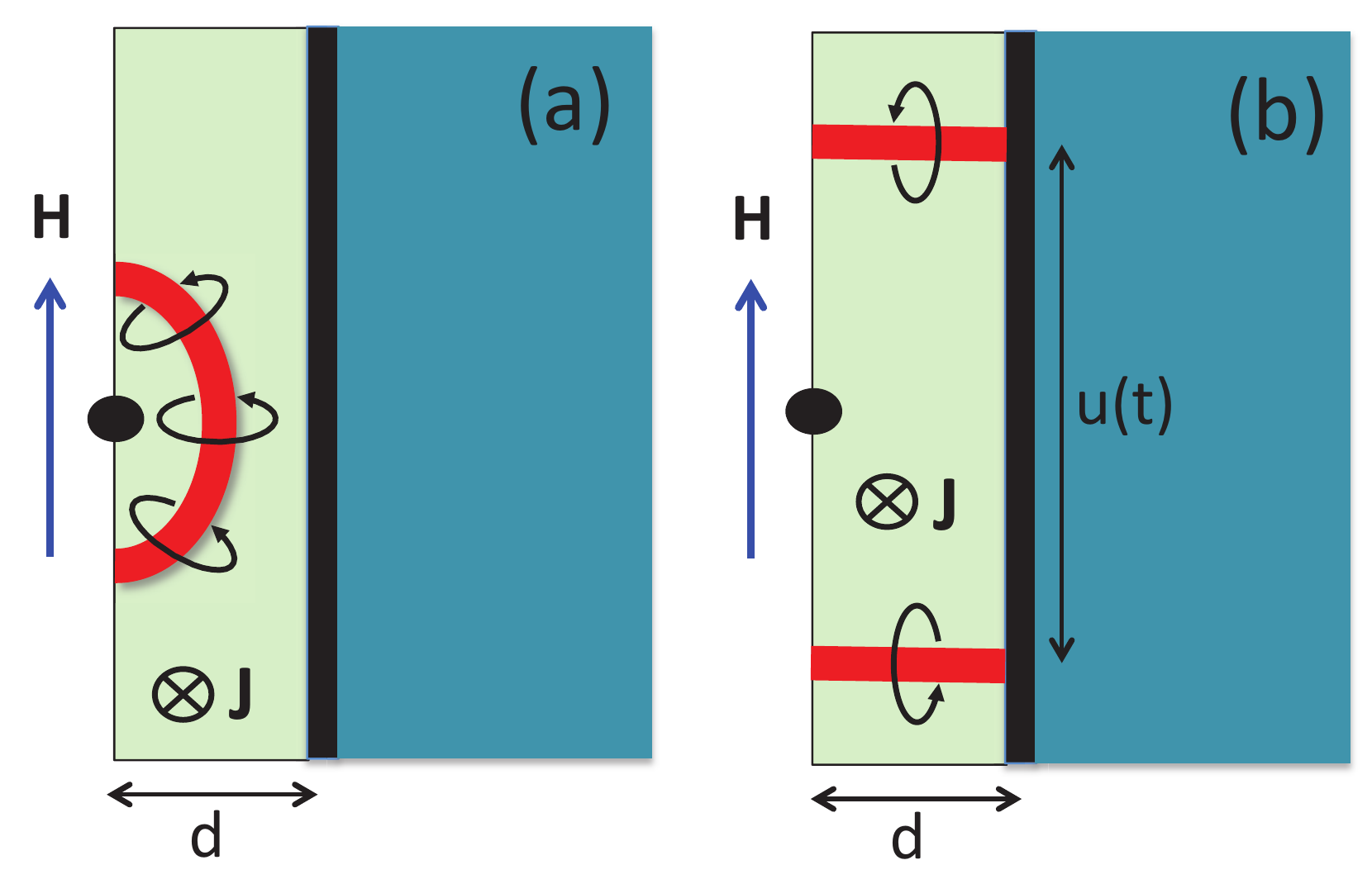}
\caption{\label{ml} Top: SIS multilayer coating of a Nb substrate. Bottom: Propagating vortex loop (a) turning into the vortex-antivortex pair (b) in the S layer with a surface defect (black dot) which lowers the field threshold of vortex penetration. Meissner current flows perpendicular to the screen. Reproduced from Ref. \cite{agsust}.}
\label{fig15}
\end{figure}

There is a strong interest in the development of superconducting resonant structures with the breakdown magnetic field $H_b$ exceeding the current state-of-the-art of Nb cavities  \cite{rec1,rec2}.  This task  requires superconductors with $H_s$ and $T_c$ larger than $H_s^{Nb}$ and $T_c^{Nb}$ to provide lower $R_s$ and higher breakdown fields in the Meissner state.  Many such materials exist \cite{srfmat,smat}, but all of them have $H_{c1}$ lower than $H_{c1}^{Nb}$, which makes them more vulnerable to penetration of vortices and high rf losses at $H_a>H_{c1}$. 
To address the low-$H_{c1}$ problem of high-$H_c$ materials, it was proposed to coat the surface of Nb resonators with multilayers of thin superconductors (S) separated by dielectric (I) layers \cite{ml}. Here the S layer material has $H_s$ higher than the superheating field $H_{s0}$ of Nb, whereas the thickness of S layers $d$ is smaller than $\lambda$ of the coating material and the thickness of I layers can be a few nm to suppress the Josephson coupling between S layers, as shown in Fig. \ref{fig15}. Such SIS structures can withstand rf fields limited by a higher superheating field of the S-layer: for instance, using Nb$_3$Sn with $B_c = 530$ mT \cite{smat} could potentially double the breakdown field as compared to Nb. In turn, the field onset of penetration of a parallel vortex in the S layer with $d\ll \lambda$ is increased because of a larger parallel 
$H_{c1}=(2\phi_0/\pi\mu_0d^2)[\ln(d/\xi)-0.07]$ in thin films \cite{aaa,stej}. Here $H_{c1}$ at $d\ll \lambda$ depends weakly on the materials properties, so getting $B_{c1}>$ 200 mT requires $d_s < 100$ nm at $\xi_s=5$ nm. Geometrical enhancement of $H_{c1}$ has been observed on films of different materials in uniform parallel fields \cite{exp1,exp2,exp3,exp4,exp5,exp6,exp7}. 

The maximum field $H_m$ screened by the S layers of total thickness $d_s\gg\lambda$ is limited by the superheating field of the S coating material \cite{ml}, for example, 
$B_s\simeq 0.84B_c=454$ mT for Nb$_3$Sn at $T\ll T_c$. At $H=H_s$ the Meissner current in top S layer becomes unstable and the magnetic barrier for penetration of vortices vanishes \cite{bean,galaiko,matricon,chapman}. However, there is an optimum layer thickness $d_m$ at which $H_m$ exceeds the superheating fields of both S-layer and the Nb substrate, as has been shown by numerical simulations of a parallel vortex in the London model \cite{kubo,kubm,cml}, numerical simulations of the GL equations \cite{cml}, and by analytical calculations of the depairing instability of Meissner currents \cite{aml}. The latter approach yields: 
\begin{eqnarray}
d_m=\lambda\ln \left( \mu +\sqrt{\mu^2+k} \right),
\label{dm} \\
H_m=\left[ H_s^2+\left( 1-\frac{\lambda_0^2}{\lambda^2}\right)H_{s0}^2 \right]^{1/2},
\label{hm}
\end{eqnarray}
where $\mu=H_s\lambda/(\lambda+\lambda_0)H_{s0}$, and $k=(\lambda-\lambda_0)/(\lambda+\lambda_0)>0$, and the subscript $0$ labels the substrate properties. The optimum thickness is due to a counterflow induced by the substrate in the S-layer which can withstand higher fields if $\lambda>\lambda_0$  \cite{agsust,kubo,aml}.

\begin{figure}[ht]
\centering
\includegraphics[scale=0.75, trim={70mm 20mm 10mm 15mm},clip]{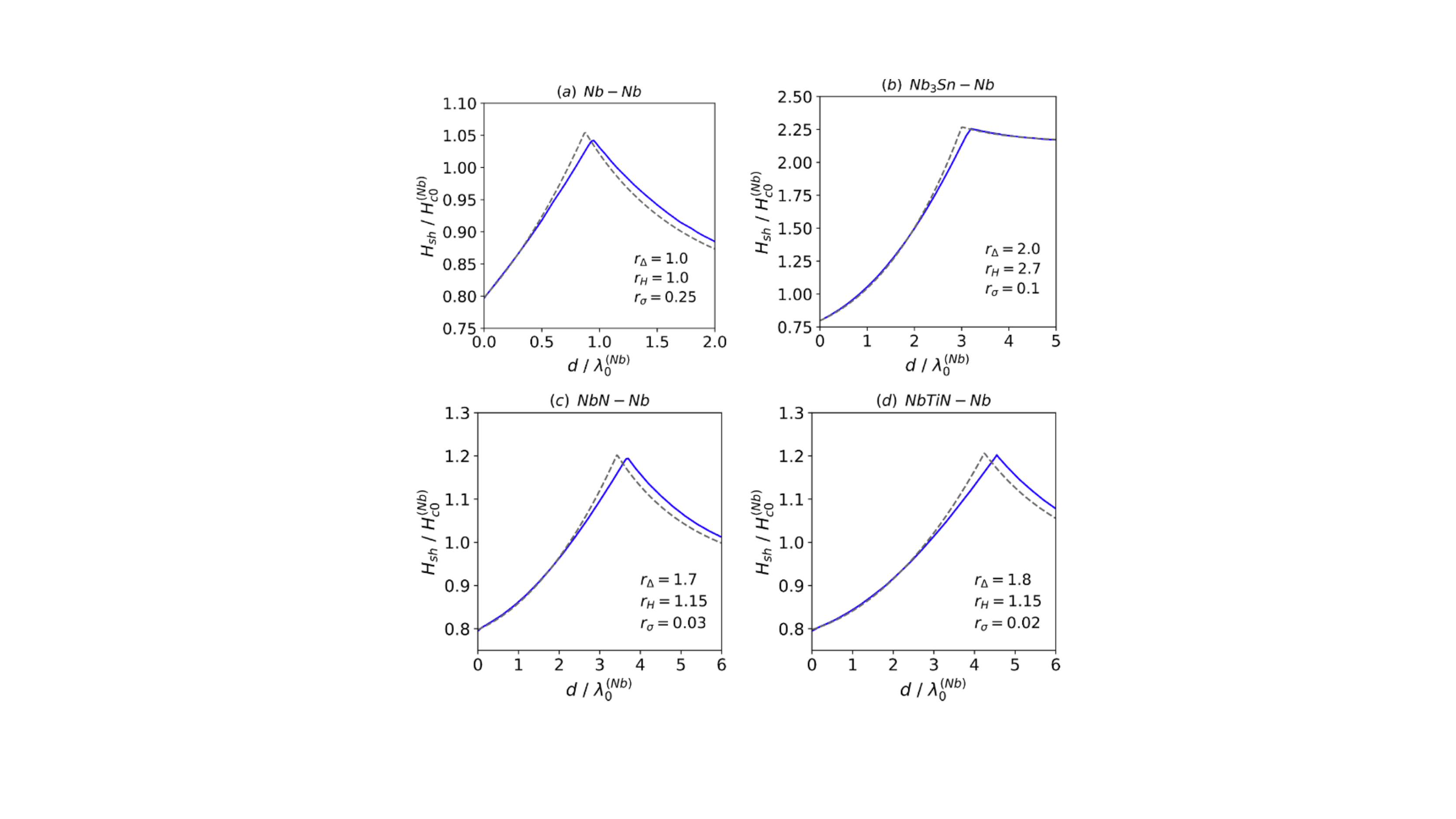}
\caption{Dc superheating fields as functions of the overlayer thickness $d$ for Nb coated with different coating materials in the dirty limit and $\kappa\gg 1$.
The solid curves are calculated by solving 
the coupled Maxwell-Usadel equations at $T\to 0$. The dashed curves show the London approximation in which the peak values 
$d_m$ and $H_m$ are given by Eqs. (\ref{dm}) and (\ref{hm}):  (a) dirty Nb-Nb structure with 
$(r_\Delta, r_H, r_\sigma) = (1, 1, 0.25)$. (b) Nb$_3$Sn-Nb
structure with $(r_\Delta, r_H, r_\sigma) = (2, 2.7, 0.1)$. (c) NbN-Nb
structure wiih $(r_\Delta, r_H, r_\sigma) = (1.7, 1.15, 0.03)$. (d) NbTiN-Nb
structure with $(r_\Delta, r_H, r_\sigma) = (1.8, 1.15, 0.02)$. Here $r_\Delta=\Delta^i(0)/\Delta_0(0)$, $r_H=H^i_{c}(0)/H_{c0}(0)$, $r_\sigma=\sigma^i_s/\sigma_{s0}$, where 
$\Delta^i(0),\, H_c^i(0),\, \sigma_s^i$ are the respective parameters of the i-th coating material at $T=0$, and the index $0$ labels the parameters of Nb. 
Reproduced from Ref. \cite{kub3}. }
\label{fig16}
\end{figure}

As follows from Eqs. (\ref{hm}) and (\ref{dm}), $H_m$ can also be enhanced by alloying the surface, where $\lambda$ is increased due to a shorter mean free path  \cite{aml}.  For instance, a dirty Nb layer with $l_i\simeq 2$ nm has $\lambda\simeq\lambda_0(\xi_0/l_i)^{1/2}\simeq180$ nm and $\xi=(l_i\xi_0)^{1/2}\simeq 9$ nm.  Taking $B_s\approx 0.84 B_c$ for $\kappa=\lambda_0/l_i=20$ in Eqs. (\ref{dm})-(\ref{hm}), yields $d_m=0.44\lambda = 79$ nm and $B_m=\mu_0H_m=1.44 B_c = 288$ mT, some $20\%$ higher than $B_{s0}=240$ mT of pure Nb. If $\lambda_0^2\ll \lambda^2$ Eq. (\ref{hm}) gives $B_m=\sqrt{B_s^2+B_{s0}^2} = 1.465B_c = 293$ mT. Therefore, the maximum screening fields can be increased by depositing thin alloyed Nb layers at the surface of clean Nb, which may also bring the benefits of the field-induced reduction of $R_s(B_a)$. Evidence of enhanced vortex penetration field by a dirty Nb/Al$_2$O$_3$ bilayer deposited onto the Nb cavity was observed in Ref. \cite{nbml}. 

Equations (\ref{dm}) and (\ref{hm}) obtained from the London theory are in good agreement with self-consistent  numerical calculations of $H_s(d)$ 
from coupled Usadel and Maxwell equations for dirty SIS multilayers \cite{kub3}.  Shown in Fig. \ref{fig16} are examples of $H_s(d)$ as functions of the overlayer thickness $d$ calculated in Ref. \cite{kub3} for dirty Nb,  Nb$_3$Sn,  NbN, and NbTiN. deposited on the Nb substrate.  Here $H_s(d)$ first increases with $d_s$ due to the counterflow effect \cite{kubo,aml}, reaches a cusp-like maximum and then decreases down to the superheating field of the S overlayer at $d\gg \lambda$.    
For the cases a, c and d, the multilayers do not give a significant gain in the superheating field relative to $H_{s0}\approx 1.2H_{c0}$ at $\kappa\approx 1$ \cite{matricon,chapman} for the clean Nb. This reflects the fact that the layer materials shown in Fig. \ref{fig16} except for Nb$_3$Sn have $H_c$ not much higher than $H_{c0}$ of Nb, and the calculations of Ref. \cite{kub3} were done in the diffusive limit $l_i<\xi_0$ and  $\kappa\gg 1$ in which $H_{s0}\simeq 0.8H_{c0}$ is about $35\%$ smaller than $H_{s0}$ at $\kappa\approx 1$. However, Nb$_3$Sn represented by Fig. \ref{fig16}b provides a significant gain in $H_m\approx 2.25H_{c0}$ relative to Nb, consistent with the proposal of Ref. \cite{ml}.

Besides the deposition of a dirty film on the surface of a cleaner superconductor, $H_m$ can also be increased by forming a dirty layer with a gradually decreasing concentration of nonmagnetic impurities as was shown in Ref. \cite{sauls} by solving the Usadel equations with an  inhomogeneous diffusivity $D(z)=D_\infty+D_1\exp(-z/L_i)$, where $L_i$ is a thickness of the dirty layer.  Yet increasing $H_m$ by producing a smooth profile of impurity concentration or direct deposition of an alloyed or a high-$B_c$ layer onto a non-ideal S surface without a dielectric interlayer \cite{kubm} does not necessarily widen the field region of the Meissner state. In the absence of I layer penetration of vortices at $H=H_m$ is impeded by the force ${\bf F}=-\nabla\varepsilon(x)$ caused by the gradient of the vortex energy $\varepsilon(x) = \phi_0^2\ln\kappa(x)/4\pi\mu_0\lambda^2(x)$.   For a smooth concentration profile, the maximum pinning force $F_p\sim \phi_0(H_{c10}-H_{c1})/L_i$ is much smaller than the pinning force $F_m\simeq H_{c10}/\xi_0$ for the S-I interface if $L_i\gg\xi$ and $\kappa\gg 1$.  For a high-$H_c$ or alloyed layer deposited directly onto Nb, the idealized sharp energy barrier due to a stepwise change in $\lambda(x)$ and $\xi(x)$ in the London model \cite{kubm,shmidt} is, in fact, weakened by the proximity effect and inter diffusion of atomic components during film deposition at high temperatures, which broaden the vortex entry energy barrier and significantly reduce the pinning force.   

In a SIS multilayer the I layers assure the necessary stability margin with respect to proliferation of vortex semi-loops penetrating at surface defects. If these expanding vortex semi-loops are not stopped, they trigger thermomagnetic flux jumps \cite{fjump,aval1,aval2,aval3,aval4,aval5,aval6}, particularly at $T\ll T_c$, where the specific heat is small.  At $H_{c1}<H_a<H_s$ the  Meissner state remains metastable due to the Bean-Livingstron barrier \cite{bean}. Many magneto-optical imaging investigations of type-II superconductors \cite{moi1,moi2,moi3,moi4} have revealed premature local penetration of vortices at grain boundaries and other materials and topographic defects at the surface \cite{gb1,gb2,gb3,gb4}. In turn, TDGL and nonlinear electrodynamic simulations have shown that  
surface defects can reduce the penetration field \cite{vodolaz,trans,mltdgl} and cause flux jets \cite{fj1,fj2} being precursors of themomagnetic avalanches.  

Materials and topographic defects at the surface \cite{cav1,cav2,Nbdef} reduce the local penetration field from $H_s$ so a lower value $H_i$ which can be close to $H_{c1}$. Figure \ref{fig15} illustrates how the multilayer not only increases $H_s$ but also blocks proliferation  of vortices: a vortex semi-loop penetrating at a small defect in the first S layer cannot not propagate further into the next S layer and then in the superconducting substrate where it can trigger a thermomagnetic avalanche. As $H(t)$ reaches $H_i$ at a week spot, a vortex semi-loop expands under the Lorentz force of Meissner current until it hits the I layer, where most part of the dissipative vortex core disappears in a loss-free flux channel connecting two short vortices of opposite polarity.  This vortex-antivortex pair expands during the positive rf cycle and contracts and annihilates as $H(t)$ changes sign. The disappearance of the most part of dissipative vortex core in the I interlayer does not happen in the case of direct deposition of a dirty or high-$H_c$ layer onto the S substrate. Thus, the SIS multilayer reduces vortex dissipation as compared to a thick Nb$_3$Sn film with $d\gg\lambda$ \cite{agsust}. TDGL simulations of penetration of straight vortices into a SIS multilayer have been performed in Ref. \cite{mltdgl}.  

Confinement of the rf power in a thin S layer inhibits expansion of multiple vortex loops in the bulk and blocks dendritic thermomagnetic avalanches that are particularly pronounced in Nb$_3$Sn, NbN or pnictides which have low $\sigma_s$ and the thermal conductivity $k$ \cite{fjump,harshani}. Yet a thin Nb$_3$Sn layer with $d\sim 100$ nm only slightly increases the thermal impedance of the cavity wall, $G=\alpha_K^{-1} + d/k_s+d_i/k_i+d_{Nb}/k_{Nb}$, where $\alpha_K$ is the Kapitza interface thermal conductance. For $d_{Nb}=3$ mm, $k_{Nb}\simeq 20$ W/mK, $\alpha_K=2$ kW/m$^2$K, the Nb$_3$Sn multilayer with $d = 100$ nm, $k_s\simeq 10^{-2}$ W/mK, and Al$_2$O$_3$ dielectric layers with $d_i=4$ nm and $k_i=0.3$ W/mK increases $G$ by $\simeq 5\%$ \cite{cav3}. A thicker Nb$_3$Sn film with $d\simeq 2-3\, \mu$m doubles $G$ and reduces the field of thermal quench, in addition to the smaller $B_{c1}$ of bulk Nb$_3$Sn with $\lambda > 65$ nm \cite{nb3sn2} as compared to $B_{c1}^{Nb}\simeq 170-180$ mT. 

Experiments on MgB$_2$, Nb$_3$Sn, NbN, NbTiN and dirty Nb as overlayers have shown an increase of the dc field onset of penetration of vortices on Nb surfaces coated with different SIS structures \cite{exp1,exp2,exp3,exp4,exp5,exp6,exp7,ml1,ml2,ml3,ml4}. However, the $Q$ factors of SIS multilayers under high-amplitude rf fields $H_a\sim H_c$ have not yet been measured. Low-field measurements of $Q$ on NbN/MgO multilayers \cite{ml3} have shown that they can have lower $R_s$ than bulk Nb. Recently Nb$_3$Sn/Al$_2$O$_3$ multilayers have been developed with a low-field $R_s$ on par with $R_s^{Nb}$ at $T>4$ K \cite{ml4}. The challenge  with the measurements of $R_s(B_a)$ at high fields on materials with $H_c>H_c^{Nb}$ is the lack of experimental techniques capable of applying a strong parallel rf field $H_a\sim H_c$ to a thin film or a multilayer without dissipative penetration of vortices.  Recently the Hall probe setups to measure a dc penetration field which quantifies the field onset of strong vortex dissipation in high-$H_c$ film coatings were developed \cite{harshani,malyshev}. Local nonlinear response of Nb surface has been probed with a  
near-field magnetic microwave microscope \cite{orip}. Yet the quadrupole resonator \cite{quad} is currently the only available techniques to measure 
a nonlinear surface resistance of large (7-8 cm in diameter) thin film multilayers at GHz frequencies.

\section{Trapped vortices} 

Vortices are detrimental to high-Q structures in which even a small number of trapped vortices can dominate rf losses at $T\ll T_c$.  Vortices can be trapped by materials defects during the cooldown of a superconductor through $T_c$ in a stray magnetic field. The field onset of penetration of perpendicular vortices in thin films $H_{c1}^\perp=(1-N)H_{c1}$ is greatly reduced by the demagnetizing factor $N\to 1$ \cite{ehb,demag}. Trapped vortices can limit $Q$ at $T\ll T_c$ in thin film quantum circuits operating at mK temperatures \cite{ph1,ph2,ph3} or resonant Nb cavities at $T<1.5$ K, and can give false signals in the search for magnetic monopoles \cite{monopole}.  Trapped magnetic flux can contribute to rf losses in different ways. In policryistals with weakly coupled grain boundaries or granular films Josephson vortices can penetrate along a network of weak links and give rise to losses at rf field amplitude $H_a$ much smaller than $H_{c1}$ \cite{jv1,jv2,jv3,jv4}. Vortices can also be generated by thermoelectric currents in the case of direct deposition of a higher $T_c$ film on top of a lower $T_c$ substrate or in Nb coatings of Cu cavities. Here temperature gradients arising upon cooling the sample produce magnetic flux when the temperature is reduced below the 
higher $T_c$ of a bimetallic structure \cite{te1,te2}. This mechanism is suppressed in SIS multilayers in which I interlayers 
effectively decouple superconducting films with different $T_c$.  Rf losses in cavities made of thin film type-I superconductors such as Al 
result from trapped flux in the intermediate state \cite{catel}.      

\begin{figure}[ht]
	\centering
	\includegraphics[scale=0.22,angle=-90]{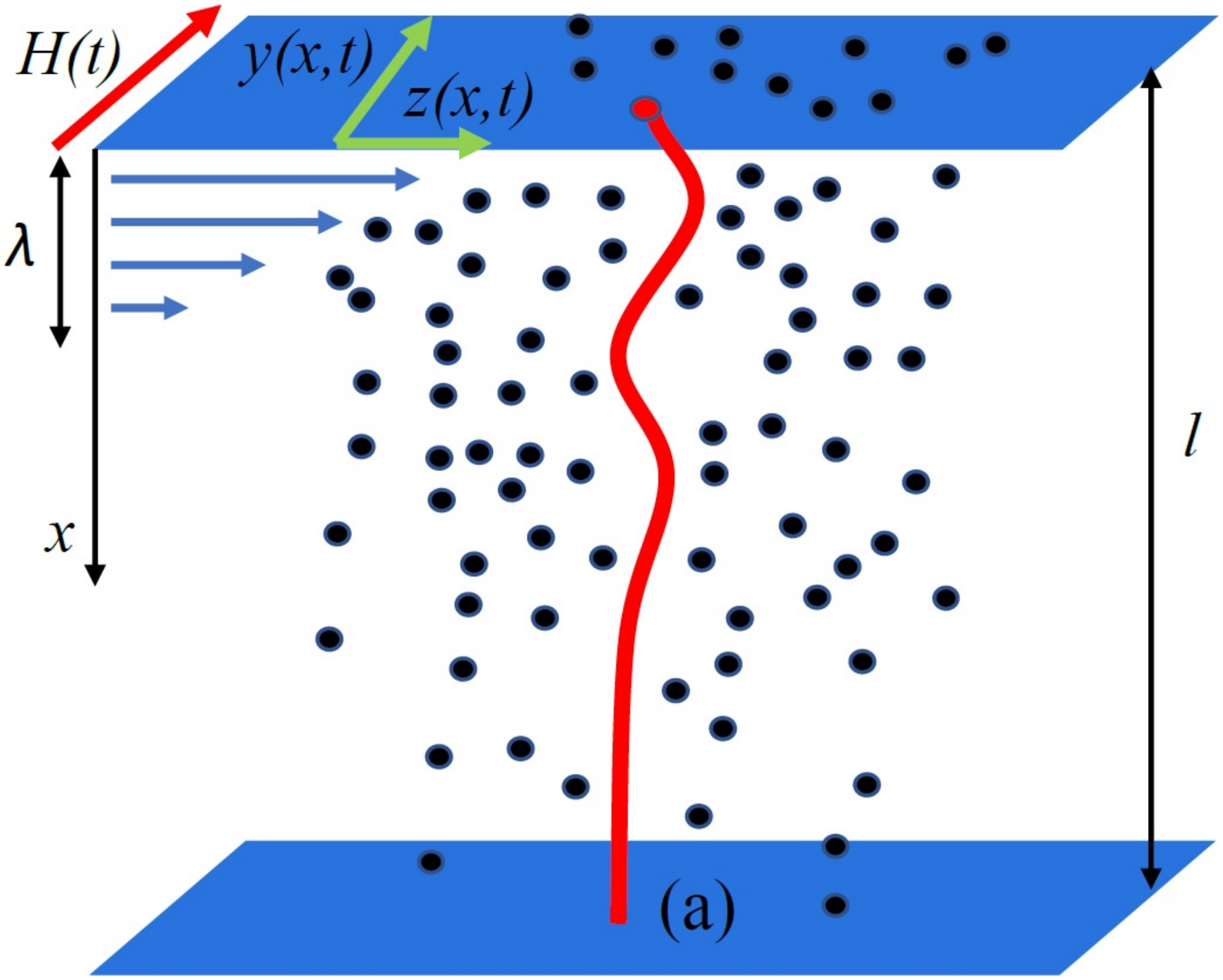}
	\\
	\includegraphics[scale=0.4, trim={30mm 80mm 40mm 65mm},clip]{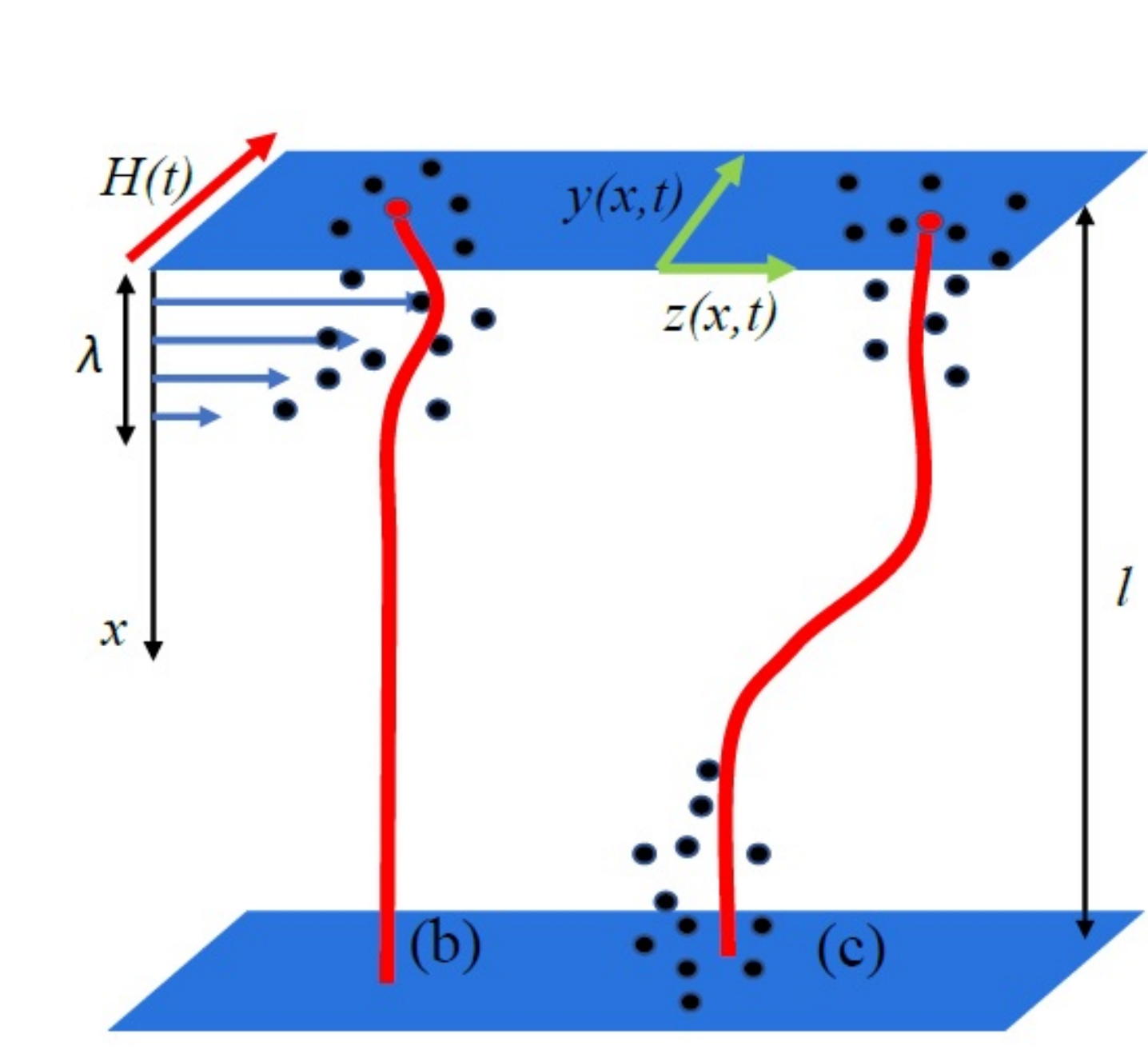}
	\caption{ A trapped vortex driven by the rf surface current for different distributions of pinning centers shown by black dots: (a) bulk collective pinning (b) surface pinning (c) cluster pinning. Reproduced from Ref. \cite{manula2}.} 
	\label{fig17}
\end{figure}

The rest of this review focuses on rf losses of Abrikosov vortices with normal cores \cite{tinkh}. Such vortices trapped by a random pinning potential of materials defects can bundle together, forming localized hotspots which have been revealed by temperature mapping of Nb cavities \cite{bak1,tmap,gc2} and thin film structures \cite{vortrf,anlage}, as well as by magnetic mapping \cite{mmap1,mmap2,mmap3}. Unlike hotspots caused by lossy materials defects, vortex hotspots can be moved or fragmented by temperature gradients produced by external heaters \cite{broom} or scanning laser \cite{gc2,laser,lsm1,lsm2} or electron \cite{esm1,esm2} beams. 
Low-field rf losses of pinned vortices have been thoroughly investigated in the literature \cite{ehb,gc2,gc1,gr,cc,ce,gol}. Nonlinear quasi-static electromagnetic response of vortices has been evaluated qualitatively for both strong pinning \cite{ce,wila} and weak collective pinning  \cite{liarte}.  Yet the rf response of a perpendicular vortex in a film  has distinctive features evident from Fig. \ref{fig17} which shows examples of a vortex trapped by randomly-distributed materials defects in the bulk (a), pinning centers segregated at the surface (b) and clusters of pins (c). Here sparse vortices in high-Q resonators are driven by the Lorentz force of surface Meissner current which causes their bending distortions extending over an elastic skin depth $L_\omega>\lambda$  \cite{gc2}. As a result, a vibrating vortex segment interacts with only a few pins, while the rest of the vortex does not move. In this case the rf response of the vortex becomes dependent on its position in a particular configuration of pinning centers, resulting in strong mesoscopic fluctuations of local $R_i$. Except for short vortices in thin films, the widely used Gittleman and Rosenblum model \cite{gr} is not applicable to the cases shown in Fig. \ref{fig17} which require numerical simulations of a vibrating elastic vortex interacting with a few pinning centers \cite{cgrass,manula1,manula2,manula3}.  

The vortex trapped in a film is driven by the Meissner current density $J(0)\sim H/\lambda$ which 
can be much higher than a depinning current densities $J_p$. For Nb at $B_a=0.1B_c\simeq 20 $ mT, we have $J(0)\sim 0.1J_d\sim 8\times 10^{11}$ A m$^{-2}$, some 2-3 orders of magnitude higher than typical depinning $J_p$ \cite{ce}. In this case the tip of the vortex is moving with a large velocity $v$ mainly determined by a balance of the Lorentz force $F_L=\phi_0J$ and the viscous drag force, $F_d=\eta(v)v$, where the vortex drag coefficient $\eta$ can depend on $v$. Here $v\simeq \phi_0J/\eta(v)$ can exceed the pairbreaking superfluid velocity of the condensate $v_c=\Delta/p_F$ at $J<J_c$ \cite{bardeen}, where $v_c\simeq 1$ km/s for Nb. Vortices moving faster than the terminal velocity of superflow which drives them have been observed on Pb and Nb-C films in which $v$ can exceed $v_c$ by $1-2$ orders of magnitude \cite{vel1,vel2}. Such high velocities may result from the Larkin-Ovchinnikov (LO) mechanism in which $\eta(v)$ decreases with $v$ as the moving vortex core becomes depleted of nonequilibrium quasiparticles lagging behind \cite{LO}.  The LO theory  predicts a nonmonotonic velocity dependence of the drag force $F_d=\eta(v)v$ which cannot balance the Lorentz force if $v$ exceeds a critical value $v_0$. The LO instability has been observed by dc transport measurements on many superconductors \cite{lo1,lo2,lo3,lo4,lo5,lo6} with typical values of $v_0\sim 0.1-1$ km/s near $T_c$. At low $T$ the LO instability is masked by heating effects which are reduced if sparse trapped vortices are driven by Meissner rf current. The velocity-dependent $\eta(v)$ and instability of flux flow can also result from overheating of moving vortices \cite{gc1,manula2,shklovsk,kunchur} or stretching the vortex core along the direction of motion revealed by the TDGL simulations \cite{vel1,tdgl1,tdgl2}. 

Addressing the variety of the observed field dependencies of $R_i(B_a)$ associated with trapped vortices \cite{tf1,tf2,tf3,tf4,tf5,tf6,tf7,tf8,tf9} require numerical simulations of a driven elastic curvilinear vortex in the case of mesoscopic pinning and a velocity-dependent $\eta(v)$. This was done in Refs. \cite{manula1,manula2,manula3}, where the vortex losses for pinning configurations shown in Fig. \ref{fig17} were calculated by solving the dynamic equation for the coordinates ${\bf u}=[u_x(x,t),u_z(x,t)]$ of the vortex moving in the $yz$ plane:
\begin{equation} 
M \frac{\partial^2 {\bf u}}{\partial t^2}+\eta(v) \frac{\partial {\bf u}}{\partial t}=\varepsilon\frac{\partial^2{\bf u}}{\partial x^2}-\nabla U(x,{\bf u})+\hat{y}\frac{\phi_0H_a}{\lambda}e^{-x/\lambda}\sin\omega t,
\label{eq1}
\end{equation}
where $M$ is the vortex mass per unit length and $\varepsilon=\phi_0^2(\ln\kappa+0.5)/4\pi\mu_0\lambda^2$ is the vortex line energy at $\kappa\gg 1$ \cite{ehb,blatter}. Equation (\ref{eq1}) represents a balance of local forces acting on a curvilinear vortex: the inertial and drag forces in the left hand side are balanced by the elastic, pinning and Lorentz forces in the right hand side. For the most efficient core pinning \cite{ehb,ce,blatter}, $U(x,{\bf u})$ can be represented by a sum of pinning centers modeled by the Lorentzian potential wells of width of the core radius $\approx \xi$ \cite{manula2,embon}:
\begin{equation} 
\!U(x,{\bf u})=-\sum_{n=1}^{N}\frac{U_n}{1+[(x-x_n)^2+|{\bf u}-{\bf r}_n|^2]/\xi^2}.
\label{eq3}
\end{equation}
Here $x_n, {\bf r}_n=(y_n,z_n)$ are the coordinates of the n-th pinning center and $U_n$ is determined by the gain in the condensation energy in the vortex core at the pin \cite{ehb,ce,blatter}. The amplitude $U_n$ defines the elementary pinning energy $u_p=\pi\xi U_n$ and the dimensionless pinning parameter $\zeta_n=2\kappa^2 u_p/\pi\varepsilon\xi$. For a dielectric precipitate of radius $r_0<\xi$, we have $u_p\sim B_c^2r_0^3/\mu_0$ and $\zeta_n\sim (r_0/\xi)^3\kappa^2$ \cite{ce}. For a single impurity with a scattering cross-section $\sigma_i$, we have  $u_p\sim B_c^2\sigma_i\xi/\mu_0$ \cite{thun} and $\zeta_n\sim \sigma_i\kappa^2/\xi^2$. In both cases $\zeta_n$ can be larger than $1$ if $\kappa\gg 1$.   

The viscous drag coefficient $\eta(v)$ can depend on $v$ at high vortex velocities. For instance, the LO model gives \cite{LO}:
\begin{eqnarray} 
\eta=\frac{\eta_0 }{1+v^2/v_0^2},
\label{LO}\\
v_0^2=\frac{D\sqrt{14\zeta(3)}}{\pi\tau_\epsilon(T)}\left(1-\frac{T}{T_c}\right)^{1/2}.
\label{vol}
\end{eqnarray}
Here $\eta_0=\phi_0^2/2\pi \xi^2\rho_s$ is the Bardeen-Stephen drag coefficient \cite{tinkh}, $D$ is the electron diffusivity, and the quasiparticle energy relaxation time $\tau_\epsilon$ is given by Eq. (\ref{tau}).  A similar dependence of $\eta(v)$ on $v$ can also result from overheating of a moving vortex  \cite{gc1,manula2,shklovsk,kunchur}. The LO model predicts a non-monotonic velocity dependence of the drag force $F_d=\eta(v)v$ which can balance the Lorentz force $F_L=\phi_0J$ only if $v<v_0$ and $F_L<\eta_0v_0/2$.  Jumps on voltage-current characteristics caused by the LO instability have been observed on many superconductors \cite{lo1,lo2,lo3,lo4,lo5,lo6} with $v_0\sim 0.1-1$ km/s near $T_c$. These experiments have shown that as $T$ decreases, $v_0(T)$ first increases near $T_c$ and then decreases at lower temperatures \cite{lo5}, consistent with Eqs. (\ref{tau}) and (\ref{vol}). No direct measurements of $v_0(T)$ at low temperatures $T\ll T_c$ have been done. 

The vortex mass $M_s\simeq 2p_F/\hbar\pi^3$ in Eq. (\ref{eq1}) results from quasiparticles trapped in the vortex core \cite{suhl}, but other mechanisms producing  $M\gg M_s$ have been suggested \cite{vm1,vm2,vm3}. For instance, $M\sim 10^2M_s$  was observed in Nb near $T_c$ \cite{golubchik}. At GHz frequencies the effect of the vortex mass on the overdamped vortex dynamics is negligible but becomes essential if the LO instability occurs. Another key characteristic of the vortex shown in Fig. \ref{fig17} is a complex penetration length $L_\omega$ of bending distortions induced by the surface rf Meissner current  \cite{agsust,gc2}:                      
\begin{equation}
L_\omega=\sqrt{\frac{\varepsilon}{k_L+i\eta\omega}},
\label{Lom}
\end{equation}
where $k_L\sim \phi_0J_p/\xi$ is the Labusch pinning spring constant~\cite{ce}. At $\omega\eta\ll k_L$  Eq. (\ref{Lom})  reduces to the Campbell penetration depth \cite{ce} or the Larkin pinning length $L_c \sim \xi (J_d/J_c)^{1/2}$ in the collective pinning theory \cite{blatter}. At  $\omega\eta\gg k_L$ Eq. (\ref{Lom}) yields the elastic skin depth $L_\omega \rightarrow (\varepsilon/\eta\omega)^{1/2}$. For Nb$_3$Sn, $L_\omega\simeq 5.15\lambda=572$ nm at 1 GHz, and $L_\omega\simeq 52\lambda=5.7\,\mu$m at 10 MHz so the rf bending distortions of the vortex can extend well beyond the field penetration depth.

The nonlinear dynamics and rf dissipation of a trapped vortex was addressed by numerical simulations of two coupled nonlinear partial differential equations (\ref{eq1}) for both $u_z(x,t)$ and $u_y(x,t)$ \cite{manula1,manula2}. In Ref. \cite{cgrass} the equation for $u_z$ was disregarded. The power of rf losses $P=R_iH_a^2/2$ per unit area from all trapped vortices is expressed in terms of the residual resistance,   
\begin{equation}
R_i=\frac{\rho_nB_0}{\lambda B_{c2}} r_i.
\label{rii}
\end{equation}
Here the dc inductance $B_0$ defines a mean areal density of trapped vortices $n_v=B_0/\phi_0$, and the dimensionless surface resistance $r_i=2p(\beta,\gamma)/\beta^2$ is proportional to the normalized power $p=P/P_0$ per vortex which depends on the reduced rf field amplitude $\beta=B_a/B_{c1}$ and frequency $\gamma=\omega/\omega_0$, where 
\begin{equation}
P_0=\frac{\lambda \omega_0\varepsilon}{2\pi},  \qquad \omega_0=\frac{2\pi B_{c1}\rho_n}{{B_{c2}\lambda^2 \mu_0}}.  
\end{equation}
These definitions of $\gamma$ and $\beta$ adopted from Ref. \cite{manula1,manula2} should not be confused with the parameters $\beta$ and $\gamma$ used in the previous sections.

\subsection{Weak rf fields}

\begin{figure}[ht]
	\centering
	\includegraphics[scale=0.27,angle=-90]{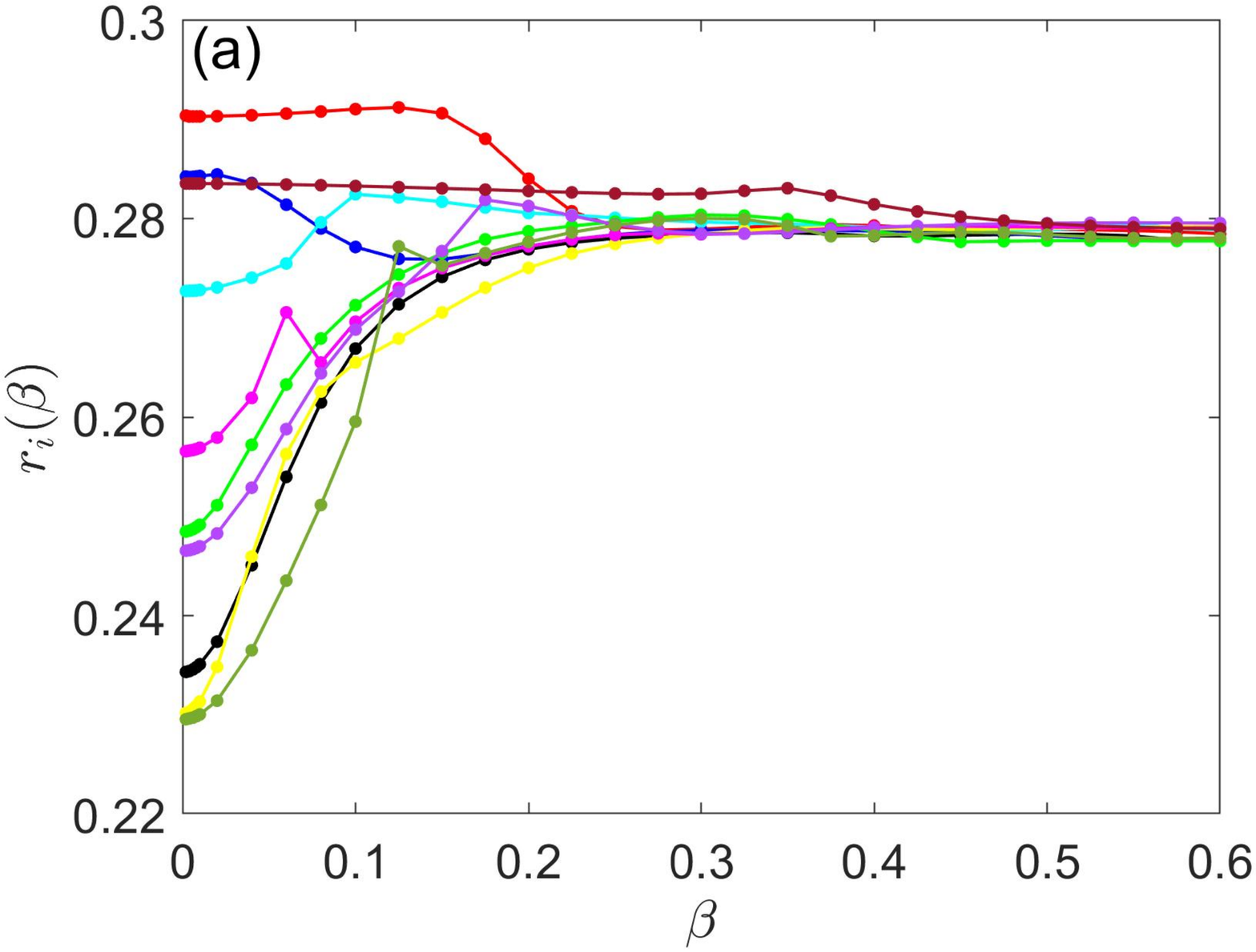} 
	\includegraphics[scale=0.27,angle=-90]{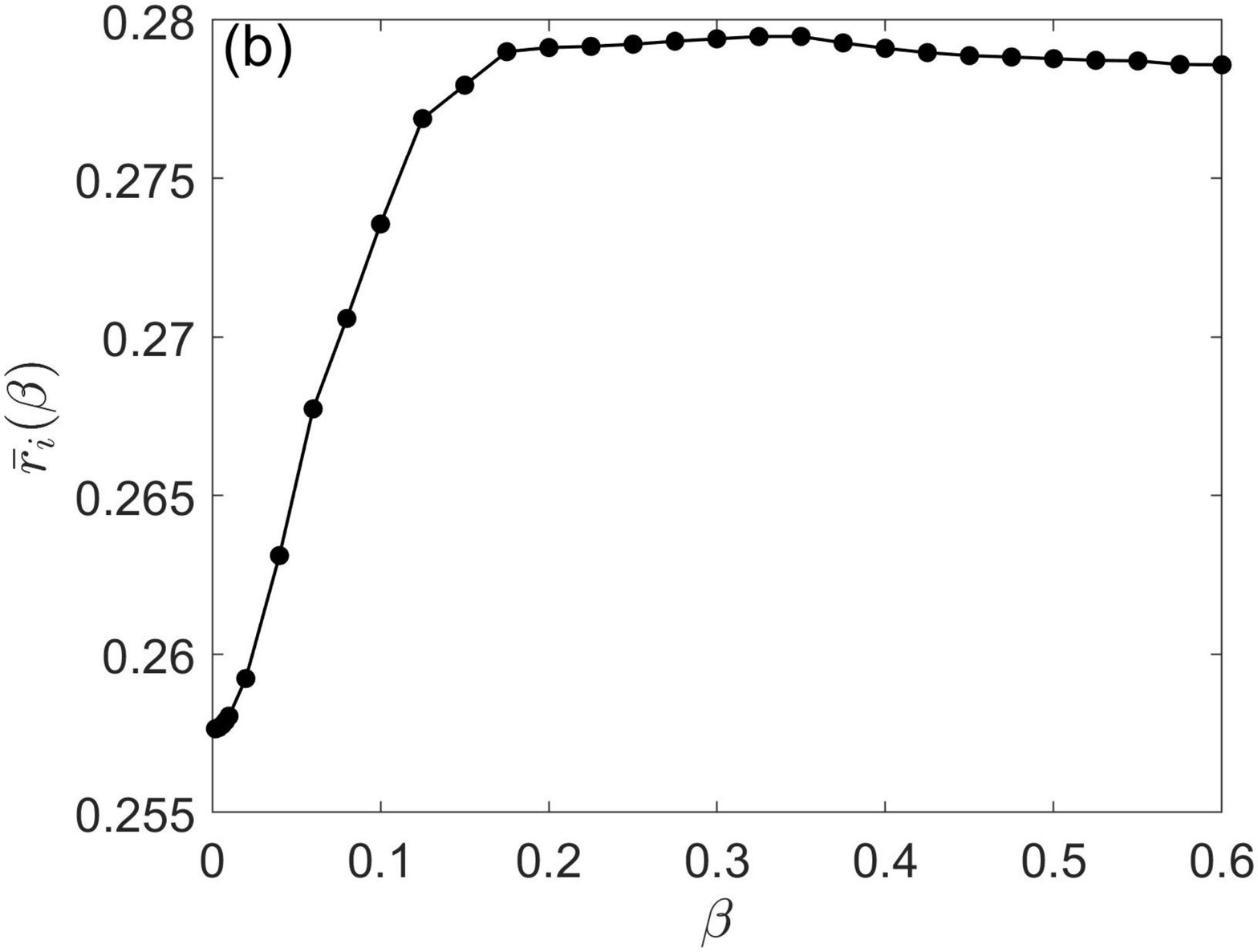} 
	\caption{The dimensionless surface resistance $r_i(\beta)$ as a function of the normalized field $\beta=B_a/B_{c1}$ in a film of thickness at $d=10\lambda$, $\nu=0.04$, the pin density $n_i=1.67 \lambda^{-3}$, $\kappa=10$ and $\zeta_n=1$: (a) ten different random distribution of pins with the same density; (b) averaged $\bar{r}_i(\beta)$. Reproduced from Ref. \cite{manula2} }
	\label{fig18}
\end{figure} 

At weak rf fields the vortex velocities are small $v\ll v_0$, so $\eta=\eta_0$ is independent of $v$ and the effect of vortex mass at GHz frequencies is negligible.  Shown in Fig. \ref{fig18} is the dimensionless surface resistance $r_i$ calculated in Ref. \cite{manula2} for pinning centers distributed randomly over the film thickness (see Fig. \ref{fig17}a).  The global surface resistance $\bar{r}$ obtained by averaging $r_i(\beta,k)$ for different random pin configurations with the same volume pin density is shown in Fig. \ref{fig18}b. The vortex moving in a particular $k$-th pinning landscape produces a unique $r_i(\beta,k)$ which can vary rather non-systematically with the rf field, reflecting many metastable positions of the curvilinear vortex in a random pinning potential.  Figure \ref{fig18}b shows the result of averaging over ten different random pin distribution with the same $n_i=1.67\lambda^{-3}$. The low-field $r_i(\beta,k)$ fluctuate strongly but converge to the same value at high fields.  Here $r_i(\beta,k)$ at low fields is strongly affected by pinning, whereas $r_i(\beta,k)$ at higher fields is mostly limited by the vortex drag and the effect of mesoscopic pinning fluctuations weakens.  The averaged $\bar{r}_i(\beta)$ shown in Fig. \ref{fig18}b first increases with $\beta$ and levels off at $\beta> 0.2$ as the low-field $\bar{r}_i(\beta)$, which mostly results from pinning hysteretic losses, crosses over to a drag-dominated $\bar{r}_i(\beta)$. A similar low-field dependence of $R_i(H)$ has been observed on Nb cavities \cite{liarte,cgrass}. 

\subsection{Microwave reduction of $R_i(B_a)$ at high fields}

The LO decrease of $\eta(v)$ with $v$ can radically change the nonlinear dynamics of a trapped vortex and $R_i(B_a)$ at high fields and frequencies  \cite{manula1,manula2}. Taking the LO effect into account raises the following questions: 
1. What happens if the tip of the vortex moves faster than $v_0$ while the rest of the vortex does not?  2. How is the LO instability affected by pinning? 3. How does  the decrease of $\eta(v)$ with $v$ manifest itself in the dependencies of $R_i$ on $B_a$, $\omega$ and the pinning strength? For a vortex segment pinned by a single strong pin, these issues were addressed in Ref. \cite{manula1}, and the effect of artificial pinning centers in films under a dc magnetic field and transport current was investigated in Refs. \cite{lop1,lop2}. The effect of LO instability is quantified by the control parameter $\alpha$ (not to be confused with $\alpha$ given by Eq. (\ref{ab})) \cite{manula1,manula2}:
\begin{equation}
\alpha=(\lambda\omega/2\pi v_0)^2.
\label{alo}
\end{equation}
For $v_0=0.1-1$ km/s, Eq. (\ref{alo}) yields $\alpha=4\times(1-10^{-2})$ at $\lambda=100$ nm and 2 GHz. The increase of $\alpha$ with $\omega$ indicates that manifestations of the LO mechanism become more pronounced at higher frequencies.

\begin{figure}[ht]
	\centering
	\includegraphics[scale=0.27,angle=-90]{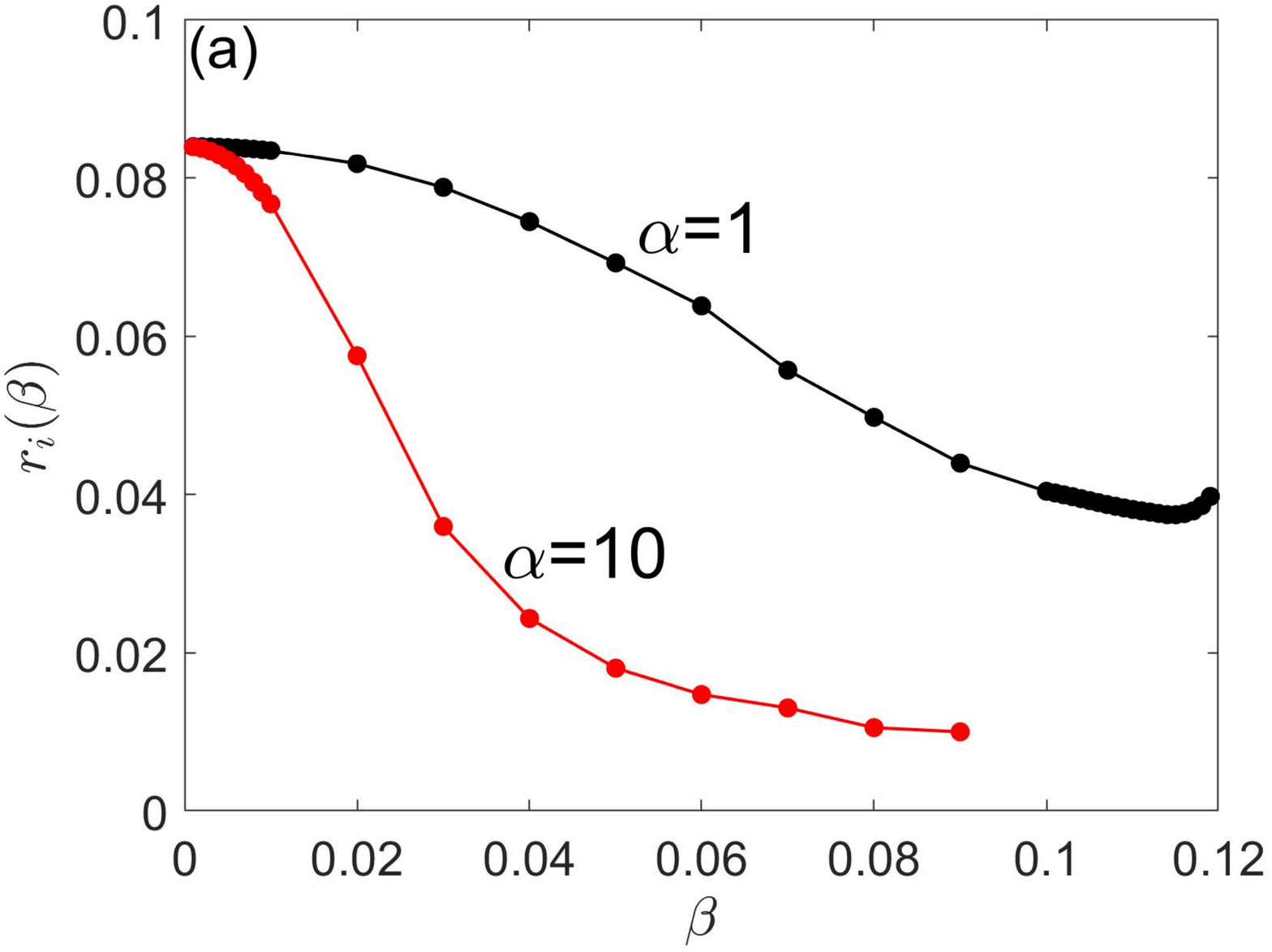}
	\includegraphics[scale=0.27,angle=-90]{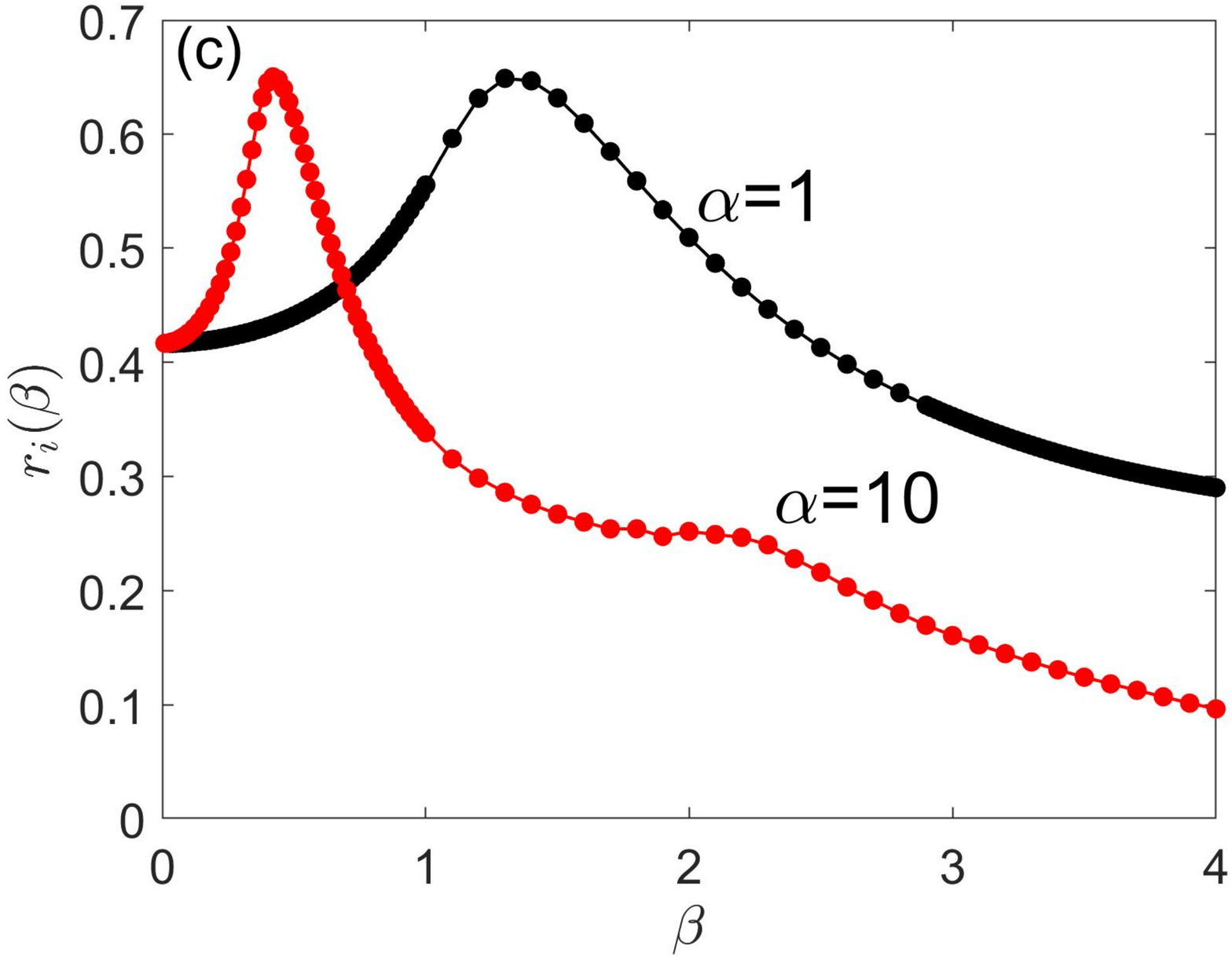}
	\caption{$r_i(\beta)$ as a function of $\beta=B_a/B_{c1}$ calculated for a film of thickness $l=10\lambda$, $\kappa=2$ in the case of bulk pinning with $\zeta_n=0.8$, $n_i=0.25 \lambda^{-3}$, $\alpha=1,10$ and: (a) $\gamma=0.01$, (c) $\gamma_0=1$. Reproduced from Ref. \cite{manula2}.}
	\label{fig19}
\end{figure} 

Shown in Fig. \ref{fig19} is $r_i(\beta)$ calculated for random bulk pinning in a film of thickness $l=10\lambda$ at $\zeta_n=0.8$, $n_i=0.25 \lambda^{-3}$, and different values of $\alpha$ \cite{manula2}.  At the lowest frequency $\gamma=0.01$ and $\alpha=1$, the elastic skin depth  $L_\omega=\lambda/\sqrt{2\pi\gamma}\approx 4\lambda$ is about half of the vortex length. In this case the surface resistance decreases with the rf field due to the LO decrease of the vortex viscosity with $v$. A  similar descending field dependence of $R_i(B_a)$ caused by the LO effect was obtained for a vortex pinned by a single defect \cite{manula1}, the results being independent of the elementary pinning force if the pin is spaced more than $L_\omega$ from the surface.

At higher frequencies the behavior of $r_i(\beta)$ changes as shown in Figs.  \ref{fig19}c. Here $r_i(\beta)$ becomes nonmonotonic, the peaks in $r_i(\beta)$ shifting to lower fields as the LO parameter $\alpha$ increases.  At the peak of $r_i(\beta)$ at $\beta=\beta_m$, the maximum velocity of the vortex tip $v_m$ becomes of the order of $v_0$, but no LO vortex jumps occur because of the restoring effect of the vortex line tension. The increase of $r_i(\beta)$ with $\beta$ at $\beta<\beta_m$ is mostly due to the increase of $L_\omega\sim[\epsilon/\eta(v)\omega]^{1/2}$ caused by the decreasing $\eta(v)$. At $\beta>\beta_m$ the elastic skin depth  $L_\omega$ becomes of the order of the vortex length and $r_i(\beta)$ decreases with $\beta$ due to the decrease of $\eta(v)$ with $v$~ \cite{manula1,manula2}. The peak velocity $v_m$ of the vortex tip increases with $B_a$ and can significantly exceed $v_0$ at high fields. 

As $\beta$ exceeds $\beta_m$, the dynamics of the vortex tip changes from a nearly harmonic oscillations at $\beta< \beta_m$ to highly anharmonic relaxation-type   oscillations at $\beta> \beta_m$, the amplitude of oscillations increasing greatly at $\beta> \beta_m$  \cite{manula2}.  Bending distortions along the vortex are mostly confined within the elastic skin depth  $L_\omega$ which increases with $v$ due to LO reduction of $\eta(v)$ and eventually becomes larger than $l$ at $v_m> v_0$.  The nonlinear dynamics of the vortex at $\beta>\beta_m$ becomes dependent on the vortex mass, the peaks in $r_i(\beta) $ shifting to higher $\beta$ as the $\omega$ increases \cite{manula1,manula2}.

The LO velocity dependence of $\eta(v)$ can produce the residual surface resistance which decreases with the rf field amplitude. Such field-induced reduction of $R_i(B_a)$ results from interplay of vortex elasticity and the LO decrease of the viscous drag with the vortex velocity. The decrease of $R_i(B_a)$ with $B_a$ can contribute to the negative $Q(B_a)$ slopes observed on alloyed Nb cavities \cite{raise1,raise2,raise3,raise4,raise5,raise6,raise7,raise8,raise9}.  Unlike the decrease of  $R_i^{TLS}(B_a)$ or the quasiparticle surface resistance with the rf field, the vortex contribution to $R_i(B_a)$ scales with the density of trapped magnetic flux.  The field-induced reduction of $R_i(B_a)$ produced by trapped vortices becomes more pronounced at higher frequencies, which appears consistent with the experiment \cite{raise7} on  N-doped Nb cavities. As was shown in Ref. \cite{manula1}, the LO mechanism can account for $R_s(B_a)\propto B_a^{-2}$ observed on Nb cavities  \cite{bak1} which cannot be explained by the weaker field dependence of $R_i^{TLS}(B_a)$. 

\subsection{Tuning $R_i$ by impurities}

The significance of vortex losses in high-Q resonators brings to focus the ways of decreasing $R_i$ by materials treatments.  The vortex losses at low fields can be reduced by increasing the volume density and strength of pinning centers.  Consider first a reduction of $R_i$ by alloying a superconductor with nonmagnetic impurities without changing the pinning defect structure. This was addressed in Refs. \cite{manula1,manula2,manula3}  by incorporating the dependencies of $\lambda=g\lambda_0$, $\xi=\xi_0/g$, and the pinning parameter $\zeta_n=2\kappa^2 u_p/\pi\varepsilon\xi=\zeta_{n0}g^5$ on the mean free path $l_i$ into Eqs. (\ref{eq1}) and (\ref{eq3}). Here the conventional factor $g=(1+\xi_0/l_i)^{1/2}$ interpolates the dependencies of superconducting properties on the mean free path \cite{tinkh}, $\zeta_n$ is evaluated for small dielectric inclusions for which $u_p\sim B_c^2r_0^3/\mu_0$ is independent of $l_i$, and the index $0$ labels the parameters of a clean material. The vortex viscosity $\eta$ and the LO parameter $\alpha$ at $l_i<\xi_0$ can be evaluated as $\eta\simeq\eta_0g^2l_i/\xi_0$ and $\alpha\simeq\alpha_0g^2\xi_0/l_i$ using Eq. (\ref{vol}) with $D=l_iv_F/3$ and $\tau_\epsilon$ being independent of $l_i$ (see also Ref. \cite{note}). Hence, alloying the material with $l_i<\xi_0$ results in the following:
\begin{itemize}
\item{Enhances the pinning parameter $\zeta_n\sim\zeta_{n0}(\xi_0/l_i)^{5/2}$ as the vortex core diameter $\xi\sim (l_i\xi_0)^{1/2}$ and the line tension $\varepsilon\sim\varepsilon_0l_i/\xi_0$ decrease. This increases the elementary pinning forces and allows a softer vortex to better accommodate the pins.}
\item{Weakly affects the Bardeen-Stephen vortex drag coefficient $\eta_0$ which becomes independent of $l_i$ at $l_i\ll \xi_0$.}
\item{Facilitates manifestations of the velocity dependence of $\eta(v)$ in the residual resistance $R_i(B_a)$ as the LO parameter $\alpha\sim\alpha_0(\xi_0/l_i)^2$ increases. }
\end{itemize}
 
\begin{figure}[h]
	\centering
	\includegraphics[scale=0.3,angle=-90]{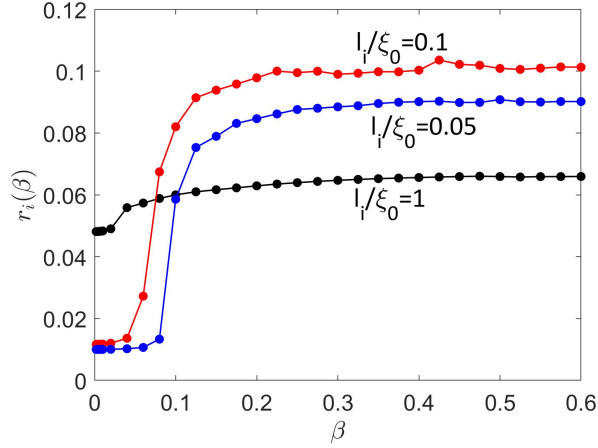}\\
	\caption{ Field dependence of $r_i(\beta) $ on $\beta=B_{a}/B_{c10}$ for a film of thickness $l=10\lambda_0$ at $\kappa_0=2$, $\zeta_{n0}=0.04$, 
	$\gamma_0=0.004$, $n_i=0.5 \lambda_0^{-3}$ and different mean free paths. Reproduced from Ref. \cite{manula3}} 
	\label{fig20}
\end{figure}

The field dependence of $r_i(B_a)$ calculated in Ref. \cite{manula3} for bulk pinning in a film with different mean free paths and the Bardeen-Stephen $\eta_0$ is shown in Fig. \ref{fig20}. The overall behavior of $r_i(B_a)$ is similar to that shown in Fig. \ref{fig18}b: the dip in $r_i(B_a)$ at low field occurs if the sheet Meissner 
current $B_a$ is smaller than the rf depinning field $B_p\sim \mu_0J_pL_\omega$.  At $B_a<B_p$ the vortex undergoes small-amplitude oscillations impeded by 
pinning which reduces rf losses. At $B_a>B_p$ the surface resistance increases significantly as the net Lorentz force exceeds 
the pinning force, and the amplitude of vortex oscillations is primarily determined by the balance of Lorentz and viscous drag forces. The transition from the pinning-dominated to viscous drag dominated regimes shifts to higher fields as pinning becomes more effective upon alloying the material. Yet alloying increases $r_i(B_a)$ at higher fields, although $r_i$ drops a bit as $l_i$ is decreased from $0.1\xi_0$ to $0.05\xi_0$. The latter reflects a nonmonotonic dependence of $R_i$ of a freely 
moving vortex segment on $l_i$: $R_i\propto l_i^{-1/2}$ at $L_\omega >\lambda$ and $R_i\propto l_i^{1/2}$ at $L_\omega <\lambda$ \cite{gc2}.       

\begin{figure}[!htb]
	\centering
		\includegraphics[scale=0.27,angle=-90]{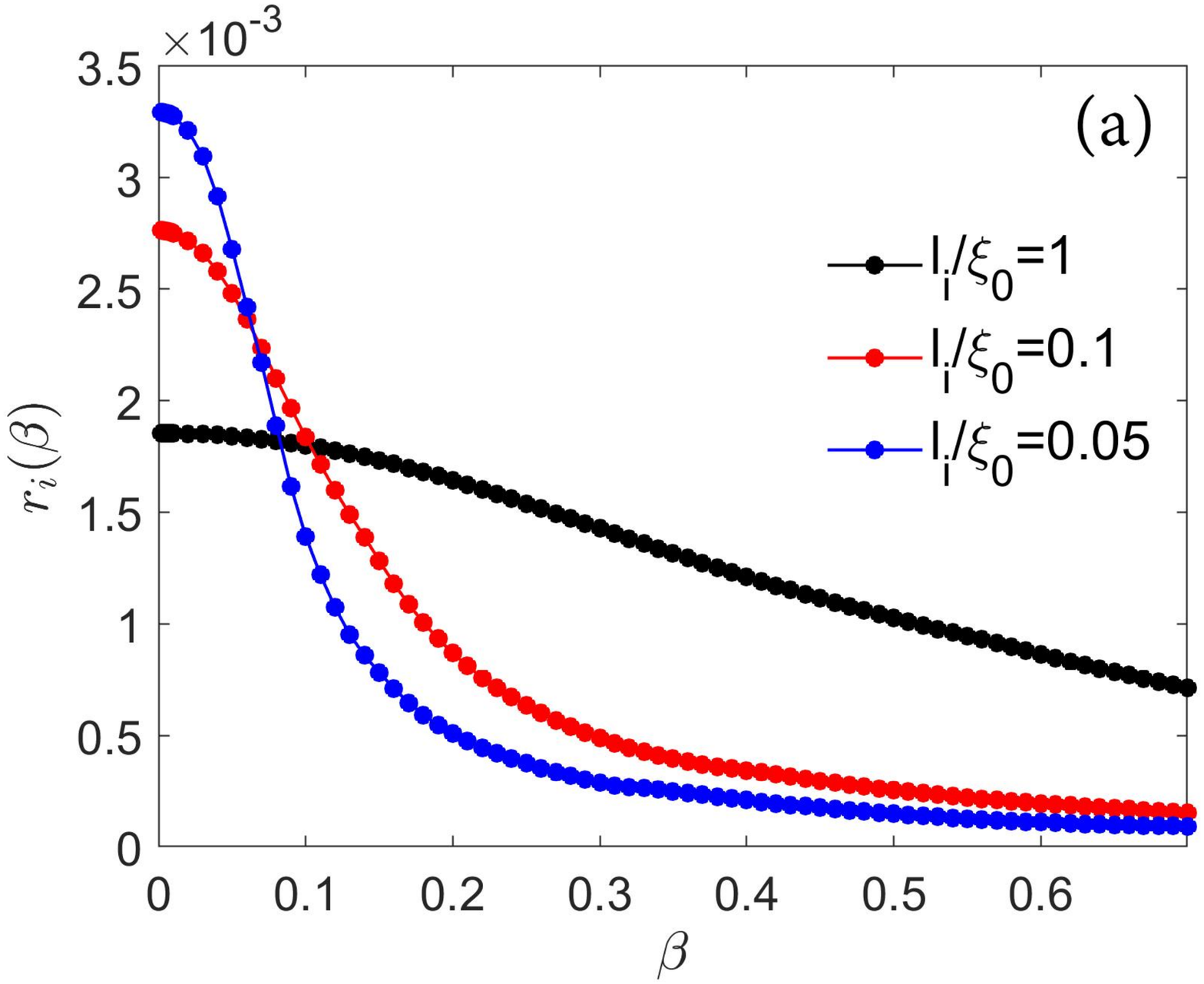}
		\includegraphics[scale=0.27,angle=-90]{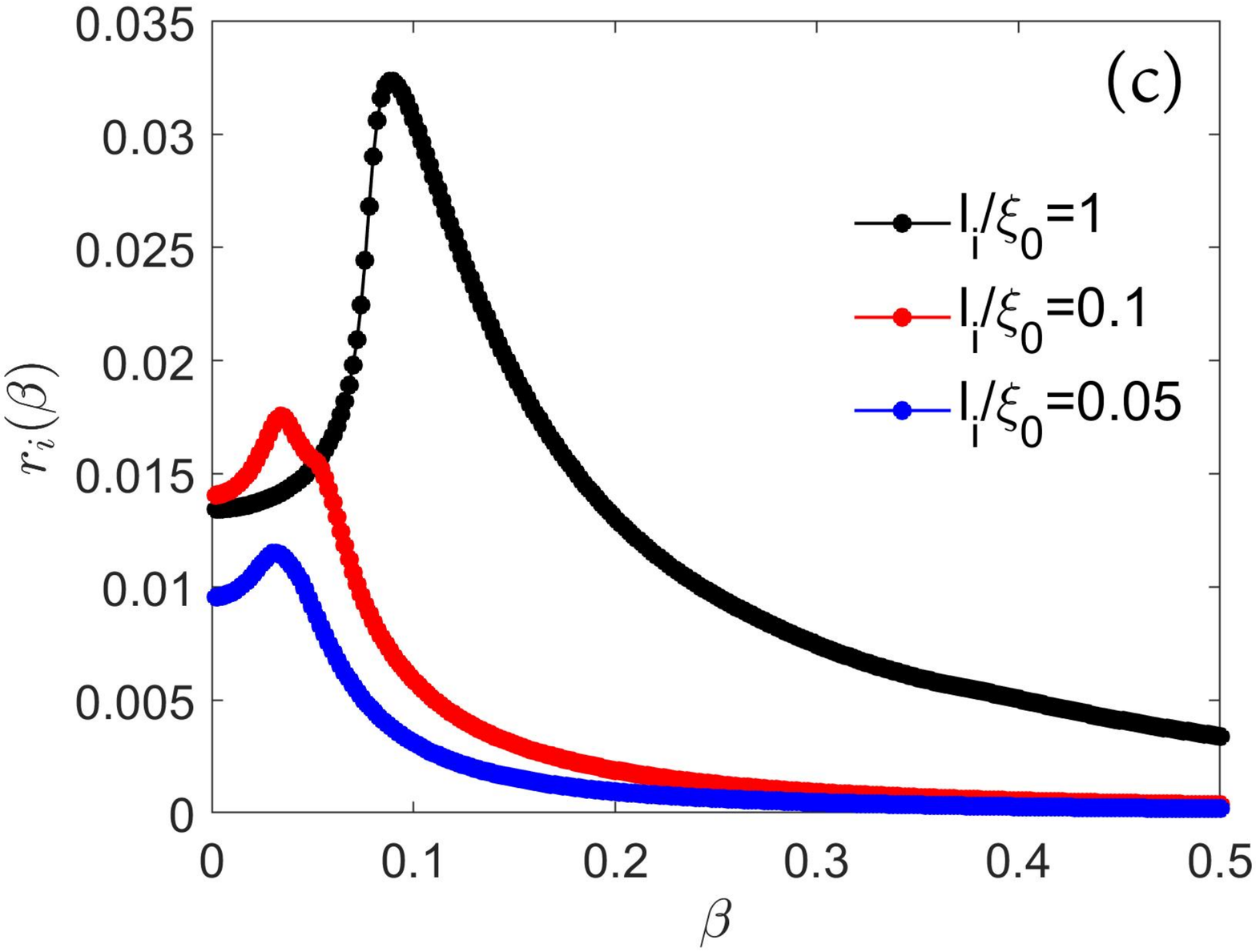}  
	\caption{The surface resistance $r_i(\beta) $ calculated for a vortex pinned by a strong pin located at $l=3\lambda_0$ for different values of $l_i/\xi_0$: (a) $ \gamma_0=0.004$, $\alpha_0=0.1$, (c) $ \gamma_0=0.4$, $\alpha_0=1000$. Reproduced from Ref. \cite{manula1}}
	\label{fig21}
\end{figure}

Figure \ref{fig21} shows the effect of the mean free path on the microwave reduction of $R_i(B_a)$ caused by the LO velocity dependence of $\eta(v)$ for a vortex pinned by a single defect spaced by $l=3\lambda_0$ from the surface \cite{manula1}. One can see that the decrease of $R_i(B_a)$ with $B_a$ becomes stronger and shifts to lower fields upon alloying the material. This reflects the increase of the LO control parameter $\alpha\propto (\omega\xi_0/l_i)^2$ as $\omega$ increases and $l_i$ decreases.  A maximum in $R_i(B_a)$ in Fig. \ref{fig21} appears at high frequencies at which $L_\omega$ becomes smaller than $l$, similar to the evolution of $R_i(B_a,\omega)$ shown in Fig. \ref{fig19}.  The descending dependence of $R_i(B_a)$ at high fields can be masked by overheating \cite{manula2}.       

\subsection{Pinning optimization}  

Low-field rf losses of trapped vortices can be reduced by making pinning more effective using designer pinning structures which have been very effective to increase $J_p$ in superconductors for dc magnets \cite{apc1,apc2}. However, the artificial pinning centers in high-Q resonators can only be used in the form of dielectric precipitates or nano pores because metallic pins (for example, $\alpha$-Ti ribbons in NbTi \cite{apc1}) can produce high rf losses. Consider an upper limit of $J_p$ for strong pinning by small dielectric precipitates or pores of radius $r_0\simeq \xi$ which chop vortex lines into short segments of length $\ell<\lambda$. If the ends of each vortex segments are fixed by the strong pins, $J_p$ is determined by the pin breaking mechanism \cite{pin,agpin} in which vortices bow out under the action of the Lorentz force and escape as the tips of two antiparallel vortex segments at the pin reconnect at the critical current density $J_p\sim \varepsilon/\phi_0\ell$ which increases with the pin density $n_p=\ell^{-3}$ \cite{ehb}. However, too many dielectric pins would block the current-carrying cross section so $J_p(\ell)$ is determined by interplay of pinning and current blocking  \cite{agpin}:   
\begin{equation}
J_p\simeq \frac{\phi_0}{2\pi\mu_0\lambda^2\ell}\ln\frac{\ell}{2\xi}\left(1-\frac{4\pi r_0^3}{3p_c\ell^3}\right).
\label{Jc}
\end{equation}
Here the factor in the parenthesis accounts for the reduction of the current-carrying cross section by dielectric pins and $p_c$ is the percolation threshold which varies from $p_c=1/2$ in two dimensions to $p_c\simeq 2/3$ in the 3D isotropic limit \cite{perc}. Interplay of pinning and current blocking yields a maximum of $J_p(\ell)$ at $\ell_m\simeq (16\pi/3p_c)^{1/3}r_0\simeq 3r_0$, an optimal volume fraction of pins $p_m\simeq 9-12\%$ and the maximum $J_{pm}\simeq (0.2-0.3)J_d$, where the spread of numerical values comes from the shapes of pins and the effect of crystalline anisotropy \cite{agpin}. The maximum in  $J_p$ at $p\sim 10\%$ was also revealed by the TDGL numerical simulation of vortices interacting with metallic pinning centers \cite{anlpin}.  The optimized pinning with $J_p\simeq 0.2J_d$ allows a superconductor with sparse trapped vortices to withstand without significant losses the applied rf field $B_m\simeq \mu_0J_{pm}\lambda\simeq 0.2B_c$. Yet even such idealized pinning structure can only provide   the rf breakdown fields much smaller than $B_b\simeq B_c$ observed on 
Nb cavities \cite{rec1,rec2}. Thus, pinning can only reduce the residual surface resistance at  $B_a\ll B_c$.

The reduction of $R_i$ by increasing the volume fraction of dielectric pins can come at the expense of higher dielectric and quasiparticle rf losses because dielectric precipitates increase the composite magnetic penetration depth $\bar{\lambda}$. This follows from the effective medium theory \cite{emt} which gives the conductivity $\bar{\sigma}=(1-p/p_c)\sigma$ \cite{perc,emt} of a composite comprised of dielectric spherical precipitates of radius $r_0$ and the volume fraction $p=4\pi r_0^3/3\ell^3$ embedded in a matrix with conductivity $\sigma$. At $T\ll T_c$, we have $\sigma_1\ll\sigma_2$ so $\sigma\to i\sigma_2=i/\mu_0\omega\lambda^2$ and $\bar{\sigma}=(1-p/p_c)\sigma=i/\mu_0\omega\bar{\lambda}^2$. Hence, $\bar{\lambda}(p)$ increases with $p$ and diverges at the percolation threshold $p=p_c$: 
\begin{equation}
\bar{\lambda}=\frac{\lambda}{\sqrt{1-p/p_c}}.
\label{eflam}
\end{equation}
According to Eq. (\ref{rs0}) and (\ref{eflam}), the surface resistance $\bar{R}_{BCS}\propto \bar{\lambda}^3$ increases with $p$. For a small fraction of dielectric pins $p\ll p_c$, we have $\bar{\lambda}=\lambda (1+3p/4)$ and
\begin{equation}
\bar{R}_{BCS}=\left(1+3\pi r_0^3/\ell^3\right)R_{BCS}.
\label{rbar}
\end{equation}
At the optimum pin spacing $\ell\simeq 3r_0$ the composite surface resistance $\bar{R}_{BCS}$ is increased by about $30\%$ relative to $R_{BCS}$.
Vortex trapping depends strongly on the sample geometry and is most pronounced in thin films in which the perpendicular $B_{c1}$ is greatly reduced. Vortex losses in thin film coplanar resonators were eliminated by producing an array of microscopic pinholes of radius $r_0>\xi$ which fully absorb the dissipative vortex and block vortex motion \cite{ph1,ph2,ph3}. Such columnar pins result in a very high depinning current density $J_p\sim J_c$ \cite{pin1,pin2}. At the same time, the pinholes can increase the TLS losses at the edges \cite{ph3}.

\section{Conclusions and outlook}

Decreasing microwave losses in high-Q superconducting structures involves dealing with interconnected mechanisms, 
tuning one of them to increase $Q$ can cause others to decrease $Q$. For example, reducing the residual surface resistance which controls the limits of $Q$ at ultra lower temperatures requires minimizing the broadening of the DOS peaks and TLS losses. By contrast, reducing $R_{BCS}(T)$ at intermediate temperatures requires engineering an optimum DOS broadening by tuning the concentration of magnetic impurities or properties of a proximity coupled metallic suboxide at the surface. In turn, the proximity effect can be used to either significantly increase or decrease the kinetic inductance $L_k$ of thin film resonators, although increasing $L_k$ by using granular Al films would also increase TLS losses in oxide inter grain contacts.  Thus, a universal optimization of rf properties of superconductors is hardly possible as different application operating in their respective ranges of temperatures, frequencies and rf fields require different superconducting materials and their treatments.  For instance, accelerator cavities operating at 2K mostly use Nb as the best compromise of many conflicting requirements outlined above. 

Reducing TLS losses is a challenging problem as the atomic origin and microscopic mechanisms of TLS are not fully understood, although O vacancies and 
O-H vacancy complexes in Nb and Al oxides seem to be viable candidates. TLS losses may be decreased by materials treatments which reduce the density of 
excess O vacancies, make the oxide layer thinner and less amorphous and reduce segregation of O vacancies at grain boundaries \cite{tlr1,tlr2,vcox}. In that regard 
the mid-T baking of Nb cavities which reduced the residual resistance below 1 n$\Omega$ \cite{bak6} may offer a new path for optimization of Nb resonators in quantum circuits. Furthermore, TLS losses in thin film resonators in transmons operating at very low fields $H\ll H_c$ can be reduced by using type-I superconductors with less complex surface oxide structure than Al or Nb. For instance, the use of tantalum thin film resonators with $T_c\approx 4.3$ K and $B_c\approx 83$ mT \cite{smat} can 
increase the $Q$ factor by 2-4 times as compared to Nb or Al resonators \cite{Ta1,Ta2}.     

The detrimental effect of trapped vortices can be mitigated by optimizing pinning to reduces vortex losses at $H_a\ll H_c$. The field region of the Meissner state can be extended by SIS multilayers which also reduce vortex losses, although the I interlayers may increase TLS losses. On the other hand, strong pinning increases the number of vortices trapped during the cooldown through $T_c$ since it prevents vortices from escaping the film as the temperature decreases and $H_{c1}(T)$ becomes larger than $H$.    Another way of reducing the losses of trapped vortices would be to get rid of them by better magnetic shielding combined with cooling the films in temperature gradients to push the maximum number of trapped vortices out by using scanning laser \cite{gc2,laser,lsm1,lsm2} or electron \cite{esm1,esm2} beams before they get stuck on pinning centers.     

In addition to their important applications, the high-Q superconducting resonators can be used to probe the fundamental limits of dissipation and dynamic superheating field in superconductors at low temperatures. These issues involve outstanding theoretical problems related to the mechanisms of subgap states and nonlinear response of nonequilibrium superconductors in strong rf fields.  Furthermore, high-Q resonators can be testbeds for probing extreme nonlinear dynamic of vortices driven by strong rf Meissner currents, particularly the terminal velocity and nonequilibrium processes in the rapidly moving vortex core which are masked by heating to a much lesser extent than in dc transport experiments.                 

\ack
This work was supported by DOE under Grant DE-SC 100387-020.


\newcommand{\newblock}{}

\end{document}